\documentclass{aa}  
\usepackage{graphicx}
\usepackage{txfonts}
\usepackage{hyperref}

\usepackage{amsmath}
\usepackage{amssymb}
\usepackage{multirow}
\usepackage{upgreek}

\newcommand{\vpeak}{V_{\rm peak}}
\newcommand{\vmax}{V_{\rm max}}

\newcommand{\sigmaLogM}{\sigma_{\rm logM}} 
\newcommand{\Mone}{M_{\rm 1}}
\newcommand{\Mmin}{M_{\rm min}} 
\newcommand{\Mcut}{M_{\rm cut}} 

\newcommand{\incomp}{ {\rm f_{ic}}}

\newcommand{\Acen}{{\rm OV}_{\rm cen}}
\newcommand{\Asat}{{\rm OV}_{\rm sat}}
\newcommand{\Bcen}{{\rm GAB}_{\rm cen}}
\newcommand{\Bsat}{{\rm GAB}_{\rm sat}}

\newcommand{\Sseg}{{\rm s}_{\rm segr}}

\newcommand{\VBcen}{\alpha_{\rm cen}}
\newcommand{\VBsat}{\alpha_{\rm sat}}

\newcommand{\sigL}{\sigma_{\rm lum}}
\newcommand{\tmerger}{t_{\rm merger}}

\newcommand{\FkP}{f_{\rm k,cen+sat}}
\newcommand{\FkC}{f_{\rm k,cen}}
\newcommand{\FkS}{f_{\rm k,sat}}
\newcommand{\FkM}{f_{\rm k,cen-sat}}
\newcommand{\betaL}{\beta_{\rm lum}}

\newcommand{\hMsun}{ h^{-1}{\rm M_{ \odot}}}
\newcommand{\hMpc}{ h^{-1}{\rm Mpc}}
\newcommand{\hGpc}{ h^{-1}{\rm Gpc}}

\newcommand{\ihMpcC}{ h^{3}{\rm Mpc}^{-3}}
\newcommand{\hkpc}{ h^{-1}{\rm kpc}}

\newcommand{\sig}{\sigma_{8}}
\newcommand{\OmM}{\Omega_\mathrm{M}}
\newcommand{\Omb}{\Omega_{\rm b}}

\newcommand{\ns}{{n_{\rm s}}}

\newcommand{\hod}{HOD}
\newcommand{\shame}{SHAMe}
\newcommand{\flamingo}{FLAMINGO}
\newcommand{\sampleA}{DESI BGS-like}
\newcommand{\sampleB}{BOSS-like}
\newcommand{\mr}{${\rm M}_{r}$}

\newcommand{\lineB}{$\rm w_{p}+\Delta \Sigma$}
\newcommand{\lineC}{$\rm w_{p}+\xi_{\ell=0,2}+\Delta \Sigma$}

\newcommand{\proj}{${w_{\rm p}}$}
\newcommand{\mono}{$\xi_{\ell=0}$}
\newcommand{\quadr}{$\xi_{\ell=2}$}

\newcommand{\lensing}{$\Delta \Sigma$}
\newcommand{\skNN}{$\rm P(N_{k})$}
\newcommand{\sCIC}{$\rm P(r_{CIC})$}
\newcommand{\sVPF}{$\rm P_0$}

\newcommand{\GalaxyEmu}{{\tt GalaxyEmu-Planck}}
\begin{document} 

    \title{Validating the clustering predictions of empirical models with the FLAMINGO simulations}

   \author{Sergio Contreras
          \inst{1}\fnmsep\thanks{E-mail: sergio.contreras@dipc.org}
          \and
          Raul E. Angulo
          \inst{1,2}\fnmsep\thanks{E-mail: reangulo@dipc.org}
          \and
          Jon\'{a}s Chaves-Montero
          \inst{3}\fnmsep\thanks{E-mail: jchaves@ifae.es}
          \and
          Roi Kugel
          \inst{4}
          \and
          Matthieu Schaller
          \inst{5,4}
          \and
          Joop Schaye
          \inst{4}
          }

   \institute{Donostia International Physics Center, Manuel Lardizabal Ibilbidea, 4, 20018 Donostia, Gipuzkoa, Spain
         \and
    IKERBASQUE, Basque Foundation for Science, 48013, Bilbao, Spain
    \and
    Institut de F\'isica d'Altes Energies, The Barcelona Institute of Science and Technology, Campus UAB, E-08193 Bellaterra, Barcelona, Spain
    \and
    Leiden Observatory, Leiden University, PO Box 9513, 2300 RA Leiden, the Netherlands
    \and
    Lorentz Institute for Theoretical Physics, Leiden University, PO box 9506, 2300 RA Leiden, the Netherlands
    }

   \date{Received July 1, 2024; accepted XXXX}

 \titlerunning{The performance of empirical models}

  \abstract
   {Mock galaxy catalogues are essential for correctly interpreting current and future generations of galaxy surveys. Despite their significance in galaxy formation and cosmology, little to no work has been done to validate the predictions of these mocks for high-order clustering statistics. }
   {We compare the predicting power of the latest generation of empirical models used in the creation of mock galaxy catalogues: a 13-parameter Halo Occupation Distribution (HOD) and an extension of the SubHalo Abundance Matching technique (SHAMe).}
   {We build \GalaxyEmu, an emulator that makes precise predictions for the two-point correlation function, galaxy-galaxy lensing (restricted to distances greater than 1$\hMpc$ in order to avoid baryonic effects), and other high-order statistics resulting from the evaluation of SHAMe and HOD models.}
   {We evaluate the precision of \GalaxyEmu~ using two galaxy samples extracted from the FLAMINGO hydrodynamical simulation that mimic the properties of DESI-BGS and BOSS galaxies, finding that the emulator reproduces all the predicted statistics precisely. The HOD showed comparable performance when fitting galaxy clustering and galaxy-galaxy lensing. In contrast, the \shame~model showed better predictions for higher-order statistics, especially regarding the galaxy assembly bias level. We also tested the performance of the models after removing some of their extensions, finding that we can withdraw two (out of 13) of the HOD parameters without a significant loss of performance.}
   {The results of this paper validate the current generation of empirical models as a way to reproduce galaxy clustering, galaxy-galaxy lensing and other high-order statistics. The excellent performance of the \shame~model with a small number of free parameters suggests that it is a valid method to extract cosmological constraints from galaxy clustering.}


\maketitle

\section{Introduction}
\label{sec:Intro}

Understanding the galaxy-halo connection is one of the biggest challenges of modern astrophysics. In the current galaxy formation paradigm, galaxies are born from the baryonic matter trapped in the gravitational potential of dark matter haloes. The evolution of galaxies is linked to the evolution of their host dark matter haloes, as both grow hierarchically. Over the past two decades, several relationships have been found between galaxy and halo properties. We have used these relationships not only to enhance our understanding of galaxy formation physics, but also to construct mock galaxy catalogues. 

With a new generation of galaxy surveys (e.g., DESI, EUCLID, LSST, 4MOST, J-PAS, SKA), the need for realistic mock galaxy catalogues is critical to correctly interpret these observations, as well as for extracting their cosmological and galaxy formation information. One of the most complex and arguably realistic ways to model galaxies is through hydrodynamic simulations (e.g., \citealt{Schaye:2015, Vogelsberger:2014b, HorizonAGN, SIMBA, TNGb, McCarthy:2018}). These simulations track the evolution of dark matter and baryonic matter simultaneously, which requires a considerable amount of computational resources compared to only simulating dark matter. This additional computational cost means that hydrodynamic simulations have a smaller volume or lower resolution compared to dark matter-only simulations. During the last year, two suites of simulations have set a new standard for large-volume hydrodynamic simulations: the MTNG suite of simulations \citep{Pakmor:2023} and the FLAMINGO suite of simulations \citep{Schaye:2023, Kugel:2023}. The largest \flamingo~simulation has a box size of more than $1.9 \hGpc$ (2.8 $\rm Gpc$), an unprecedented volume for this type of simulation.

Unfortunately, the number of simulations of these two suites is not enough to fulfil the required number of mocks needed for the upcoming generation of surveys. Other alternatives to model galaxies, such as semi-analytical models (SAM, e.g., \citealt{ Croton:2016, Lagos:2018, Stevens:2018, Henriques:2020, Perez:2023}) and semi-empirical models ( e.g., \citealt{Moster:2018, Behroozi:2019}), offer a valid alternative to following the evolution of galaxies. Because of their lower computational requirements, these models can be run on dark matter-only simulations in a fraction of the time a hydrodynamic simulation takes to run, offering one of the best ways to study galaxy formation and evolution on really large scales. Unfortunately, these models can still be too slow to be run on large cosmological volumes, and they do not always provide enough flexibility in their astrophysical prescriptions, which limits their predictive power.

Another alternative for modelling galaxies is empirical models. These models use fundamental relations from the galaxy-halo connection to populate galaxies in the haloes and subhaloes of dark matter-only simulations. Although empirical models are not entirely capable of tracking the evolution of galaxies, they do possess the adaptability to accurately reproduce the galaxy clustering of different kinds of galaxy samples in a highly precise way (e.g., \citealt{Guo:2015a, ChavesMontero:2016}). These models can populate large cosmological volumes on a standard laptop or a small computer cluster in minutes if not seconds. Two of the most popular empirical models today are the halo occupation distribution and the subhalo abundance matching technique.

The Halo Occupation Distribution (HOD, \citealt{Jing:1998a, Benson:2000, Peacock:2000, Berlind:2003, Zheng:2005, Zheng:2007, C13, Guo:2015a, C17}) characterises the mean number of galaxies that populate a halo as a function of the halo mass. This technique was developed to interpret galaxy clustering (e.g., \citealt{Zehavi:2011, Guo:2015}) and later expanded into the creation of mock galaxy catalogues (e.g., \citealt{Grieb:2016}). Initially, this method relied on three to five parameters to define its parametric shape \citep{Zehavi:2005, Zheng:2005} and was meant to be used along with the halo model (e.g., \citealt{Berlind:2002, Berlind:2003}). However, it has since evolved to populate dark matter-only simulations and include most of the processes we know affect galaxy clustering, such as assembly bias \citep{Hearin:2016, Xu:2021b}, velocity bias \citep{Guo:2015a}, and satellite segregation \citep{Yuan:2018}. One of the primary advantages of the HOD technique is that it typically only requires information from dark matter haloes, not subhaloes or merger trees. These requirements imply that low-resolution simulations can be used to generate HOD mocks, contrary to other mock galaxy algorithms.

Another popular technique to create mock galaxy catalogues is the SubHalo Abundance Matching (SHAM, \citealt{Vale:2006, Conroy:2006}). In its original form, this technique matched the most massive/fastest-rotating haloes to the most massive/brightest galaxies. Later improvements to the SHAM technique used subhaloes instead of haloes by calculating the subhalo's peak mass or rotational velocity during its evolutionary history. The model had a single free parameter that added scatter between the subhalo and galaxy properties. With these two inclusions, the SHAM can reproduce accurate galaxy clustering predictions in real and redshift space \citep{ChavesMontero:2016}. The reason for this technique's success is that it uses the positions and velocities of the dark matter simulation's subhaloes to assign the positions and velocities of satellite galaxies, which is not possible in a standard HOD. This improvement, however, comes at the additional cost of only being able to create SHAM mocks using intermediate/high-resolution simulations.

Empirical models are used to interpret clustering data (e.g., \citealt{Zehavi:2011, Guo:2015}), characterise different galaxy samples (\citealt{C13, Yuan:2022b}), create mock galaxy catalogues (e.g., \citealt{Grieb:2016}), examine galaxy assembly bias effects (e.g., \citealt{Zehavi:2018, Salcedo:2022}) and constrain cosmological parameters (e.g., \citealt{Cacciato:2013, More:2015, AEMULUS3, Miyatake:2022, Yuan:2022b, AEMULUS5}). Aside from their numerous applications, few studies have attempted to compare their accuracy to more complex (and potentially realistic) mock galaxy samples, such as those predicted by hydrodynamic simulations. These simulations provide us with galaxy clustering statistics that are not subject to observational uncertainty and have a defined cosmology, implying that any differences between the predictions of the empirical models will be due to limitations in their implementation. 

In this paper, we test the performance of two state-of-the-art empirical models: a 13-parameter HOD (based on the work of \citealt{Yuan:2022b}) and a 5-parameter extension of the SHAM technique (SHAMe, \citealt{C21a, C21c}). We compare their clustering predictions (the projected correlation function and the multipoles of the correlation function) with those of the \flamingo~suite of simulations. We also expand this comparison to include higher-order statistics like galaxy-galaxy lensing, count-in-cylinders, the void probability function, and k-nearest neighbours, as well as derivate statistics like the galaxy occupation number and galaxy assembly bias. To facilitate the comparison with \flamingo, we build an emulator called \GalaxyEmu, capable of reproducing all these statistics in a fraction of a second for both empirical models. We find that both models perform well in reproducing most of these statistics. These findings will help validate both empirical models while determining their accuracy in each of these statistics.

The outline of this paper is as follows: Section~\ref{sec:simulations} presents \flamingo~and the dark matter simulations used in this work. The empirical models, the clustering statistics we compute, and \GalaxyEmu~are described in Section~\ref{sec:models}. The main results of this paper, the measurement of galaxy clustering, galaxy-galaxy lensing, and the rest of the statistics are presented in Section~\ref{sec:results}. In Section~\ref{sec:simple}, we test the importance of the different extensions of both empirical models when reproducing galaxy clustering and galaxy-galaxy lensing. We finish by presenting our conclusions in Section~\ref{sec:summary}.

Unless otherwise stated, the standard units in this paper are $\hMsun$ for masses and $\hMpc$ for distances. All logarithm values are in base 10.

\section{Numerical simulations \& galaxy samples}
\label{sec:simulations}

In this section, we first describe the \flamingo~suite of simulations used to test our models (Sect.~\ref{sec:FLAMINGO}) and then the galaxy samples we extract from these simulations (Sect.~\ref{sec:sample}). Finally, we describe the dark matter-only simulation we used to build our mocks (Sect.~\ref{sec:Planck1Gpc}).

\subsection{The FLAMINGO simulations}
\label{sec:FLAMINGO}

We used galaxy samples from the \flamingo~suite of hydrodynamic simulations \citep{Schaye:2023, Kugel:2023, McCarthy:2023, Braspenning:2024, Elbers:2024, Broxterman:2024} to test the performance of our empirical models. This is one of the largest suites of hydrodynamic simulations available today. The simulations were run using the SWIFT code \citep{Schaller:2024}, a highly efficient gravity and smoothed particle hydrodynamics (SPH) solver using the SPHENIX SPH implementation \citep{Borrow:2022}. The simulation includes most of the astrophysical processes we expect to affect galaxies in the Universe, such as radiative cooling and heating, stellar and AGN feedback, chemical enrichments, and others. In this paper, we mainly used the galaxies from the largest box of FLAMINGO, with a box length of $1906.8 \hMpc$ ($\rm 2800 Mpc$) and a resolution of $2\times 5040^3$ particles (same number of dark matter and initial gas particles), equivalent to a mass of $3.85\ 10^9 \hMsun$ and $7.29\ 10^8 \hMsun$ for the dark matter and the initial gas particles, and $2800^3$ neutrino particles. The cosmology of these simulations is $\OmM$ = 0.306, $\Omb$ = 0.0486, $\sig$ = 0.807, $\ns$ = 0.967, $h$ = 0.681, and $\Omega_{\nu} = 1.38\ 10^{-3}$, with this last value being the neutrino matter density parameter ($\Omega_{\nu} \sim \sum m_{\nu}c^2/(93.14h^2\ eV)$). The simulation used partially fixed initial conditions, effectively reducing its cosmic variance \citep{Angulo:2016}. 
Unless we specify otherwise, we will refer to this simulation as ``FLAMINGO''. 

In addition to the largest hydrodynamic simulation, we also used a set of the \flamingo~simulations run with the same cosmology and resolution but a smaller volume and different astrophysics implementations. These simulations have a box length of $681 \hMpc$ ($\rm 1000 Mpc$) and $2 \times 1800^3$ particles (the same amount of dark matter and initial gas particles) and $1000^3$ neutrino particles (i.e., the same resolution as the larger volume). We used these simulations to test the validity of our main findings for other galaxy formation prescriptions. In addition to a simulation with the same calibration as the biggest hydrodynamic run, which we will refer to as the ``fiducial'' model, this sample has two other groups of simulations. The first group comprises six distinct simulations, calibrated similarly to the fiducial model but incorporating a modified observed cluster gas fraction and/or observed stellar mass function in the calibration process. These simulations are ``$\rm fgas+2\sigma$'', ``$\rm fgas-2\sigma$'', ``$\rm fgas-4\sigma$'', ``$\rm fgas-8\sigma$'', ``$\rm M^{*}-\sigma$'' and ``$\rm M^{*}-1\sigma\ fgas-4\sigma$''. Their names indicate the number of sigmas the observed gas fraction ($\rm fgas$) and the observed stellar mass function ($\rm M^{*}$) were shifted during the calibration \citep{Kugel:2023}. Instead of using the thermal AGN model in the fiducial method, the remaining simulations employ jet-like AGN feedback \citep{Husko:2023} instead of a thermally driven AGN feedback \citep{Booth:2009}. These simulations are called ``$\rm Jet$'' and ``$\rm Jet\ fgas-4\sigma$'', with the last case also including a shift in the observed cluster gas fraction by 4 sigmas (see table 2 of \citealt{Schaye:2023} for a list of all these simulations). 

The haloes and substructures of the simulations are identified using the \textsc{VELOCIraptor} subhalo finder algorithm \citep{Elahi:2019}. This approach first identifies the haloes using a 3D Friend-of-Friend algorithm ({\tt FOF}) with a linking length of $l$ = 0.2 of the mean interparticle separation \citep{Davis:1985}. Substructures are then identified within haloes by performing an iterative 6D {\tt FOF} search in phase space and including all particle types except neutrinos. Finally, galaxy properties are computed using the Spherical Overdensity and Aperture Processor (SOAP), a tool developed for the FLAMINGO project. This paper uses the galaxy properties computed by SOAP in apertures of 50 $\rm kpc$, excluding other substructures and unbound particles.

\subsection{Target galaxy samples}
\label{sec:sample}

\begin{figure}
\includegraphics[width=0.49\textwidth]{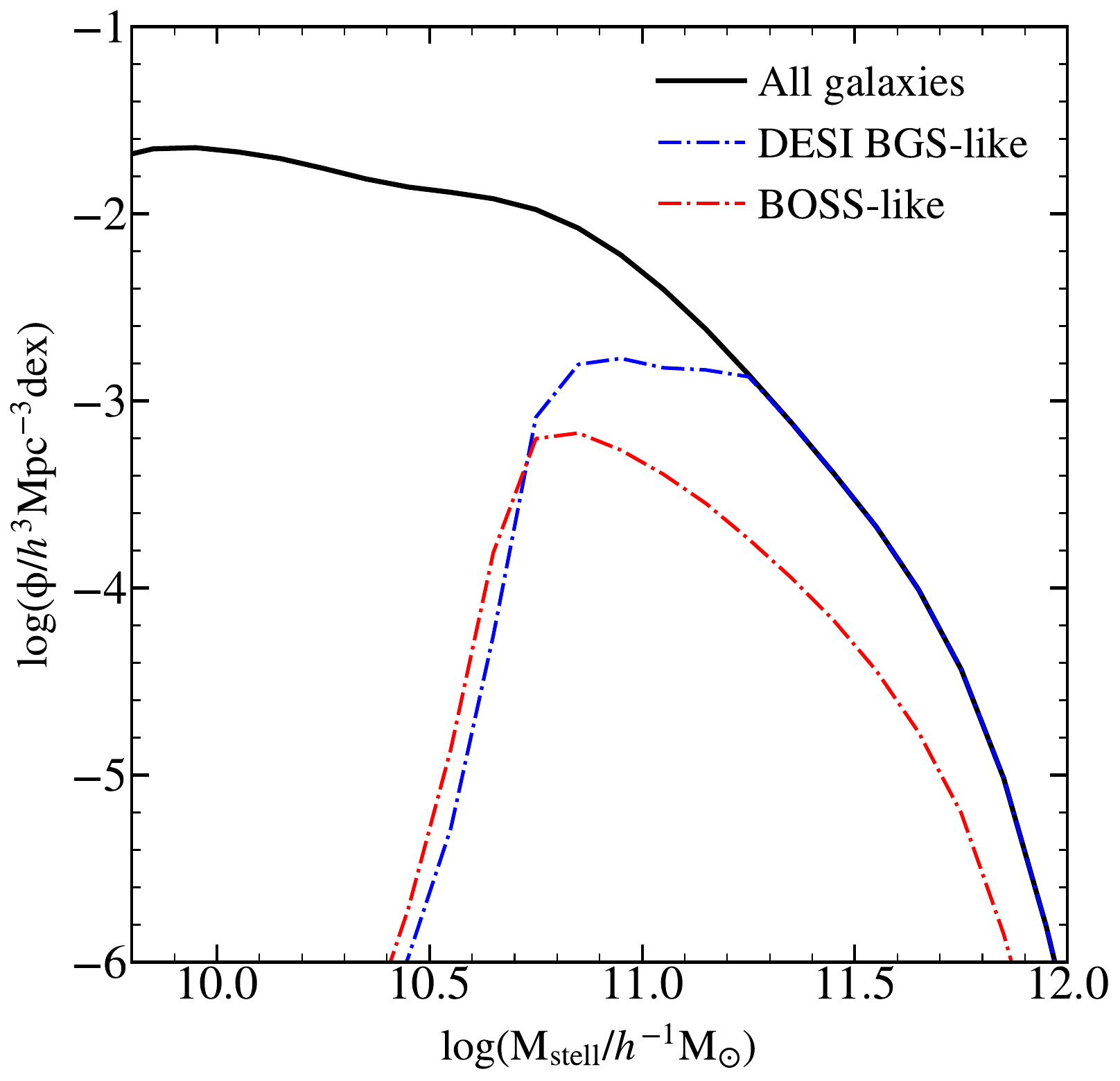}
\caption{The stellar mass function of the \flamingo~simulation at $z=0.4$. The solid black line represents all the galaxies from the simulation, while the blue and red lines represent the galaxies from the \sampleA~and \sampleB~samples. As explained in section~\ref{sec:sample}, the \sampleA~sample is complete in \mr~with a number density of $10^{-3}$ $\ihMpcC$ while the \sampleB~sample is incomplete and red with a number density of $10^{-3.5}$ $\ihMpcC$.}  
\label{fig:smf}
\end{figure}

Using the galaxies of \flamingo, we built two different samples that resemble the selection function of two galaxy surveys. The first selection is an incomplete red sample that resembles the BOSS-LOWZ galaxies \citep{Dawson:2013}. We select the galaxies of \flamingo~following the selection procedure described in \cite{Chaves:2023}. This approach selects the most massive LRGs based on the galaxies' colour and magnitude. This sample is built at $z = 0.4$ and has a number density of approximately $3.16\ 10^{-4}\ihMpcC$. We refer to this selection as the \sampleB~sample.

The second galaxy sample is complete in \mr~with a number density of $10^{-3}\ihMpcC$ also at $z = 0.4$. This number density is similar to the one we expect from the DESI-BGS \citep{Hahn:2023} galaxies at $z = 0.4$. While the mean redshift (number density) of the DESI-LRG is expected to be lower (higher), this sample allowed us to have a more straightforward comparison with the \sampleB~galaxies while using a number density allowed by the resolution of \flamingo. We call this the \sampleA~sample. 

We used both samples with the main \flamingo~simulation, while we only used the \sampleA~galaxies when using the suite of simulations with different astrophysical prescriptions. We facilitate the comparison between samples with different colour distributions. We would like to emphasise that this paper does not aim to make precise observational samples but rather to show the performance of the empirical models when reproducing galaxies obtained using different selection functions.

Fig~\ref{fig:smf} shows the stellar mass function of the \sampleA~and \sampleB~samples of the \flamingo~simulation. As expected, the galaxies from \sampleA~are complete until a specific stellar mass value, and then their abundance decreases (due to the scatter in the stellar mass-\mr~relation), while the \sampleB~sample is incomplete for all stellar masses.

\subsection{The Planck-1Gpc simulation}
\label{sec:Planck1Gpc}

In this paper, we run our empirical models on a pair of gravity-only simulations with a ``Planck~cosmology''\footnote{$\OmM$ = 0.3089, $\Omb$ = 0.0486, $\sig$ = 0.8159, $\ns$ = 0.9667 and $h$ = 0.6774}. The pair of simulations were run with opposite initial Fourier phases, using the procedure of \cite{Angulo:2016}, which suppresses cosmic variance by up to 50 times when compared to a simulation with random phases for the same volume. The simulations have a volume of $(1024 \hMpc)^3$, a resolution of $3072^3$ particles, and were carried out using an updated version of {\tt L-Gadget3} \citep{Angulo:2012}, a lean version of {\tt GADGET} \citep{Springel:2005} used to run the Millennium XXL simulation \citep{Angulo:2012} and the Bacco Simulations \citep{Angulo:2021}. This implementation enables on-the-fly identification of haloes and subhaloes using a Friend-of-Friend algorithm and uses an extended version of {\tt SUBFIND} \citep{Springel:2001} that better identifies substructures by considering their past history. In addition, this version of {\tt SUBFIND} provides properties that are non-local in time, such as the maximum circular velocity ($\vmax \equiv {\rm max}\sqrt{GM(<r)/r}$) attained by a subhalo during its evolution, $\vpeak$. The code is also capable of identifying orphan substructures, i.e.~satellite structures with known progenitors that the simulation cannot resolve but that are expected to exist in the halo, by tracing the most bound particle of all subhalos after these cannot be identified.

\section{Galaxy Models}
\label{sec:models}

\begin{table*}
\caption{The description of the \shame~and \hod~parameters. }  
\label{table:mock_param}
  \begin{tabular}{l|l}
  \multicolumn{2}{c}{SHAMe} \\
    \hline
    \hline
      Parameter & Description \\
    \hline
$\sigL$ & The scatter between the subhalo property ($\vpeak$) and the luminosity of the galaxies  \\ 
$\tmerger$ & Dimensionless free parameter that effectively regulates the number of orphan galaxies \\
$\FkC$ & Assembly bias parameter for central galaxies \\
$\FkS$ & Assembly bias parameter for satellite galaxies \\
$\betaL$ & The timescale parameter that determines how long satellite galaxies survive inside their host haloes \\
\hline
\multicolumn{2}{c}{ } \\
\multicolumn{2}{c}{HOD} \\
    \hline
    \hline
      Parameter & Description \\
    \hline
$\Mmin$ & The halo mass at which half of the haloes are populated by a central galaxy  \\ 
$\sigmaLogM$ & The smoothness of the transition from zero to one central galaxy per halo\\
$\Mone$ & The mass at which there is, on average, one satellite galaxy per halo \\
$\Mcut$ & The minimum halo mass for hosting satellites \\
$\alpha$ & The power-law slope of the halo mass/number of satellite galaxies relation \\
$\incomp$ & Incompleteness parameter \\
$\VBcen$ & Velocity bias parameter for central galaxies \\
$\VBsat$ & Velocity bias parameter for satellite galaxies \\
$\Acen$ & Occupancy variation parameter by halo concentration for central galaxies \\
$\Asat$ & Occupancy variation parameter by halo concentration for satellite galaxies \\
$\Bcen$ & Assembly bias parameter for central galaxies \\
$\Bsat$ & Assembly bias parameter for satellite galaxies \\
$\Sseg$ & Satellite segregation parameter \\
\hline
  \end{tabular}
\end{table*}

In this section, we first describe the empirical models we use to create the mocks: the SHAMe model (Sect.~\ref{sec:SHAMe}) and the HOD model (Sect.~\ref{sec:HOD}). The description of the clustering statistics we measure from \flamingo~and the empirical models are presented in Sect.~\ref{sec:clustering}. We describe the process of fitting the data of \flamingo~in Sect.~\ref{sec:fit}.
We finalise by presenting \GalaxyEmu~in Sect.~\ref{sec:GalaxyEmu}.

\subsection{SHAMe}
\label{sec:SHAMe}

The {\bf S}ub-{\bf H}alo {\bf A}bundance {\bf M}atching {\bf e}xtended model (SHAMe) developed by \cite{C21a, C21c} is an extension of the standard SHAM approach \citep{Vale:2006, Conroy:2006, Reddick:2013, C15, Guo:2016, ChavesMontero:2016, Lehmann:2017, Dragomir:2018,  Hadzhiyska:2021b, Favole:2022} which includes a more sophisticated treatment of the satellite galaxies as well as a flexible level of galaxy assembly bias. 

As in the standard SHAM approach, we begin by matching a subhalo property (in this case, $\vpeak$) to the expected luminosity function. We include a log-normal noise element in the matching process to reproduce the expected scatter between $\vpeak$ and the luminosity of galaxies. For galaxy samples selected according to a number density (instead of a cut in luminosity), this scatter only fulfils the purpose of assigning a physical unit to the value of the scatter of the SHAMe model; the choice of luminosity function has no significant effect on galaxy clustering statistics \citep{C21c}. 

The number of satellite galaxies is set by a free parameter that removes substructures that have been satellites for long periods of time. This process intends to reproduce the disappearance of satellite galaxiesIn this case, we into the intra-cluster medium before they can merge with their central galaxy. 
The number of orphan satellite galaxies is controlled by a free parameter that is compared with the ratio of the time since accretion and the dynamical friction timescale (based in Eq. 7.26. of \citealt{BT:1987}). If this ratio exceeds our threshold, we assume the orphan merged with its central galaxy. 

The final step in the SHAMe implementation involves incorporating additional galaxy assembly bias \citep{Croton:2007}, which is the impact of the evolutionary history of their host haloes on galaxy clustering. This effect is introduced using two free parameters (one for centrals and one for satellites), which modulate the luminosity of the galaxies based on their large-scale bias. To define this bias, we used the individual bias-per-object of the galaxies \citep{Paranjape:2018}, corresponding to the cross-correlation between a given point in space and the dark matter density field. We computed this property by measuring the power spectrum around each object on scales between $0.08 < {\rm k/}\ h\ {\rm Mpc^{-1}} < 2$. 

The positions and velocities of the galaxies in \shame~are equal to the positions and velocities of their assigned subhaloes. The velocity of the subhaloes is computed as the average of the inner 10\% of the particles. We describe all the \shame~parameters in the top part of Table~\ref{table:mock_param}. For a more detailed description of this model, we refer the reader to \cite{C21c, C23c}

\subsection{Halo Occupation Distribution}
\label{sec:HOD}

The HOD describes the average number of galaxies that populate haloes as a function of halo mass, $\langle N(M_{\rm h}) \rangle$. This relation can be parametrized independently for centrals ($\langle N_{\rm cen}(M_{\rm h})\rangle$) and satellites ($\langle N_{\rm sat}(M_{\rm h})\rangle$), where its parametric form is motivated by the occupation distribution of semi-analytical models and hydrodynamic simulations  \citep{Zheng:2005}. 

In this paper, we use the base model of \cite{Zheng:2005}, along with the model extensions presented in \cite{Yuan:2022b}. These extensions include the relocation of satellite galaxies to the inner or outer part of the halo (satellite segregation, 1-free parameter), the modulation of the velocities of galaxies (velocity bias, 2-free parameters), galaxy assembly bias via environment (2-free parameters) and concentration (also known as occupancy variation, 2-free parameters); and incompleteness due to the selection of galaxies (1 free parameter). 

We describe all the \hod~parameters in the bottom part of Table~\ref{table:mock_param}. A more detailed description of these extensions is present in Appendix~\ref{sec:HOD_extra} (see also \citealt{Yuan:2022b}). While this HOD model is based on the work of \cite{Yuan:2022b}, there are two differences with their implementation. The first is that we assumed that the mean number of galaxies is equal to the sum of the mean number of centrals and satellites, $\langle N_{tot} \rangle = \langle N_{cen} \rangle + \langle N_{sat} \rangle$, while \cite{Yuan:2022b} used $\langle N_{tot} \rangle = \langle N_{cen} \rangle (1 + \langle N_{sat} \rangle)$. The later approach aims to suppress the occupation of satellite galaxies in haloes without a central. Still, it has recently been shown that it is expected to find haloes with satellites and no centrals \citep{Jimenez:2019, Chaves:2023}. Nonetheless, we do not expect significant differences in the performance of this model due to this different implementation.

The second difference is the environmental property used to add assembly bias, which is, in our case, the individual bias per object described in the previous section (different from the local overdensity used by Yuan et al.). During this study, we found that different environmental properties significantly impact the quality of the clustering prediction, especially for the galaxy-galaxy lensing signal. The individual bias-per-object computed on the scales we mentioned before was the property that gave the best performance. In future work, we will compare the performance of different environmental properties and their impact on galaxy assembly bias (Contreras et al., in prep.).

\subsection{Clustering statistics}
\label{sec:clustering}
This section describes the clustering statistics we measure from the \flamingo~simulations and the empirical models. These statistics are the projected correlation function (\proj), the multipoles of the correlation function ($\xi_\ell$), the galaxy-galaxy lensing (\lensing), the k-nearest neighbour cumulative distribution function, the counts-in-cylinder, and the void probability function. We average each statistic over three lines of sight to reduce the uncertainty in our calculations. All calculations include redshift space distortion effects. In addition, we compute two derivate statistics normally inferred from galaxy clustering: galaxy assembly bias and halo occupation number.

\subsubsection*{Projected correlation function}
\label{sec:wp}
The projected correlation function is obtained by integrating the 2-point correlation function, $\xi(r_{\uppi}, r_\mathrm{p})$, over the line-of-sight, 
\begin{equation}
    w_{\rm p} = 2 \int_{0}^{\pi_{\rm max}} \xi( r_{\uppi}, r_\mathrm{p}) \mathrm{d} r_{\uppi},
\label{eq:wp}
\end{equation}
\noindent with $\pi_{\rm max}=30\ \hMpc$ the maximum depth used in this work. The computations were done using the public code {\tt corrfunc} \citep{Corrfunc1, Corrfunc2}.

\subsubsection*{Multipoles of the correlation function}
\label{sec:xil}

To measure the multipoles of the correlation function, we first measure the 2-point correlation function in bins of $s$ and $\mu$, where $s^2 = r^2_{\rm p} + r^2_{\uppi}$ and $\mu$ is the cosine of the angle between $s$ and the line-of-sight. We compute this statistic once again with {\tt corrfunc}. The multipoles of the correlation functions are then defined as follows: 
\begin{equation}
    \xi_{\ell} = \frac{2 \ell+1}{2} \int^{1}_{-1} \xi(s,\mu)P_\ell(\mu)d\mu
\end{equation}
\noindent where $P_{\ell}$ is the $\ell$-th order Legendre polynomial.
In this paper, we only use the monopole (\mono) and quadrupole (\quadr) since higher orders are noisier and do not provide significant additional information.

\subsubsection*{Galaxy-galaxy lensing}
\label{sec:dS}

A foreground mass distribution induces a shear signal on background sources that varies with the lens-source pair's separation. The signal stacked from a sufficient number of lenses is proportional to the excess surface density
\begin{equation}
    \Delta\Sigma(r_\perp) = \overline{\Sigma}(\leq r_\perp) - \Sigma(r_\perp),
\end{equation}
where $\Sigma$ is the azimuthally-averaged surface mass density and $\overline{\Sigma}(\leq r_\perp)$ is the mean surface density within the projected radius $r_\perp$. We estimate the azimuthally-averaged surface mass density using
\begin{equation}
    \Sigma(r_\perp) = \Omega_\mathrm{m} \rho_\mathrm{crit} 
    \int_{-r_\parallel^\mathrm{max}}^{r_\parallel^\mathrm{max}} \xi_\mathrm{gm}(r_\perp, r_\parallel)\,\mathrm{d}r_\parallel,
\label{eq:ds}
\end{equation}
where $r_\parallel$ and $r_\perp$ refer to the projected distance along and perpendicular to the line-of-sight from the lens galaxy, $r_\parallel^\mathrm{max}$ is the integration boundary, $\xi_\mathrm{gm}$ is the galaxy-matter three-dimensional cross-correlation, $\Omega_\mathrm{m}$ and $\rho_\mathrm{crit}$ are the matter and critical density of the Universe, respectively, and the mean surface mass density within a radius $r_\perp$ is
\begin{equation}
    \overline{\Sigma}(\leq r_\perp) = \frac{2}{r_\perp^2}\int_0^{r_\perp} \Sigma(\tilde{r})\,\tilde{r} \,\mathrm{d}\tilde{r}.
\end{equation}

We measure $\xi_{\rm gm}$ using {\tt corrfunc}, with $r_\parallel^\mathrm{max}=30 \hMpc$, and a subsampled version of the matter density field diluted by a factor of $\approx 3000$ for the mocks and 100 for \flamingo~(we checked lower dilute factors, finding almost identical results). 

To account for the different cosmologies, we apply a correction factor equal to the ratio of the \lensing~for \shame~mocks on scaled simulations \citep{Angulo:2010, C20} with the cosmologies from \flamingo~and Planck. The parameters used to build the mocks were obtained by fitting simultaneously \proj, \mono, \quadr~and \lensing~from the \sampleA~and \sampleB~samples. 
The correction changes the amplitude of the lensing signal by $\approx 2\%$, which did not end up affecting our results in a significant way.

\subsubsection*{k-Nearest Neighbour}
\label{sec:kNN}
The k-Nearest Neighbour Cumulative Distribution Function (kNN-CDF, \citealt{Banerjee:2021}) measures the cumulative distribution function of distances of the galaxies to a set of volume-filling Poisson distributed random points to the k-nearest data points (\skNN). We computed this statistic for k = 1, 2, 3, 4 and 5 but limited them to showing only k = 2 and 5 to facilitate the comparison between the different samples. The number of random points we used equals ten times the number of galaxies for each sample. The computations were done using the {\tt KDTree} function from {\tt scipy} \citep{scipy}.

\subsubsection*{Counts-in-cylinder}
\label{sec:cic}

The counts-in-cylinder (CIC, \sCIC) statistic measures the probability of having ``N'' galaxies inside a cylinder centred in each galaxy with a radius ``$\rm r_{CIC}$'' and a half-depth of $r_{\pi,1/2}$. We characterize this probability by showing the probability of having fewer than ``N'' for a fixed radius (i.e., $\rm P(N_{CIC})(< N_{CIC})$). We measure this statistic for different radii (${\rm r_{CIC} = 1,\ 2,\ 4,\ 7}\ \&\ 10\hMpc$). For simplicity, we only show the results for ${\rm r_{CIC}} = 1,\ \&\ 10\hMpc$. The counts-in-cylinder were computed using {\tt HALOTOOLS} \citep{halotools}.

\subsubsection*{Void Probability Function}
\label{sec:vpf}

The void probability function (VPF, \sVPF) measures the probability of finding a void (i.e., no objects from a galaxy sample) of radius ``r'' at any point in the simulation. We compute $\rm P_0(r)$ from a set of volume-filling Poisson distributed random points. The number of random points equals five times the number of galaxies in each sample. We again compute $\rm P_0(r)$ using {\tt HALOTOOLS}. This statistic contains the same information as the kNN-CDF with k = 1.

\subsubsection*{Galaxy assembly bias}
\label{sec:gab}

Unlike previous statistics, galaxy assembly bias can not be measured directly from the spatial distribution of galaxies. This is why it is difficult to get definite proof of the existence of this bias in the real universe. To estimate the amount of galaxy assembly bias from a mock or a simulation, we used the ``shuffling technique'' developed by \cite{Croton:2007}. The method consists of creating a version of a mock without assembly bias and measuring the difference in their 2-point correlation function. These mocks without assembly bias are produced by shuffling the galaxy positions among haloes of the same mass, erasing any dependence of the galaxy population with any halo property besides halo mass (see \citealt{C19, C21b} for more details about the creation of the shuffle mock). The level of assembly bias is then quantified as
\begin{equation}
    \rm b^2_{gab} = \xi/\xi_{shuffled},
\end{equation}

\noindent with $\xi$ the correlation function of the original galaxy sample, $\xi_{\rm shuffled}$ the correlation function of the shuffled sample and $\rm b_{gab}$ the level of assembly bias. We compute $\xi_{\rm shuffled}$ as the average of two independent shuffle runs to reduce the noise of this measurement. To ease the comparison with the literature, we will refer to galaxy assembly bias as the ratio of the correlation functions $\xi/\xi_{\rm shuffled}$ and not the square root of this ratio. The correlation function was computed with {\tt CORRFUNC}.

\subsubsection*{Halo occupation number}
\label{sec:hon}

The halo occupation number measures the average number of galaxies that populate haloes as a function of halo mass, $\rm \langle N(M_h) \rangle$. This statistic is usually called halo occupation distribution, but we will avoid referring to it as that to avoid confusing it with the HOD mocks. We computed $\rm \langle N(M_h) \rangle$ by measuring the ratio between the number of galaxies and the number of haloes in bins of 0.1 dex in halo mass.

We correct the halo masses of the halo occupation number of \flamingo~to account for the effects of the different cosmologies between the simulations and the impact of baryons on the halo mass function. We do this using a similar approach as \cite{C15}, which consists of matching the cumulative halo mass function by creating a displacement vector as a function of the halo mass ($\rm \Delta log(M_h)$). We do this to facilitate the comparison between samples with different halo mass functions.

\subsection{Fitting data}
\label{sec:fit}

We use two techniques to determine the best-fitting parameters for our empirical models. When we only want to find the mock parameters that best match the target sample, we use the Particle Swarm Optimisation (PSO) algorithm. To predict the likelihood of a range of parameters being a good representation of a target sample, we use the Markov Chain Monte Carlo (MCMC) algorithm.

The PSO tracks a group (or swarm) of particles in the parameter space in which we want to evaluate our model. After assigning each particle a random velocity, they will feel an attraction (or acceleration) to two different points in the parameter space: the point where each individual particle has the smallest difference with the target sample and the point where the swarm of particles has the smallest difference with the target sample. We calculate the likeness of the samples by computing their $\chi^2$. The algorithm stops when the velocities of all particles become small, and the best-fitting value does not change significantly for at least 100 steps. In this paper, we will use the public code {\tt PSOBACCO}\footnote{\url{https://github.com/hantke/pso_bacco}} \citep[see][]{Arico:2021a}.

An MCMC is used to generate a probability distribution from our parameter space. We use this distribution to evaluate the parameter space of the empirical models that successfully reproduced the statistics of \flamingo. The empirical models' performance is evaluated by computing the $\chi^2$ between the samples. This paper employs the Affine Invariant Markov Chain Monte Carlo Ensemble sampler {\tt emcee} \citep{emcee} public code\footnote{https://emcee.readthedocs.io/en/stable/}. For the \shame~(\hod) model, we ran 1,000 (2,000) chains of 10,000 (20,000) steps each, which included a 1,000-step burn-in phase. This combination of chains and steps is ideal for emulator-based MCMC, which is extremely efficient when computing a large number of points at the same time.

We used MCMC to predict extended clustering statistics (kNN-CDF, CIC, VPF, halo occupation number, and galaxy assembly bias) when fitting galaxy clustering and galaxy-galaxy lensing, as explained in Sect.~\ref{sec:results}. Following \cite{C23b, C23c}, we extract a random set of points from the MCMC chains after burn-in (1,000 points) and compute the extended statistics for all points. The median, 16th, and 84th per cent of the distribution for each statistic will represent their expected distribution when fitting galaxy clustering and galaxy-galaxy lensing.

The PSO and MCMC depend on estimating the $\chi^2$ between the clustering predictions of our empirical models and \flamingo. We calculate $\chi^2$ as

\begin{equation}
\label{eq:chi2}
\chi^2 = (V_{\rm FLAMINGO}-V_{\rm mock})^T C_v^{-1} (V_{\rm FLAMINGO}-V_{\rm mock}).
\end{equation}

$V_{\rm FLAMINGO}$ and $V_{\rm mock}$ are vectors that combine all the statistics we want to fit (for example, the projected correlation function and the galaxy-galaxy lensing) from the \flamingo~simulation and the empirical models. The covariance matrix ($\rm C_{\rm v }$) is computed using $10^3$ jackknife subsamples ($7^3$ for the \flamingo~suite of simulations with different astrophysics). Each volume is equal to the total volume minus a cubic subvolume equal to $1/10^3$ (or $1/7^3$) of the entire volume \citep{Zehavi:2002, Norberg:2008}. The covariance matrix is then computed as:

\begin{equation}
{\rm C_{\rm v }}({\boldsymbol V}_{i},{\boldsymbol V}_{j}) = 
\dfrac{N-1}{N} \sum_{l=1}^{N} ({\boldsymbol V}^l_i-\langle{\boldsymbol V}_i\rangle) ({\boldsymbol V}^l_j-\langle{\boldsymbol V}_{j}\rangle).
\end{equation}
To compensate for the larger volume of \flamingo~compared to the Planck-1Gpc simulations, we scaled the error by the square root of the ratio between the simulation volumes. We did not perform this for the smaller boxes of \flamingo~ because they are smaller than our dark matter-only pair of simulations. Following \cite{C23b}, we add a 5\% signal in the diagonal of the covariance matrices to account for various model and observational systematics; e.g., emulator errors, shotnoise, and differences in the cosmologies of the simulations. We replace the 5\% error on the quadrupole of the correlation function by adding $\rm 1.3/r^2$, which is similar to the errors of BOSS for a similar selection sample as in our sample. The range of scales we computed the covariance matrix is 100 $\hkpc$ to $100 \hMpc$ for the projected correlation function, monopole and quadrupole (up to $70 \hMpc$ for the \flamingo~simulations with different astrophysical implementations) and between $1 \hMpc$ to $30 \hMpc$ for the galaxy-galaxy lensing. We remind the reader that the goal of this paper is not to obtain the exact errors of a specific survey but rather to provide a fair comparison of the performance of two state-of-the-art galaxy population models when reproducing different galaxy populations. While the results differed slightly, our conclusions remained unchanged when we changed the covariance matrix definition or the range of scales used to fit our data.

\subsection{GalaxyEmu-Planck}
\label{sec:GalaxyEmu}

One of the primary advantages of empirical models is their computational efficiency. An empirical model takes only a few minutes to run, whereas a standard hydrodynamic simulation takes millions of CPU hours. Reading a dark matter simulation, creating a mock and computing all the statistics mentioned in the previous section takes less than 10 minutes (on an 8-threaded CPU task). While efficient, this time is significant when performing multiple fits or Monte Carlo analyses. This is why we developed a suite of emulators capable of performing all of the aforementioned statistics: projected correlation function, multipoles of the correlation function, galaxy-galaxy lensing, k-nearest neighbours cumulative distribution function, counts-in-cylinder, void provability function, galaxy assembly bias, and halo occupation number. 

We call this suite emulator `` \GalaxyEmu'' it was built using $\approx 30,000$ \shame~and \hod~mocks between $z=0$ and $z=0.8$, computing all our statistics for seven number densities between $0.0001\ \ihMpcC$ and $0.00316\ \ihMpcC$ (equally separated in logarithmic space, which includes the two number densities used in this work, 0.000316 and 0.001 $\ihMpcC$). We compute each statistic by averaging the results from two mocks with different random seeds (which effectively reduces the shotnoise from our mocks to less than 0.5\%) and from both Planck-1Gpc simulations (paired phases), which significantly reduces the cosmic variance of our statistics (see Sect.~\ref{sec:Planck1Gpc} for more details). We compute the 2D correlation functions $\xi(r_{\uppi}, r_\mathrm{p})$ and $\xi_\mathrm{gm}(r_\perp, r_\parallel)$ (from eqs. \ref{eq:wp} and \ref{eq:ds}) up to $100 \hMpc$ and $240 \hMpc$, even though we end up integrating up to $\pi_{\rm max} = 30 \hMpc$. These additional ranges, combined with the wider range of redshifts and number densities, will allow this emulator to be compared to observational samples such as KiDS, DESI, EUCLID, and so on.

One of the aims of this paper is to show the performance of empirical models for different samples of galaxies. This is why we chose two galaxy samples with different selection functions (\sampleA~and \sampleB~galaxies) and why we tested our results on the \flamingo simulations with different astrophysical prescriptions. To ensure that our emulator covers these and any other selection samples, we used a broad range of \shame~\& \hod~parameters. For \shame~we based our constraints on \cite{C23b}, where we tested this model against extreme physical modifications of the {\tt L-Galaxies} semi-analytical model (e.g., \citealt{Henriques:2020}). For the \hod, we used a range of parameters from \cite{Yuan:2022b} and extended some parameters when it was necessary to fit all the astrophysical variations of \flamingo~easily. 

To reduce the number of free parameters in the \hod, we fixed the target number density (${\rm n_{den}}$) to the one we want to measure. This constraint allowed one degree of freedom to be removed from the mock. We fix all the parameters that impact the number density except $\Mmin$ and find the value that produces a number density of the HOD equal to our target number density. This number density is equal to the integral of the HOD times the halo mass function, $\phi(\rm M_{h})$: 

\begin{equation}
\int \phi(\rm M_{h}) \langle {\rm N(M_{h}}, \Mmin) \rangle {d M_{h}} = {\rm n_{den}}.
\label{Eq:HOD_minim}
\end{equation}

We use the Newton–Raphson root-finding algorithm to find the value of $\Mmin$. We check the precision of this minimisation even when using galaxy assembly bias and occupancy variation parameters, finding excellent agreement with the target number density. 

To use a single range of parameters for all HOD mocks and to avoid a non-physical combination of parameters, we vary $\Mone/\Mmin$, which is normally used to characterise the HODs \citep{Zheng:2005, Zehavi:2005, C17}, and $\rm log(\Mcut/\Mmin)/log(\Mone/\Mmin)$, which is not expected to be lower than zero (i.e., $\Mcut > \Mmin$) and cannot be above one (i.e., $\Mcut < \Mone$). This parameterisation enabled all parameter combinations to be valid for any number density. Although this parametrisation reduces the number of free parameters in the \hod~model to 12, we will still refer to it as a 13-parameter HOD for clarity.

For the \shame~model we normalised the value of $\betaL$ ($\betaL^*$) by dividing the value for the lookback time when the mock is created ($\approx 13.8 Gyr$). A value of zero indicates that the model has no satellites at the current time, while a value of one indicates that all non-orphan galaxies in the simulation are used. The assembly bias parameters are expressed as the sum ($\FkP$) and difference ($\FkM$) of the parameters to facilitate the training.

The parameter space we used for \shame~model are:
\begin{eqnarray*}
\label{eq:par_range}
\sigL                 &\in& [0, 1.8],\\ 
\log(\tmerger)        &\in& [-1.5, 1.2],\\
\FkP                  &\in& [-0.5, 0.5],\\
\FkM                  &\in& [-0.5, 0.5],\\
\betaL^*    &\in& [0,1].
\end{eqnarray*}

The parameters used for the \hod~model are:

\begin{eqnarray*}
\label{eq:par_range}
\sigmaLogM &\in& [0, 1.5],\\
\rm log(\Mone/\Mmin) &\in& [0, 2],\\
\rm \frac{log(\Mcut/\Mmin)}{log(\Mone/\Mmin)}&\in& [0,1],\\
\alpha &\in& [0.5,1.4],\\
f_{\rm ic} &\in& [0.2,1],\\
\alpha_c &\in& [0,0.7],\\
\alpha_s &\in& [0.5,1.5],\\
\rm GAB_{cen} &\in& [-0.5,0.5],\\
\rm GAB_{sat} &\in& [-0.5,0.5],\\
\rm OV_{cen} &\in& [-0.5,0.5],\\
\rm OV_{sat} &\in& [-0.5,0.5],\\
{\rm s_{\rm segr.}} &\in& [-0.7,0.7].
\end{eqnarray*}

Similar to \cite{Angulo:2021, Arico:2021a, Arico:2021b, Pellejero:2023, Zennaro:2023, C23b, C23c}, we use a feed-forward Neural Network to build our emulators. The architecture is made up of three fully connected hidden layers, each with 200 neurons, and a Rectified Linear Unit activation function for the multipoles of the correlation function and the galaxy-galaxy lensing signal. The remaining statistics are calculated using two fully connected layers. Each statistic has a separate network. We explored alternative configurations and obtained comparable results.

The neural networks were trained with the Keras interface of the Tensorflow library \citep{tensorflow}. We used the Adam optimisation algorithm with a 0.001 learning rate and a mean-absolute error loss function. Our dataset is divided into distinct groups for training and validation. A single Nvidia Quadro RTX 8000 GPU card took around 30 minutes per number density/statistic to process the training set, which contains 90\% of the data. On an 11-year-old Intel i7-3820, evaluating one emulator takes $\approx 6$ milliseconds, whereas evaluating batches of 1,000,000 samples takes $\approx 22$ seconds.

Following \cite{C23c}, we directly trained the projected correlation function, monopole, quadrupole, hexadecapole (not used in this work), and galaxy-galaxy lensing. This means that we do not compress this information (such as polynomial fit coefficients) in order to simplify neural network training. 

The new statistics included in this work can be easily represented by a functional form, which allows us to represent them with a set of coefficients that we trained to produce our neural network. We described the various parameterizations done to all these statistics in Appendix~\ref{sec:compr}.
We evaluated the performance of \GalaxyEmu~for these statistics and found excellent agreement across all redshifts and number densities.

\section{Reproducing the clustering statistics of the FLAMINGO simulation}
\label{sec:results}

With \GalaxyEmu, we can predict any clustering statistics mentioned in the previous sections in milliseconds. We are now focusing on quantifying how well \GalaxyEmu~can reproduce the statistics from \flamingo. We used two different fits for each galaxy sample and empirical model: (i) the projected correlation function and galaxy-galaxy lensing (statistics we can obtain from a photometric survey), which will be represented as dotted lines in most of the plots unless specified otherwise; (ii) the projected correlation function, the monopole and quadrupole of the correlation function and galaxy-galaxy lensing (statistics we can obtain from a spectroscopic survey). These results will be shown in solid lines unless specified otherwise.

We used fits to the galaxy clustering and galaxy-galaxy lensing to predict values for the rest of the statistics. We do this to (a) reduce the number of fits performed, making the results easier to interpret, and (b) because, for the rest of the statistics, there is a limited amount of observational data or covariance matrices to compare to. Furthermore, the galaxy assembly bias and halo occupation number cannot be directly measured through observation; they must be inferred. The predicted values for these statistics are derived from the best fit using MCMCs as described in Sect.~\ref{sec:fit}. 

In Sect.~\ref{sec:fit_01}, we show the prediction of our mock models for the statistics we fit, and in Sect.~\ref{sec:fit_02}, we will show the constraints for the rest of the parameters.

\subsection{Fitting galaxy clustering and galaxy-galaxy lensing}
\label{sec:fit_01}

\begin{figure*}
\includegraphics[width=0.50\textwidth]{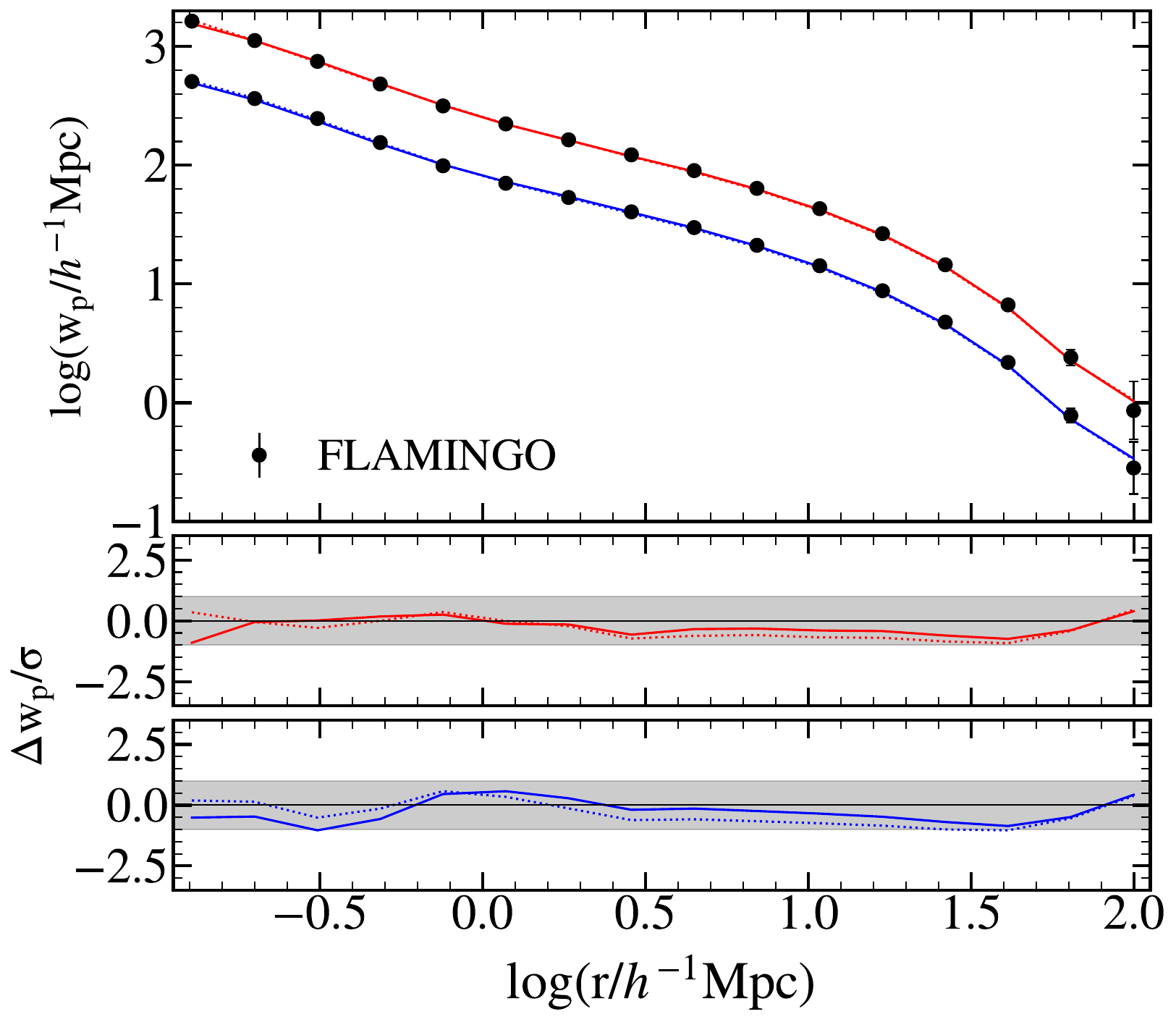}
\includegraphics[width=0.50\textwidth]{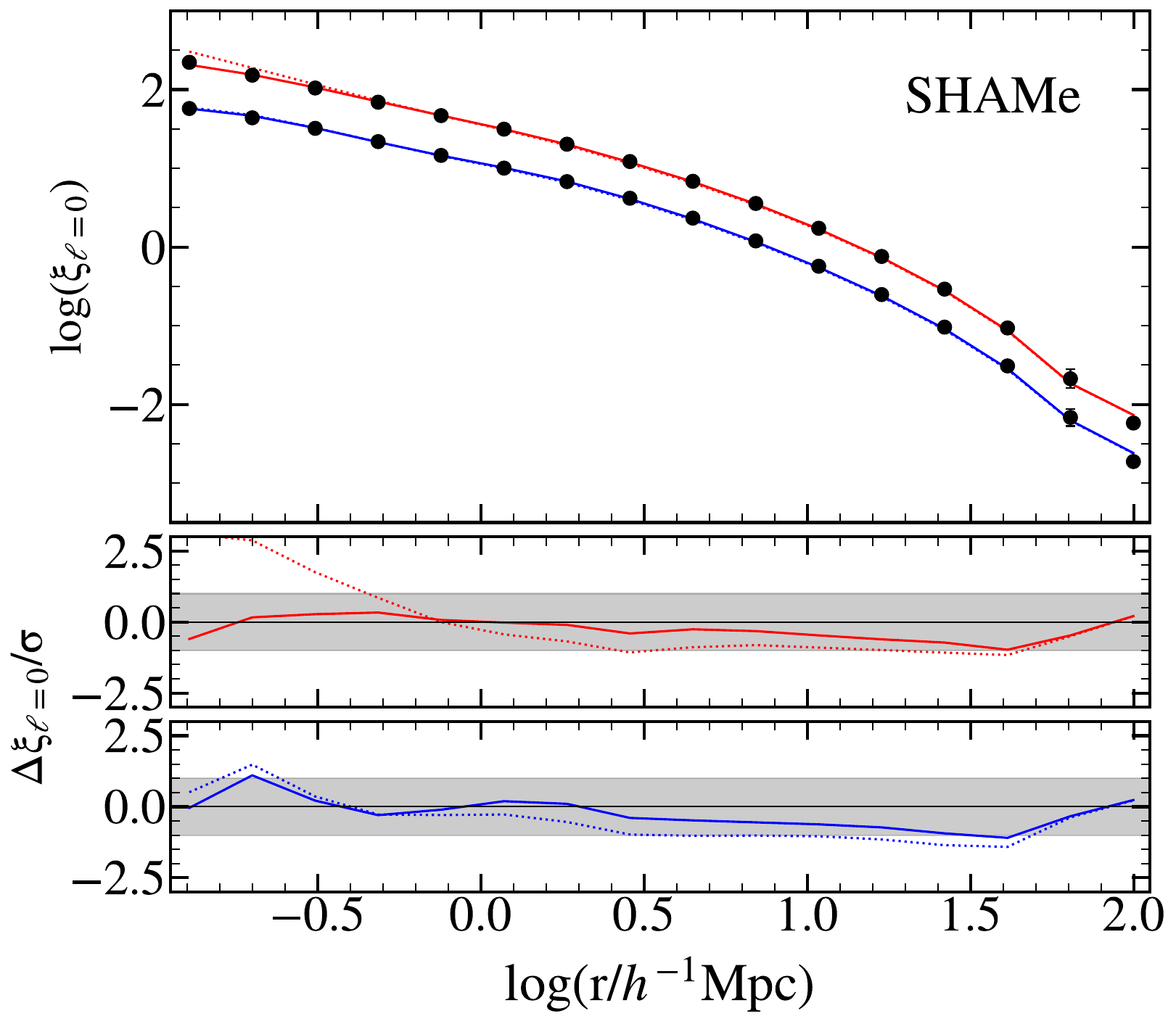}
\includegraphics[width=0.50\textwidth]{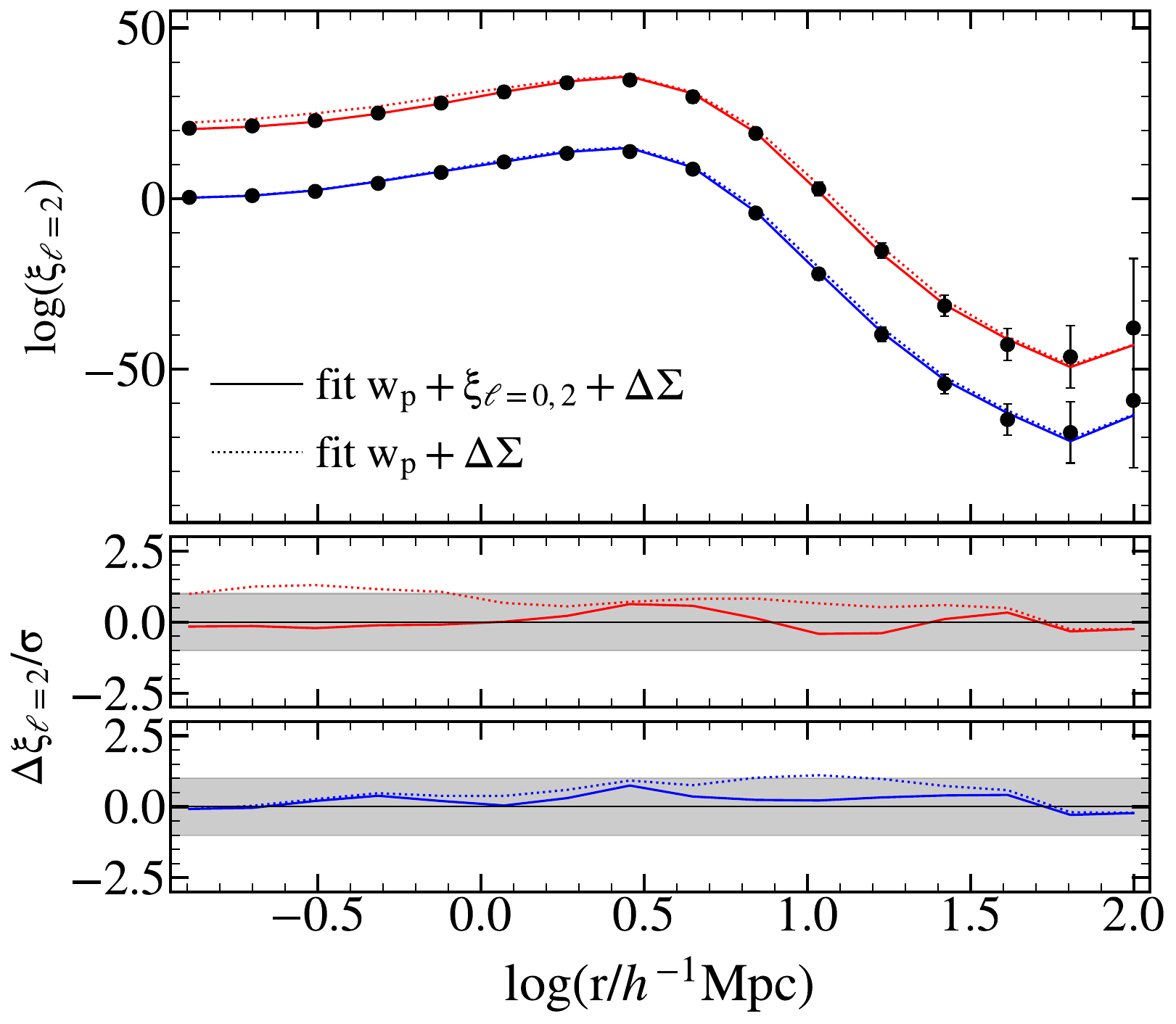}
\includegraphics[width=0.50\textwidth]{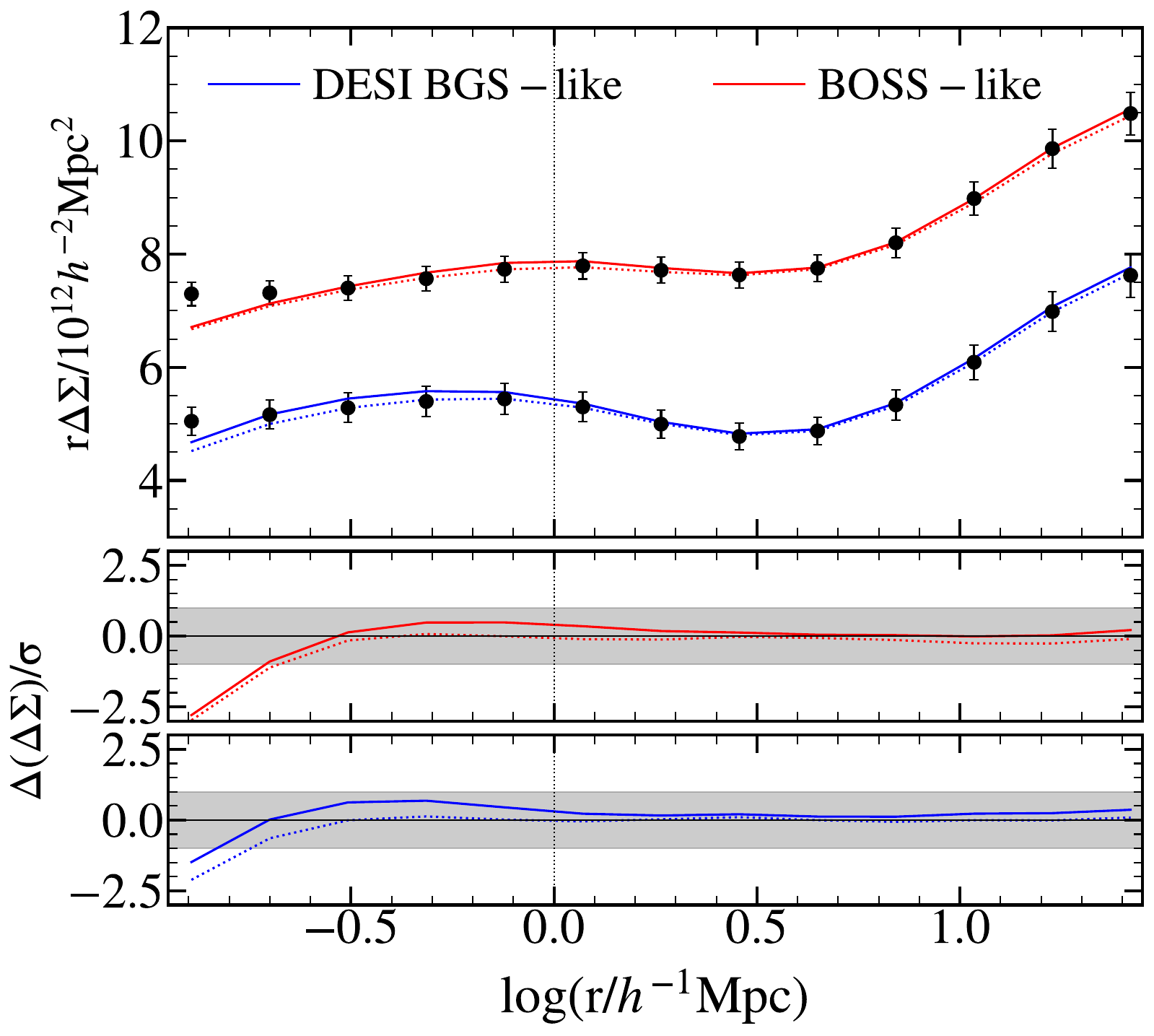}

\caption{The galaxy clustering (\proj, top left panel; \mono, top right panel; \quadr, lower left panel) and galaxy-galaxy lensing (\lensing, lower right panel) for the \flamingo~simulation (symbols) and the \shame~model (lines). The results for the \sampleB~sample, shown in red, are displaced along the y-axis for clarity. The different line styles represent the statistics used to fit \flamingo. The solid line corresponds to the best-fitting model when fitting to the projected correlation function, the monopole and quadrupole of the correlation function, and the galaxy-galaxy lensing. The dotted line corresponds to the best-fitting model when fitting only to the projected correlation function and the galaxy-galaxy lensing. The bottom subpanels show the difference between the empirical model and \flamingo, normalised by the error in each point. The fits are performed for all scales shown for the galaxy clustering statistics and for scales above $1\ \hMpc$ for galaxy-galaxy lensing (denoted by a vertical line in the bottom right panel) to avoid baryonic effects.
}  
\label{fig:stat_fit_shame}
\end{figure*}

\begin{figure*}
\includegraphics[width=0.50\textwidth]{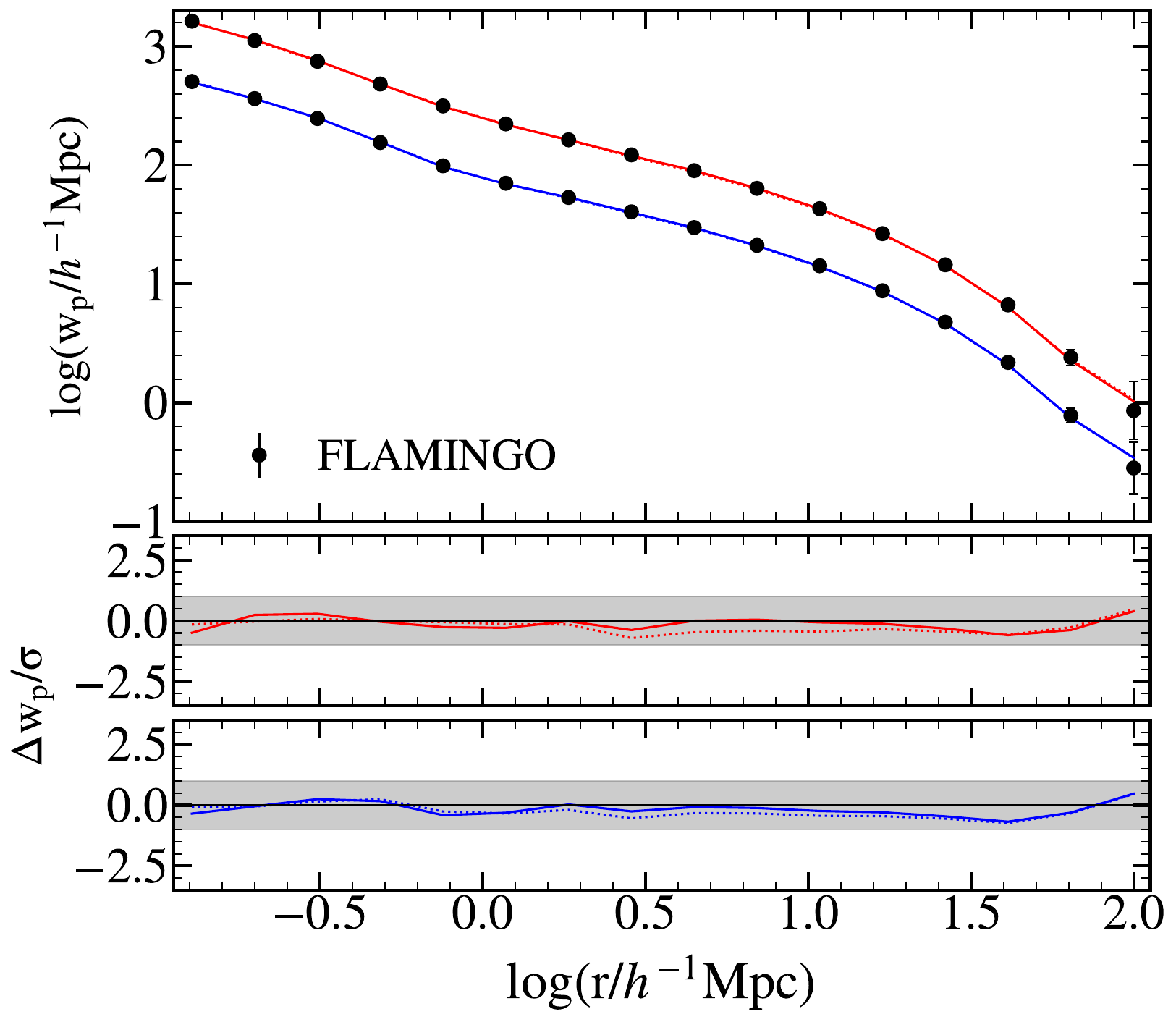}
\includegraphics[width=0.50\textwidth]{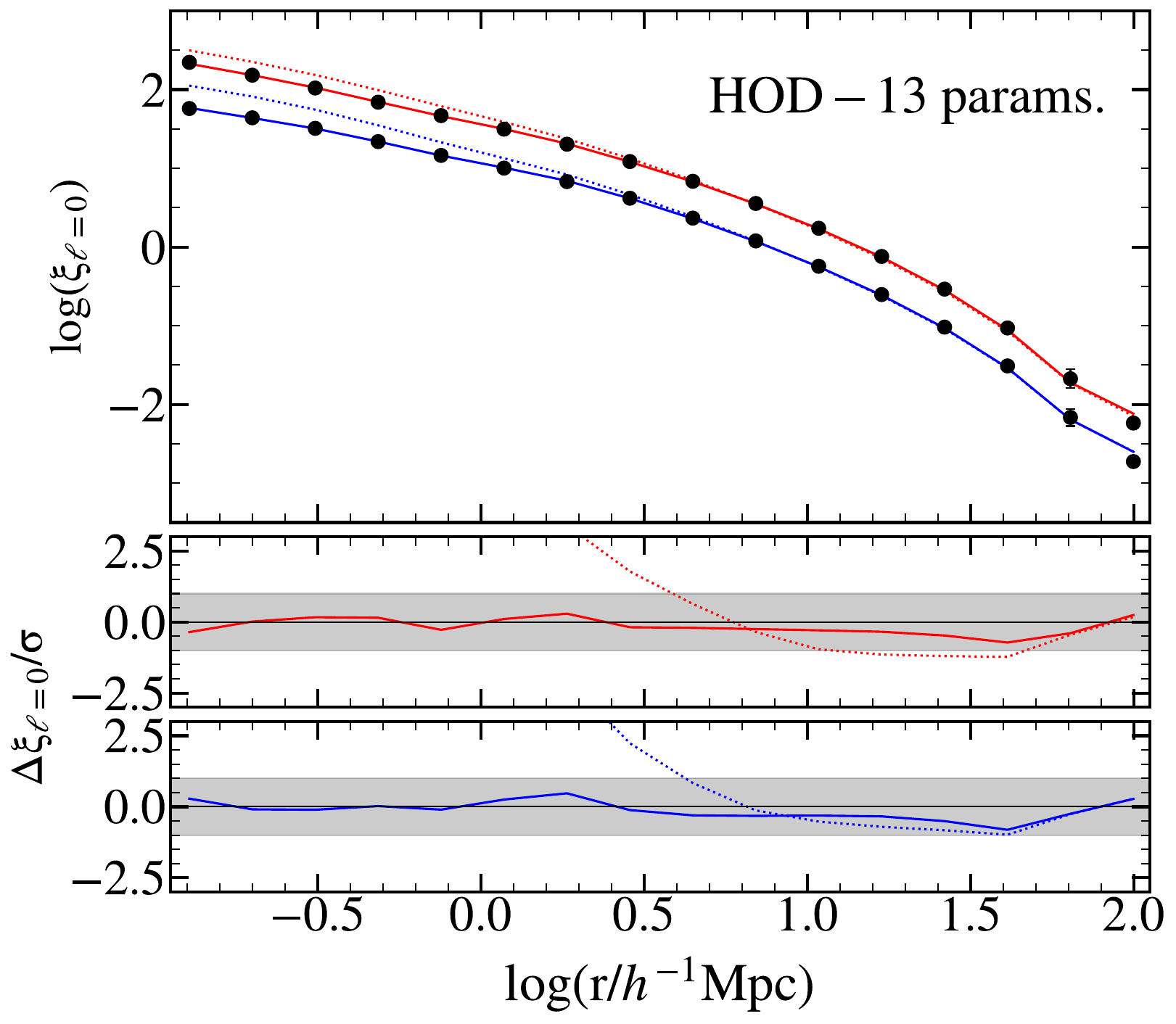}
\includegraphics[width=0.50\textwidth]{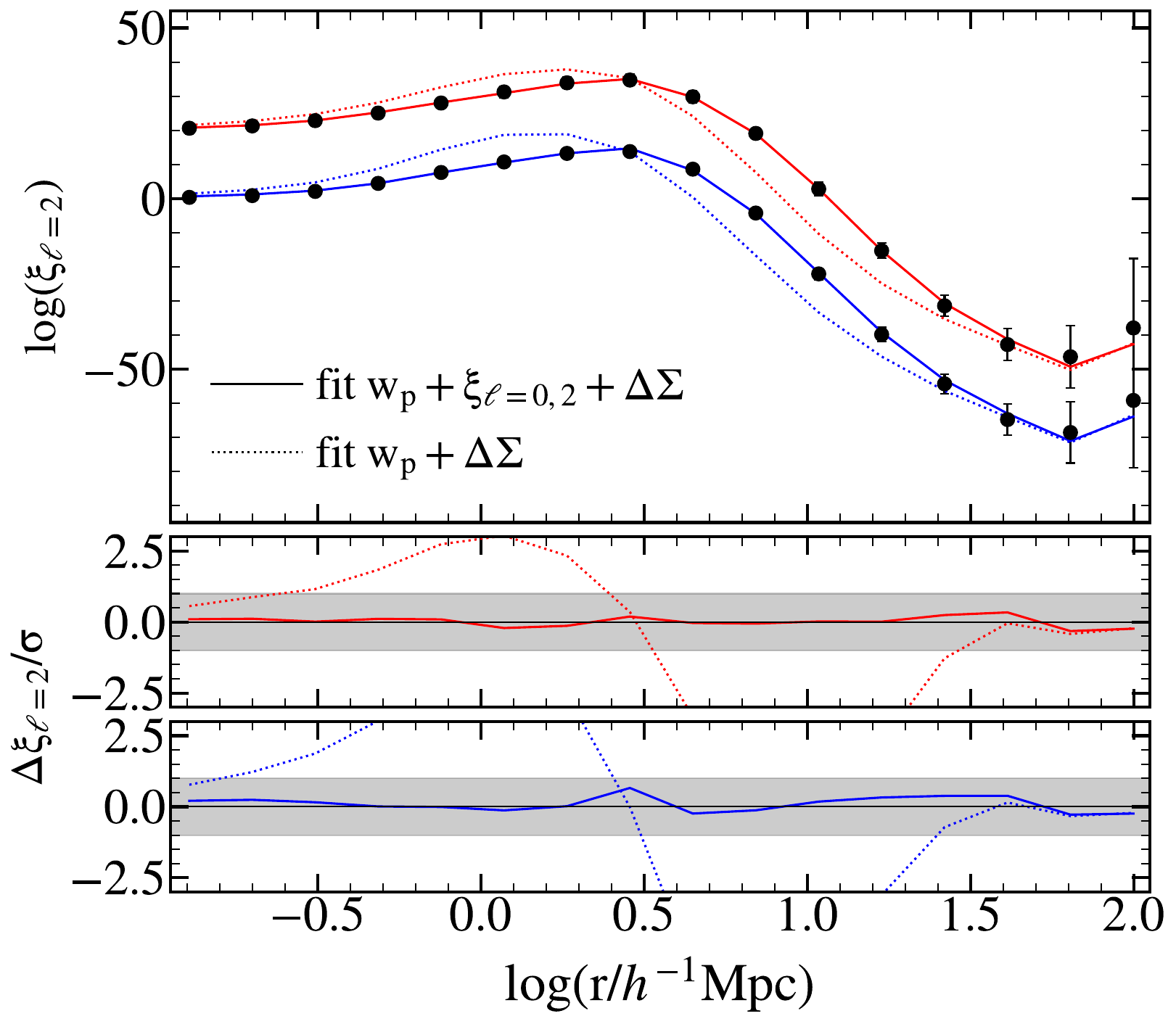}
\includegraphics[width=0.50\textwidth]{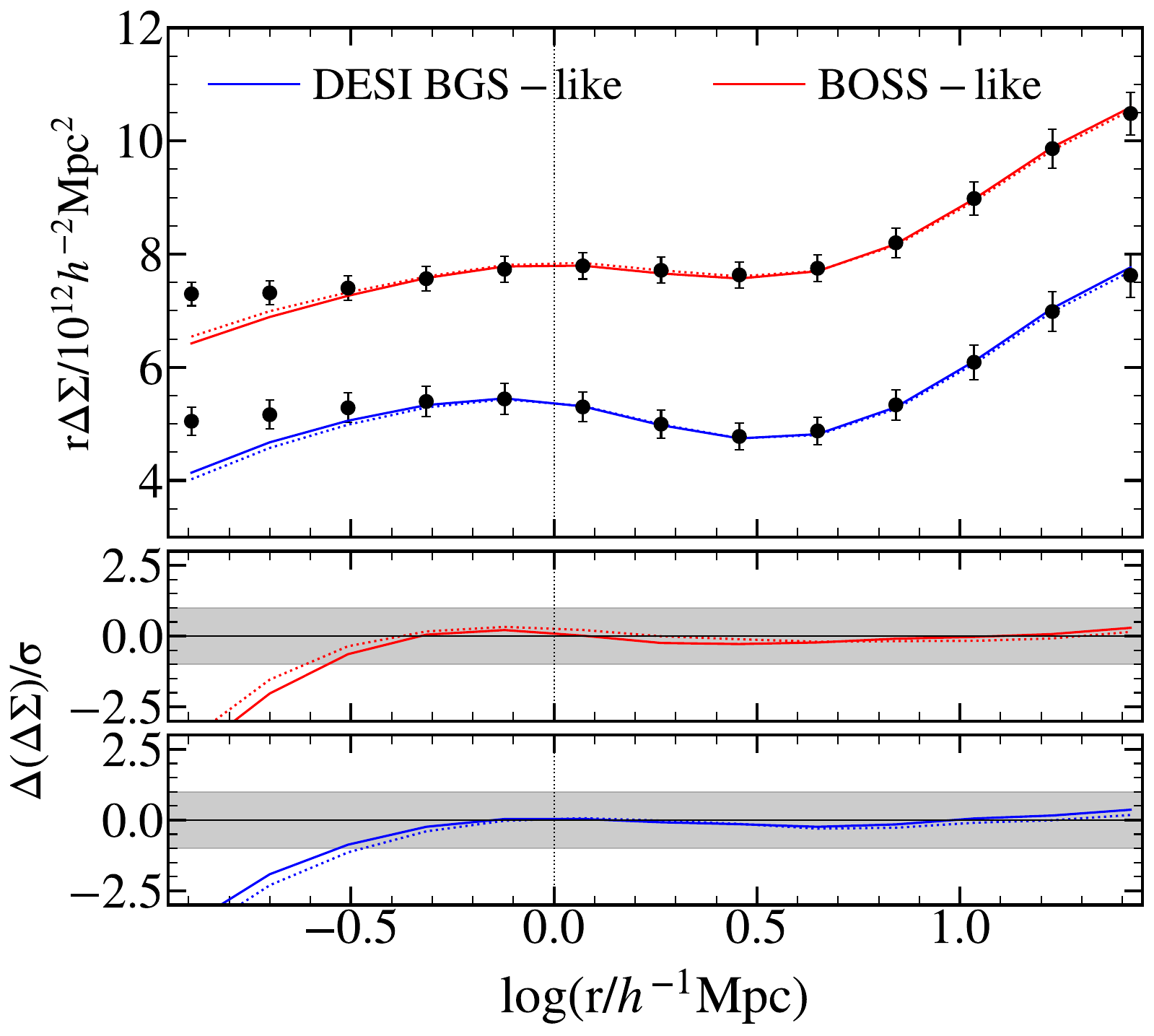}

\caption{Similar to Fig.~\ref{fig:stat_fit_shame}, but for the \hod~model.}  
\label{fig:stat_fit_hod}
\end{figure*}

\begin{figure*}
\includegraphics[width=0.5\textwidth]{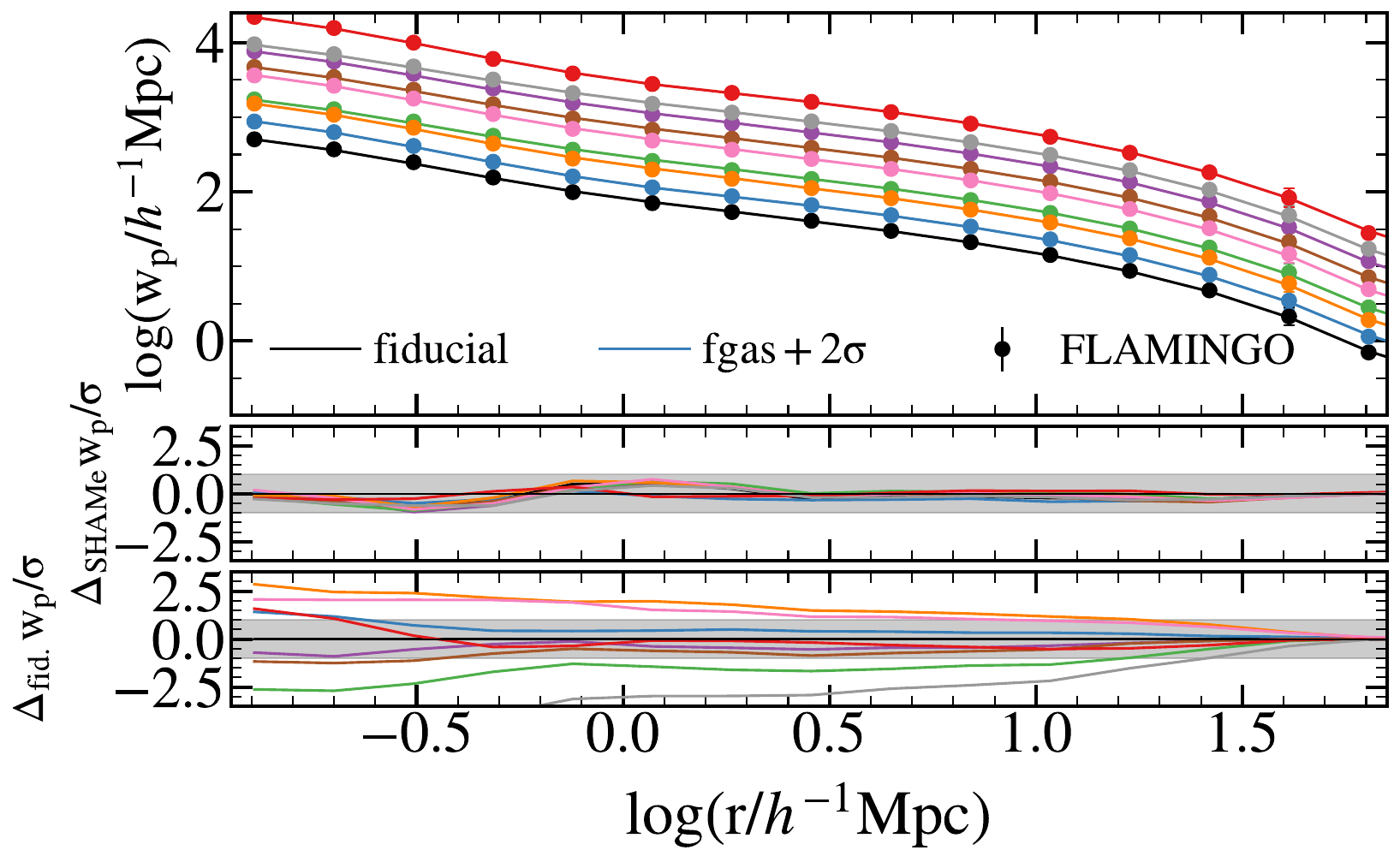}
\includegraphics[width=0.5\textwidth]{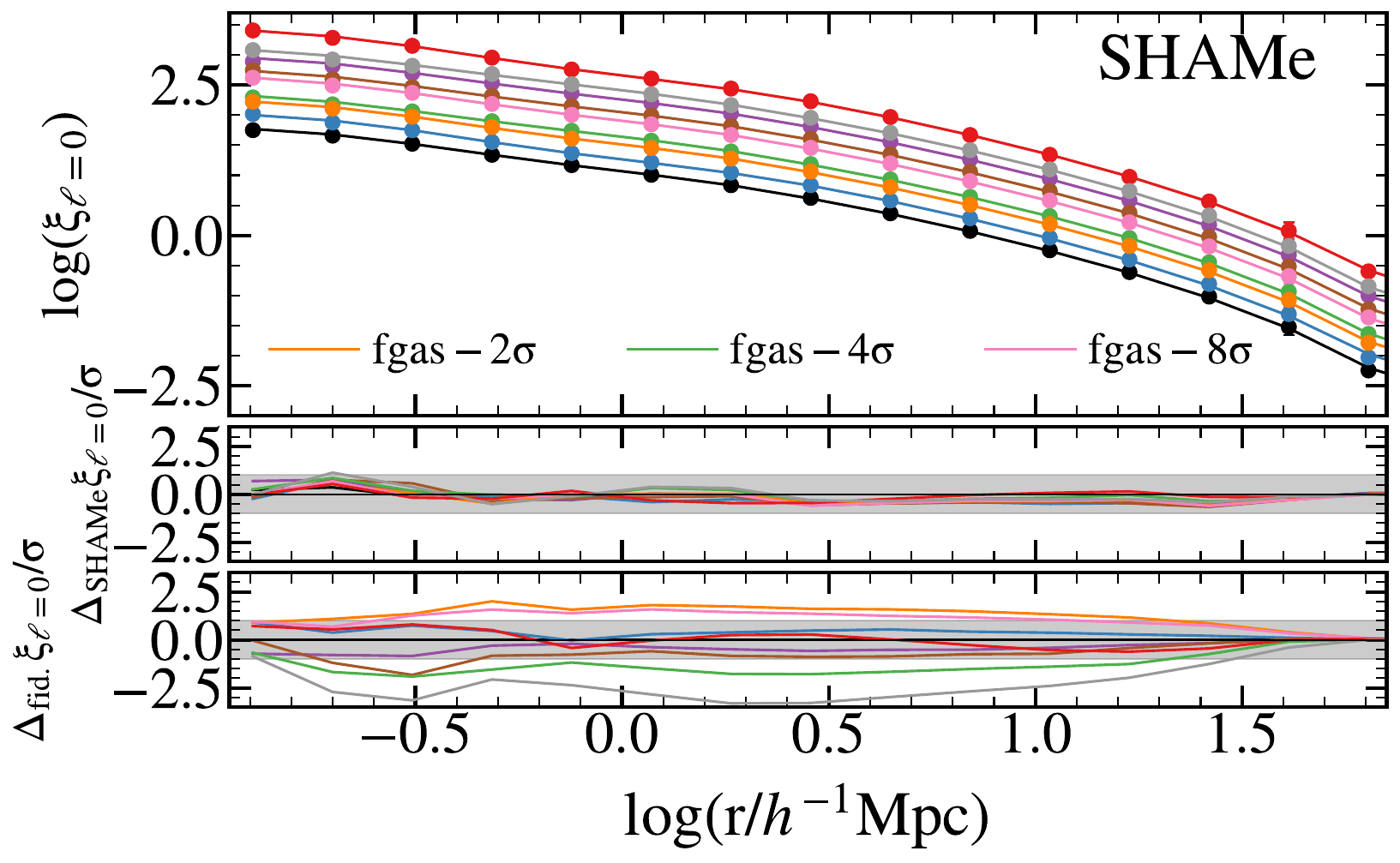}
\includegraphics[width=0.5\textwidth]{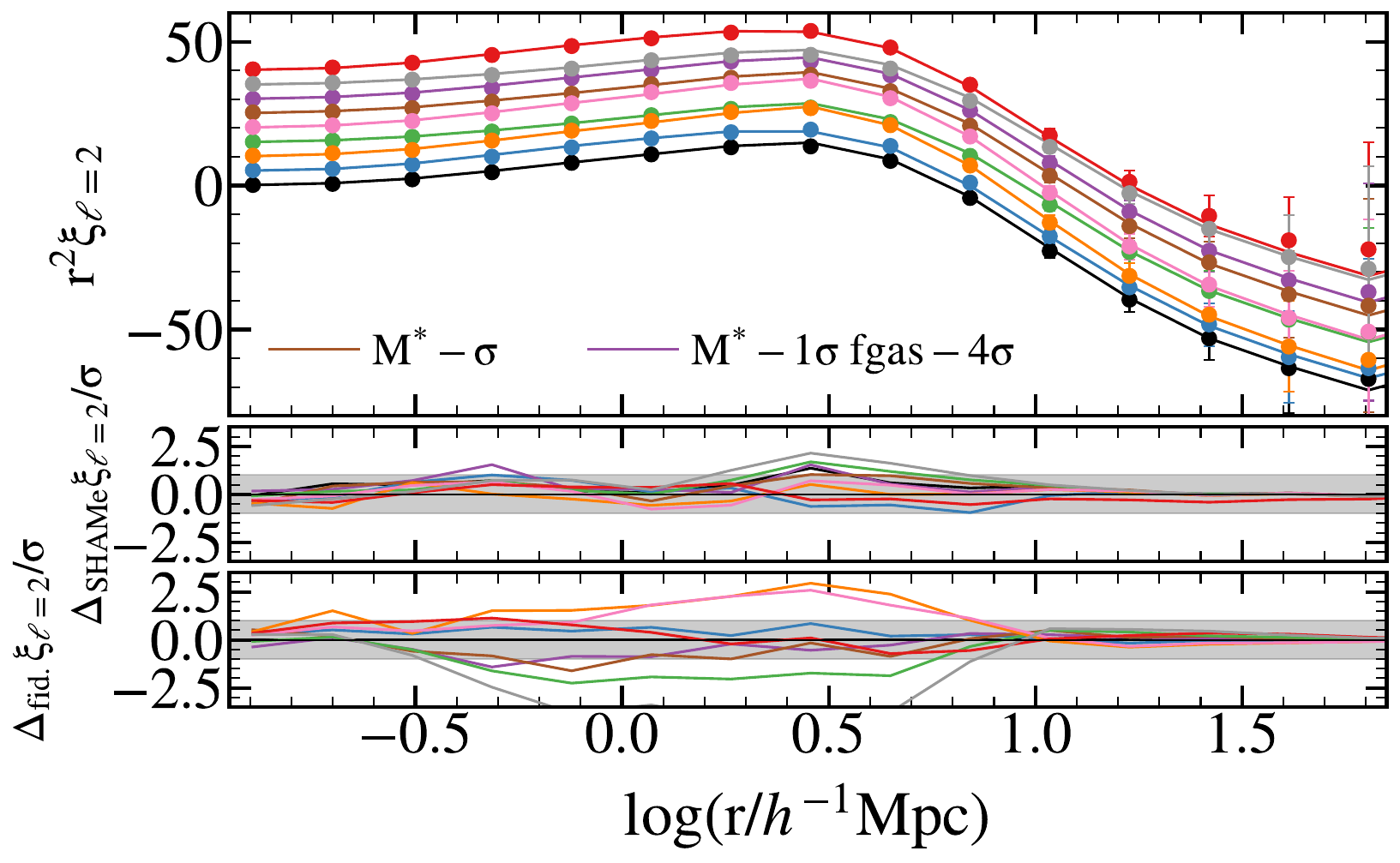}
\includegraphics[width=0.5\textwidth]{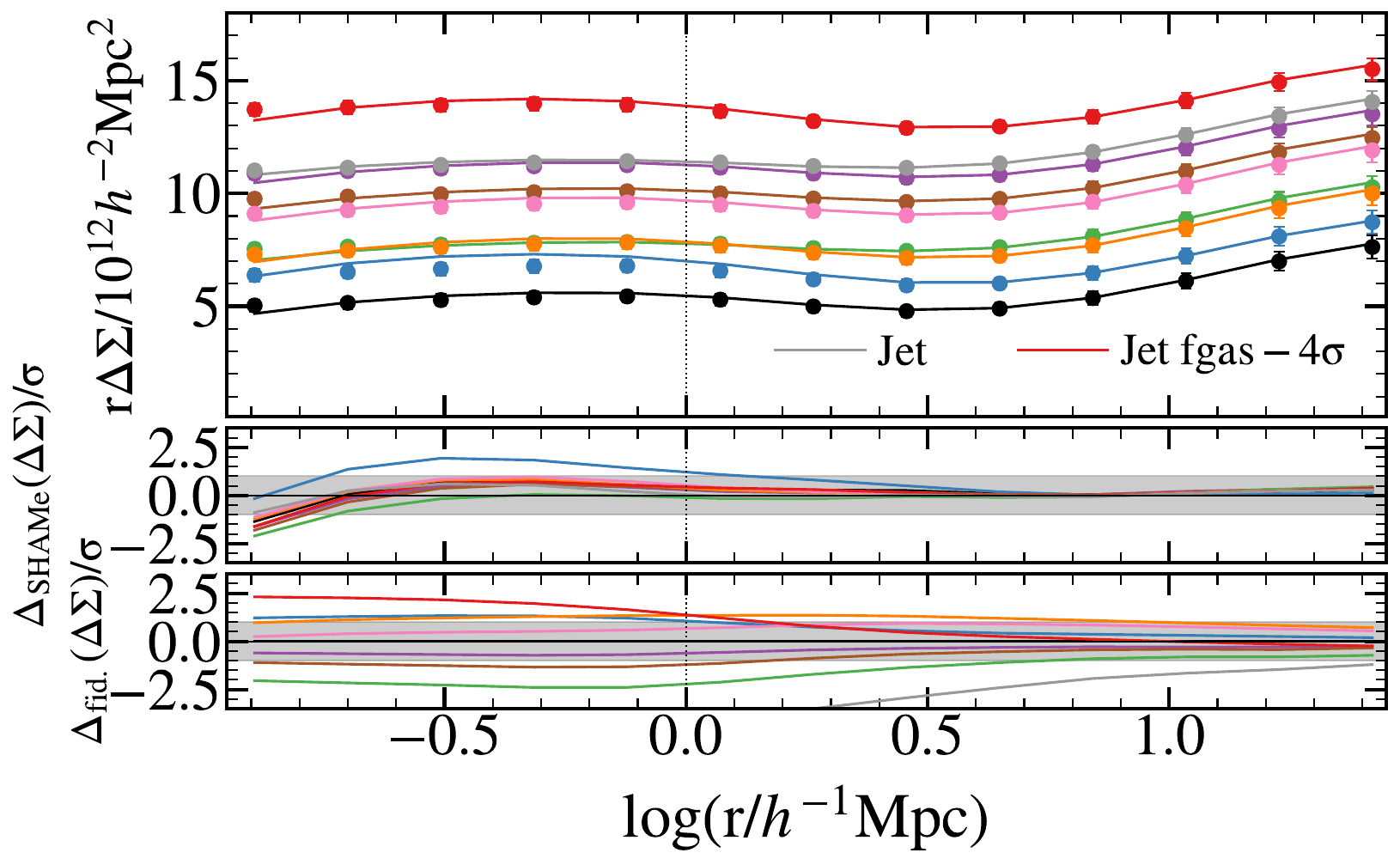}

\caption{The galaxy clustering (\proj, top left panel; \mono, top right panel; \quadr, lower left panel) and galaxy-galaxy lensing (\lensing, lower right panel) for the \flamingo~suite of simulations with different astrophysical implementations (symbols) and the \shame~model (lines). The different colours represent the various astrophysical implementations, as labelled. All samples are displaced along the y-axis for clarity. Only the results from the \sampleA~sample are shown in the plot. The middle subpanel shows the difference between the empirical model and \flamingo, normalised by the error. The bottom subpanel shows the difference between the\flamingo~model variations and the fiducial \flamingo~model. The fits are performed for all scales shown for the galaxy clustering statistics and for scales above $1\ \hMpc$ for galaxy-galaxy lensing.}  
\label{fig:MultiModel_SHAMe}
\end{figure*}

\begin{figure*}
\includegraphics[width=0.5\textwidth]{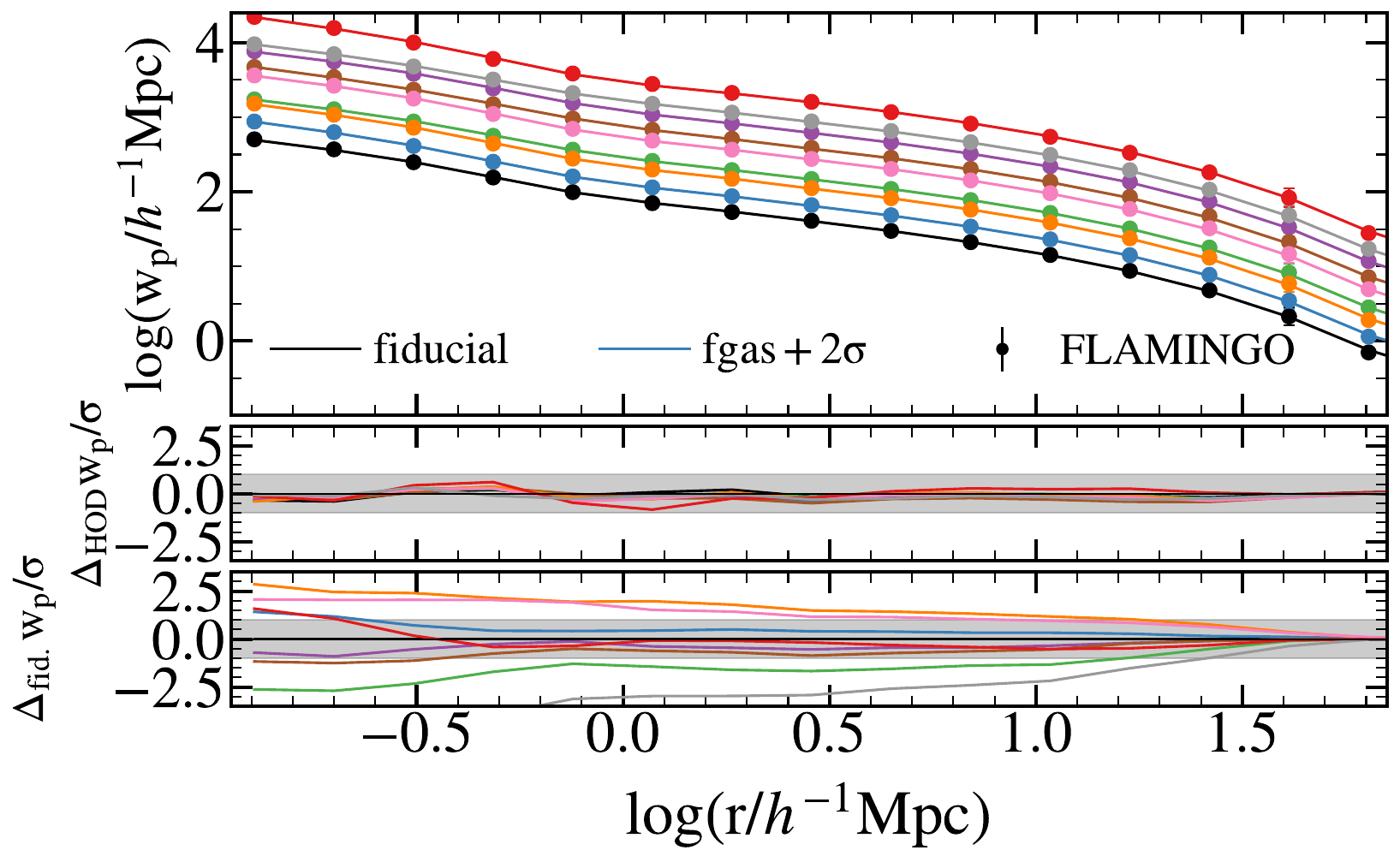}
\includegraphics[width=0.5\textwidth]{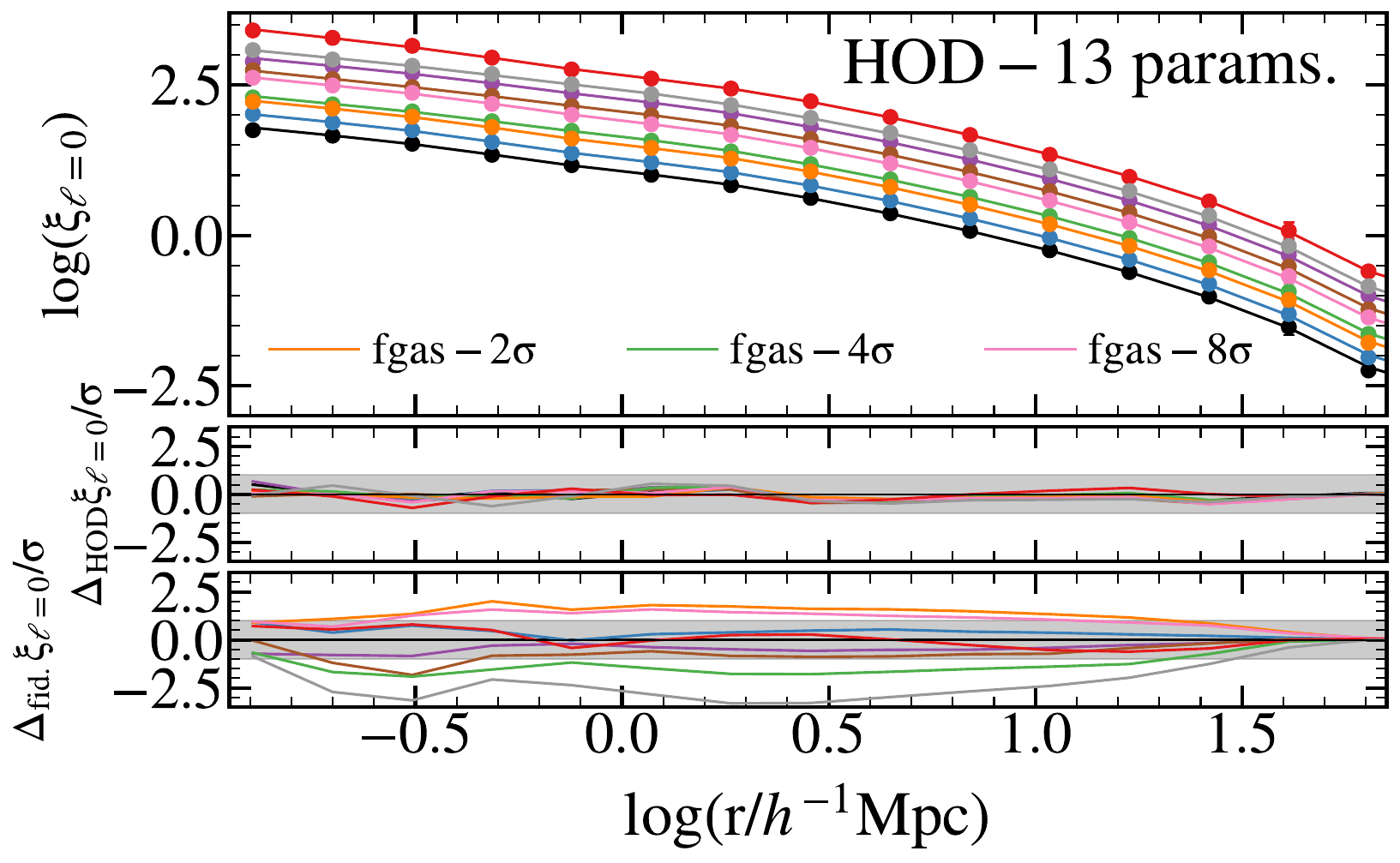}
\includegraphics[width=0.5\textwidth]{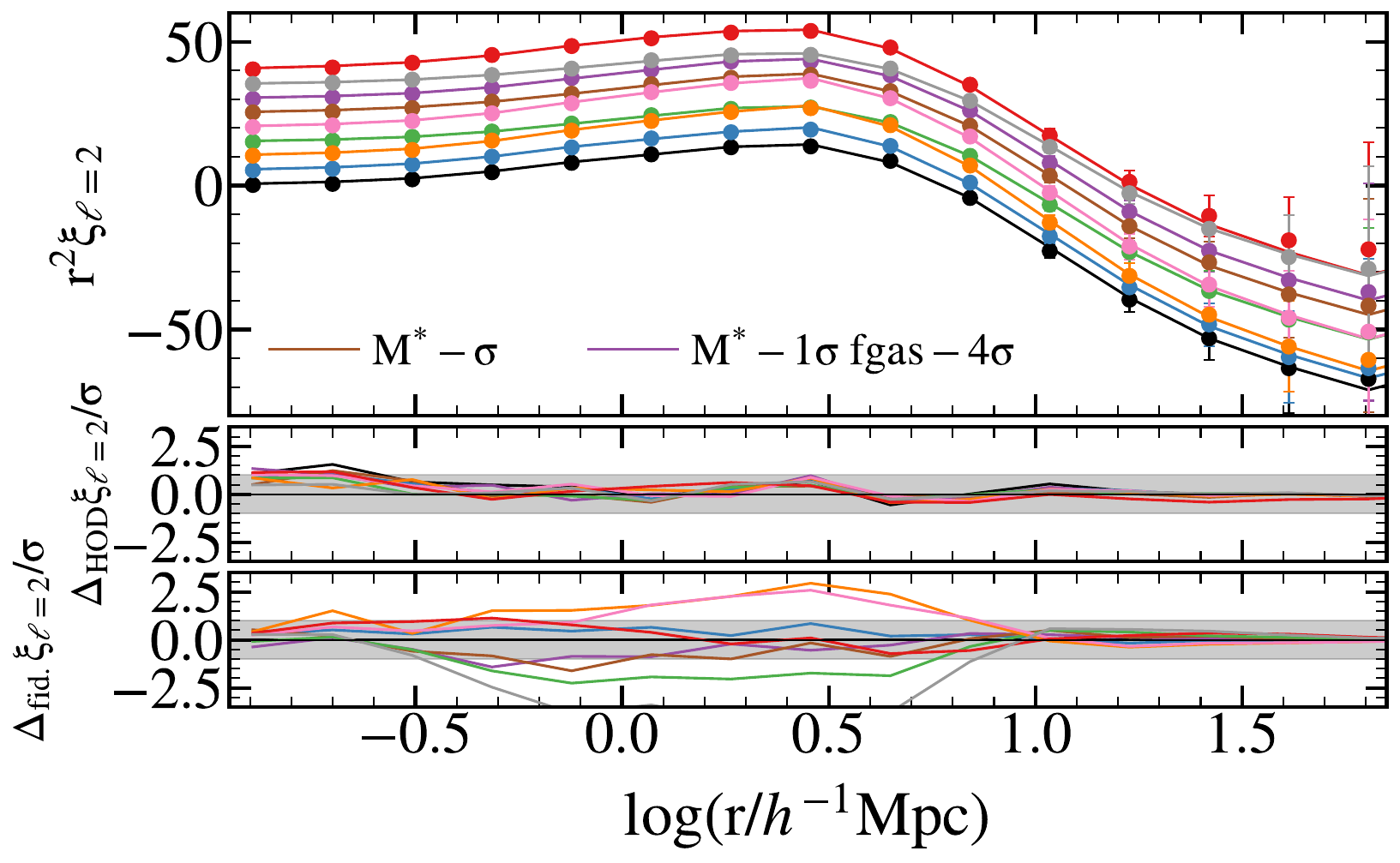}
\includegraphics[width=0.5\textwidth]{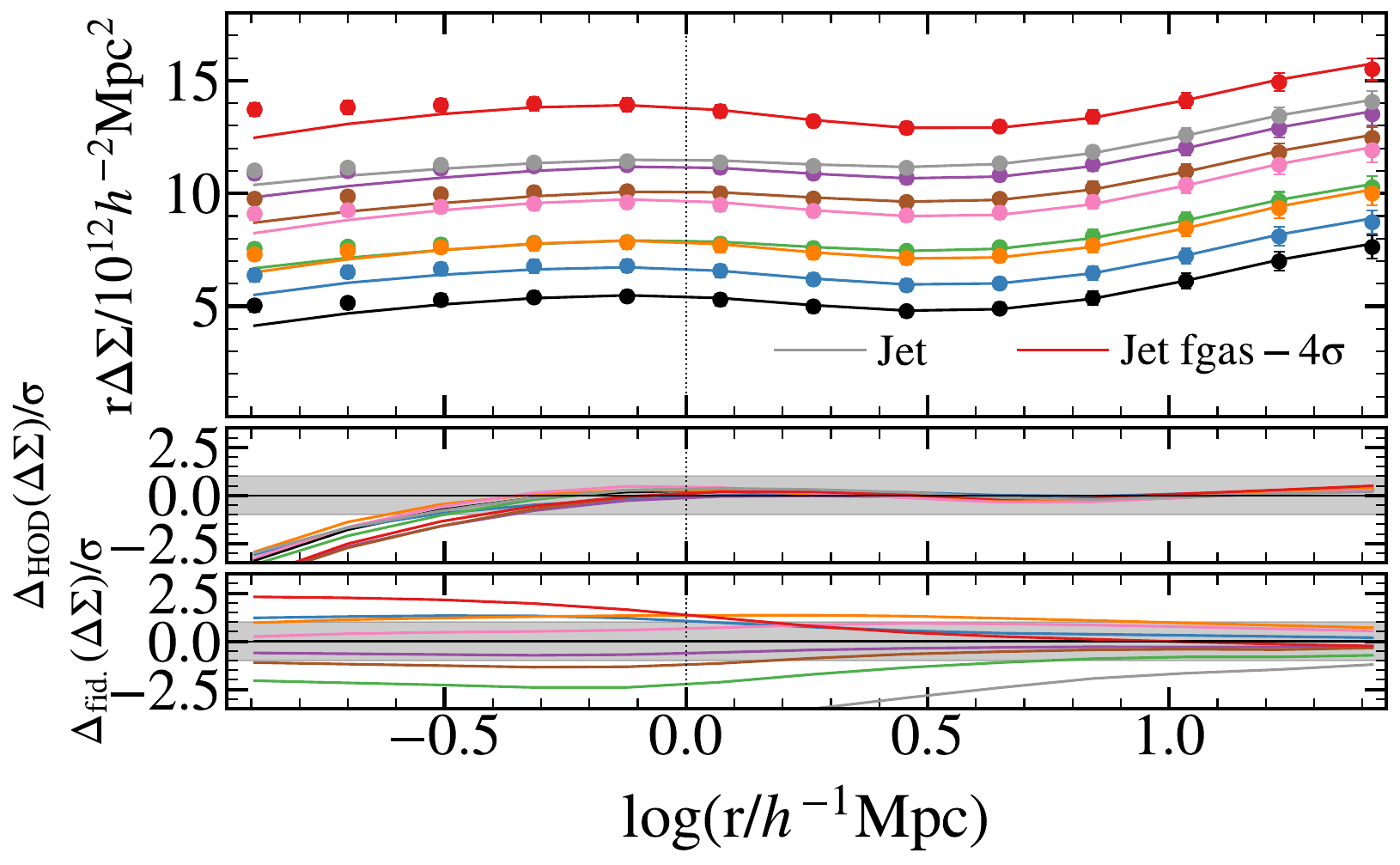}

\caption{Similar to Fig.~\ref{fig:MultiModel_SHAMe}, but for the \hod~model.}  
\label{fig:MultiModel_HOD}
\end{figure*}

The projected correlation function, monopole, quadrupole, and galaxy-galaxy lensing of the \flamingo~simulation and the best fitting \shame~and \hod~models are shown in Figs.~\ref{fig:stat_fit_shame}~\&~\ref{fig:stat_fit_hod}. We fit galaxy clustering for all scales between 100 $\hkpc$ and 100 $\hMpc$, i.e. a three-orders of magnitude window. We use 100 $\hkpc$ since it is usually the lower scale used in observational clustering measurements, and both models reproduce the \flamingo~suite of simulations well on these scales. We set the 100 $\hMpc$ limit to avoid the BAO scales, which we will explore in a follow-up work. We fit galaxy-galaxy lensing from scales above 1 $\hMpc$ to avoid the impact of baryonic effects, which are not modelled by any of our mocks (see appendix~\ref{sec:bar_effects} for more details). Each figure shows the results from two galaxy samples, the \sampleA~galaxies in blue and the \sampleB~galaxies in red. Along the y-axis we have displaced the predictions for the \sampleB~galaxies to improve the clarity of the plot. The largest panels show the value for each statistic, while the bottom panels show the differences between the mock predictions and the \flamingo~data normalised by the square root of the diagonal of the covariance matrix used for the fits (which we will refer to from now on as the data error).

We find excellent agreement across all statistics, galaxy samples, and empirical models. The \hod~model performs slightly better than the \shame~model but with differences within the statistical error bars. We find only a minor improvement when fitting the projected correlation function and galaxy-galaxy lensing (dotted line) but at the expense of a significant discrepancy in the multipole predictions, particularly for the HOD model.

One of the most exciting results is the simultaneous fit of galaxy clustering and galaxy-galaxy lensing for the \hod~model. Until recently, most attempts to reproduce galaxy clustering and galaxy-galaxy lensing from the BOSS galaxy survey were unsuccessful, with a constant gap in the galaxy-galaxy lensing measurements compared to the observed ones at small scales (an effect known as the lensing-is-low problem, \citealt{Leauthaud:2017}). While there were explanations for this effect linked to a problem with the fiducial cosmology (suggesting a lower value of the $\rm S_8 = \sigma_8\sqrt{\OmM/0.3}$ cosmological parameter), \cite{Chaves:2023} found that the most probable origin of this problem were the assumptions made by some mocks models (such as the lack of galaxy assembly bias, the functional form of the HOD, the way satellite galaxies distribute within the halo among others). As we will show in Sect.~\ref{sec:simple}, more simple HOD models are incapable of simultaneously reproducing galaxy clustering and galaxy-galaxy lensing, presenting a signal compatible with the Lensing-is-Low problem. Recently, a more complex \hod~model was able to fit these two statistics simultaneously \citep{Paviot:2024}. The \shame~model also proves to be capable of simultaneously fitting these two statistics \citep{C23c}. 

In addition to the main \flamingo~simulation, we fit all the different astrophysical implementations mentioned in Sect.~\ref{sec:FLAMINGO}. Because of the smaller volume of these simulations, we limited the clustering measurements to scales up to $\approx 70 \hMpc$, which is approximately $\approx 10\%$ of the simulation box size. We only looked at the \sampleA~sample, since it has a higher number density and does not rely on colours to select the galaxies. 

We show the galaxy clustering and galaxy-galaxy lensing measurements for the \flamingo~suite of simulations along with the best fitting \shame~and \hod~model in figures \ref{fig:MultiModel_SHAMe} \& \ref{fig:MultiModel_HOD}. The empirical models shown in the figures are those that better reproduce the projected correlation function, the multipoles of the correlation functions and the galaxy-galaxy lensing (the solid lines in Fig.~\ref{fig:stat_fit_shame}~\&~\ref{fig:stat_fit_hod}). The top panels show the statistics for all models; the middle panels show the difference between the \flamingo~variations and the mocks, normalised by the error; and the bottom panel displays the relative difference between the \flamingo~variations and the fiducial run, normalised by the error (black line). 

The empirical models reproduce clustering statics equally well for all the astrophysical variations of \flamingo, regardless of whether they target an extreme astrophysical variation. Only small differences between the fits can be found for the \shame~model in the quadrupole at scales of ${\rm log(r}/\hMpc) \approx 0.5$. This agreement is especially impressive for galaxy-galaxy lensing. We anticipated that due to differences in the baryonic implementations (particularly for the most extreme models), some models could end up being difficult to fit.  

These results demonstrate the accuracy of the current generation of empirical models and how \GalaxyEmu~can reproduce galaxy clustering and galaxy-galaxy lensing for most target galaxy samples.

\subsection{Predicting higher-order statistics}
\label{sec:fit_02}

\begin{figure*}
\includegraphics[width=0.32\textwidth]{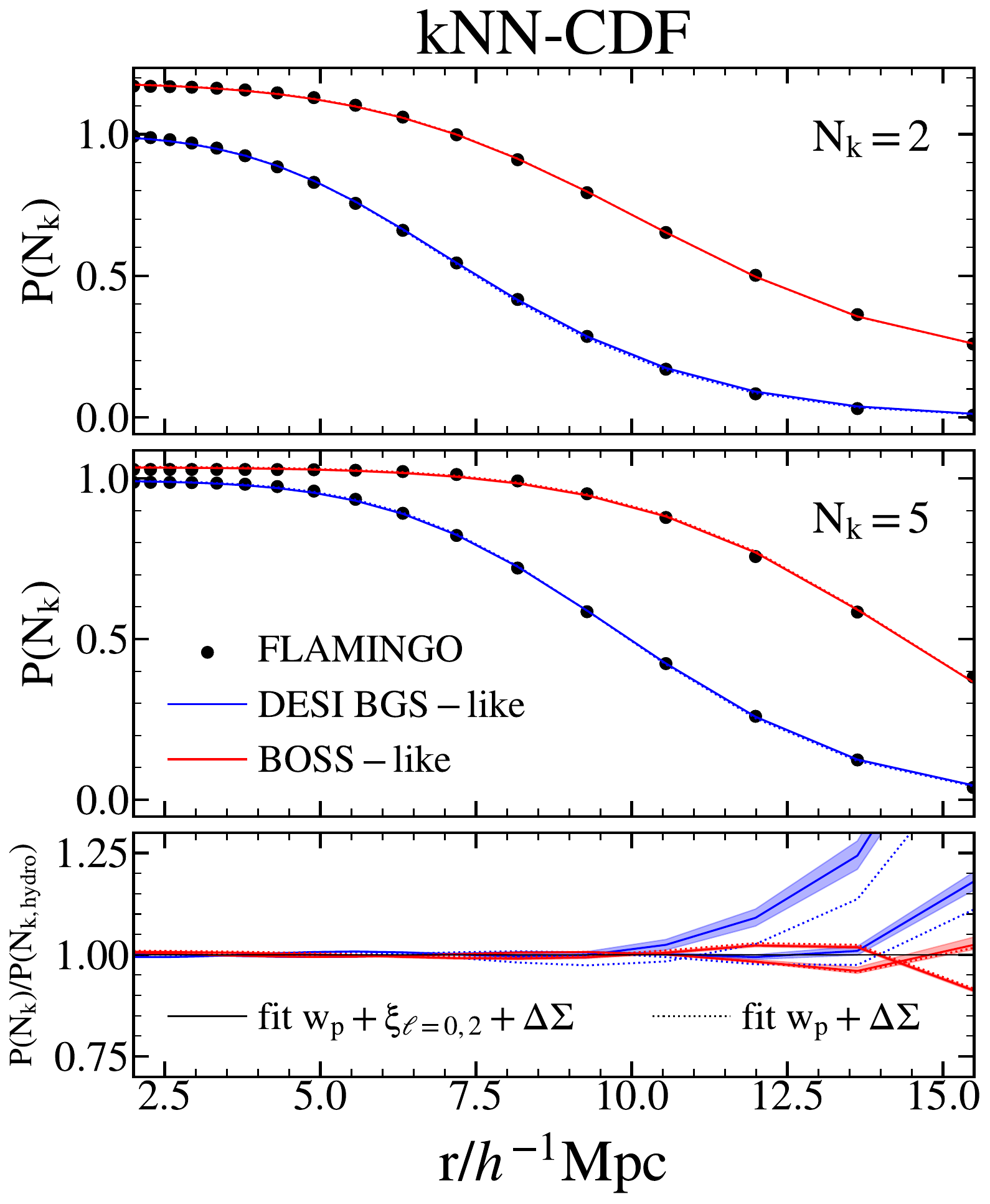}
\includegraphics[width=0.33\textwidth]{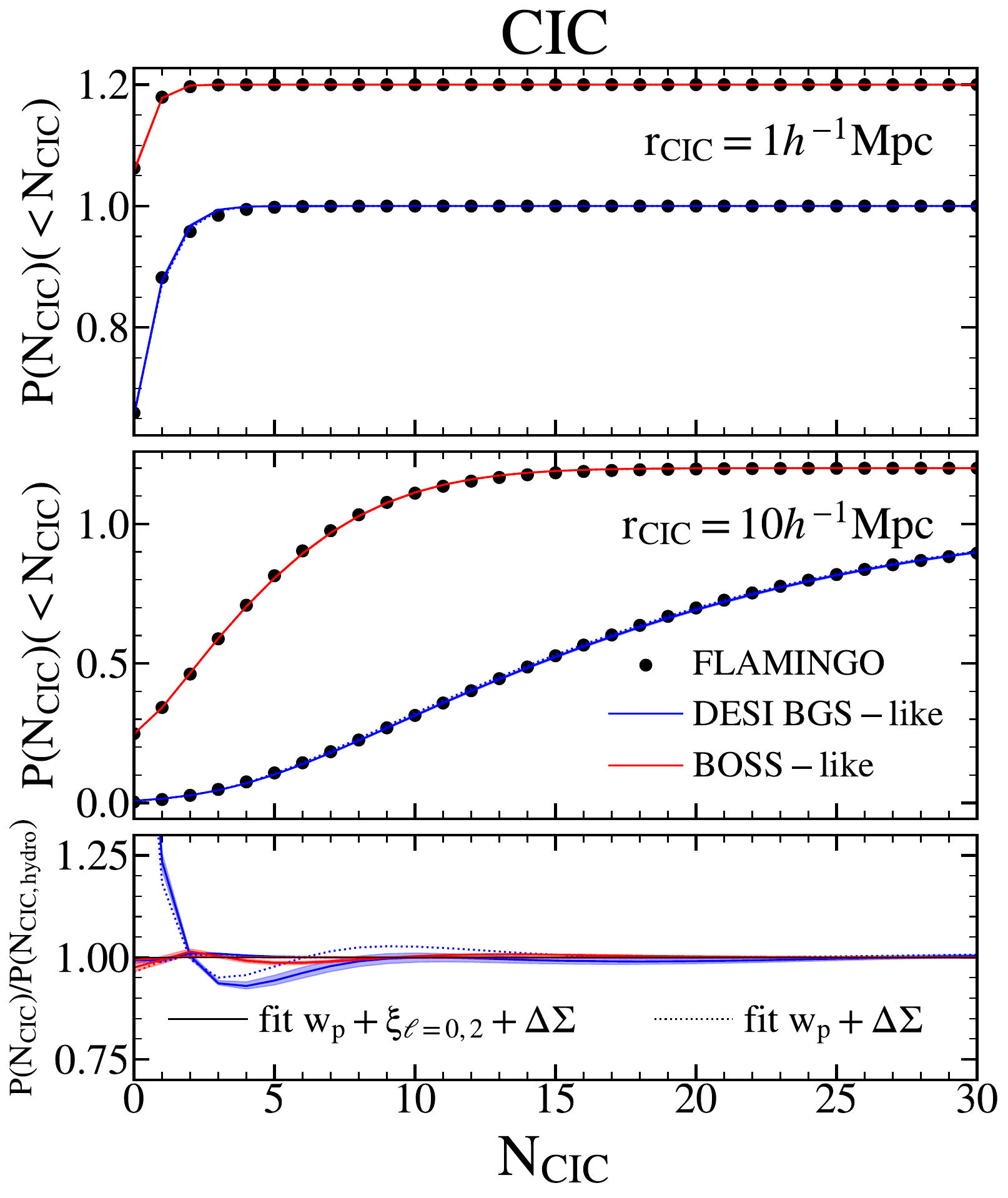}
\includegraphics[width=0.335\textwidth]{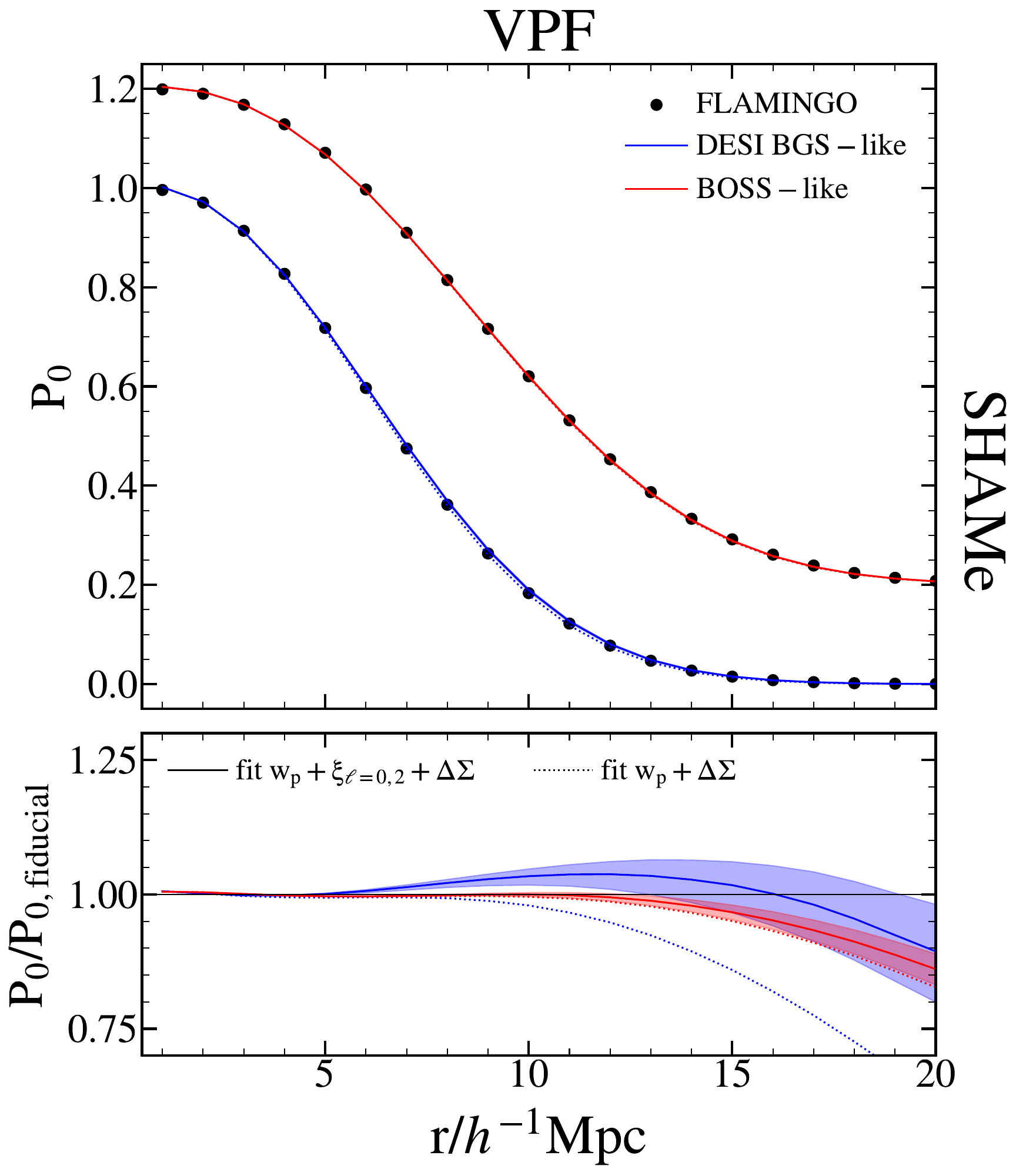}
\caption{The k-nearest neighbour cumulative distribution functions (kNN-CDF, left panel), count-in-cylinder (CIC, middle panel), and void probability function (VPF, right panel) of the \flamingo~simulation (symbols) and the \shame~model (lines). The results for \sampleB, shown in red, are displaced along the y-axis to facilitate the comparison. The different line styles represent the statistics used to fit \flamingo~as labelled. The lines represent the median of 1000 MCMC points when fitting the projected correlation function and the galaxy-galaxy lensing (dotted lines), as well as the projected correlation function, multipoles of the correlation function, and galaxy-galaxy lensing (solid lines). It is worth noting that none of the statistics shown in this figure were used in the empirical model fitting process. The shaded region corresponds to the 16th and 84th percentile of the distribution (only shown for the solid line). Only the predictions for $\rm N_k$ = 2 and 5 are shown for the kNN-CDF. Similarly, the CIC only shows the predictions for $\rm r_{CIC}$ = 1 and 10 $\hMpc$. The bottom panels show the ratio between the empirical models and \flamingo.}  
\label{fig:ext1_shame}
\end{figure*}

\begin{figure*}
\includegraphics[width=0.32\textwidth]{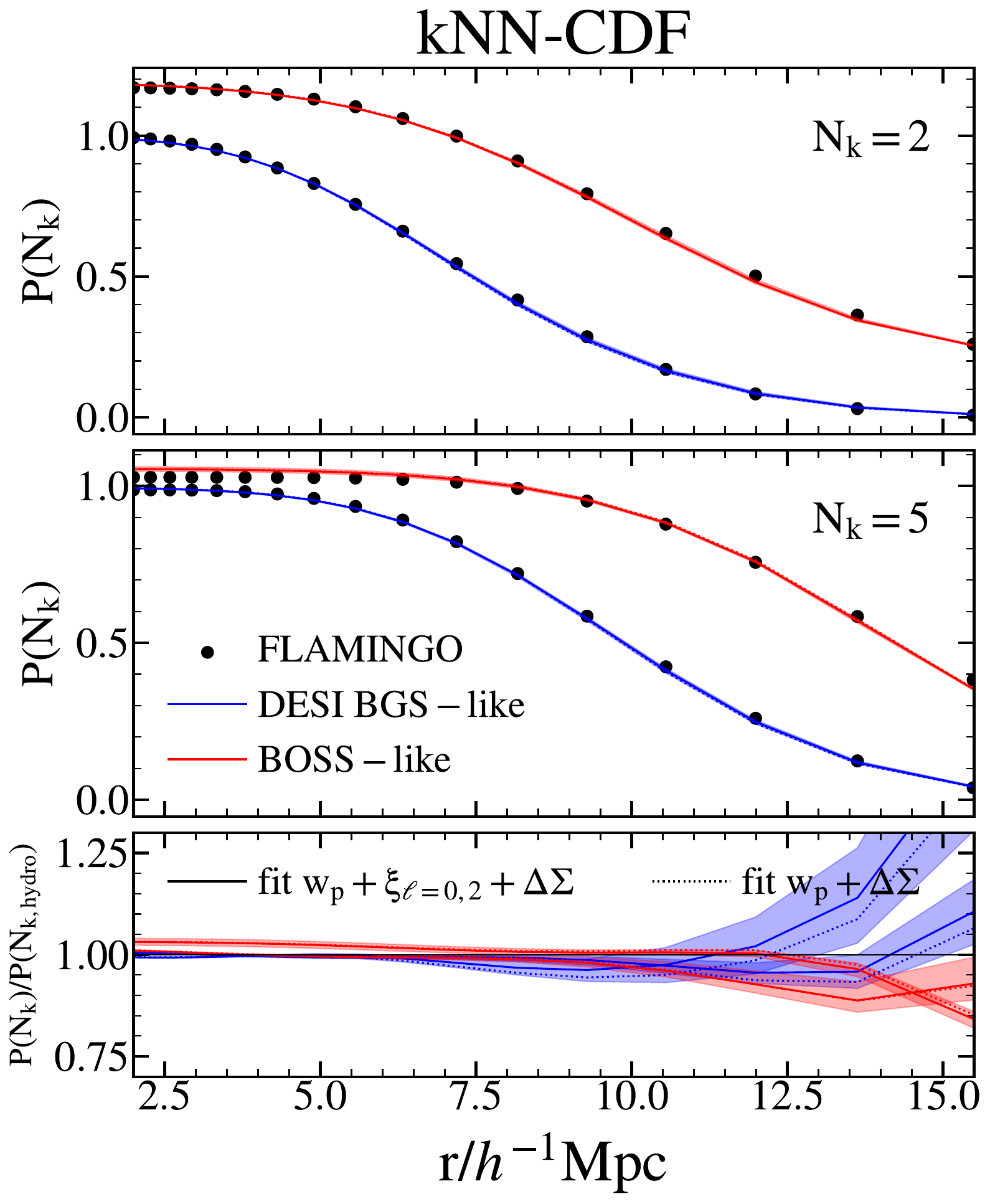}
\includegraphics[width=0.328\textwidth]{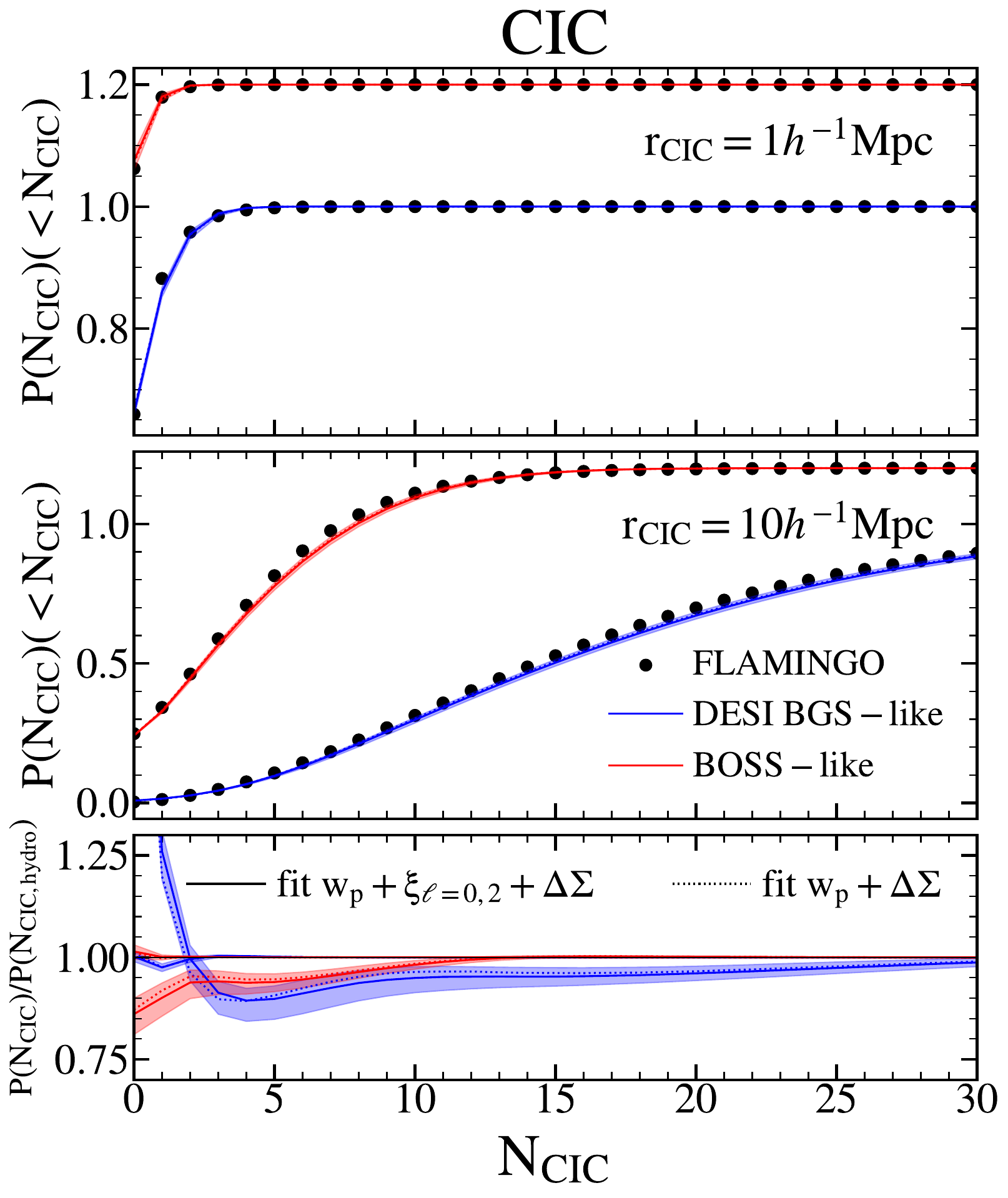}
\includegraphics[width=0.347\textwidth]{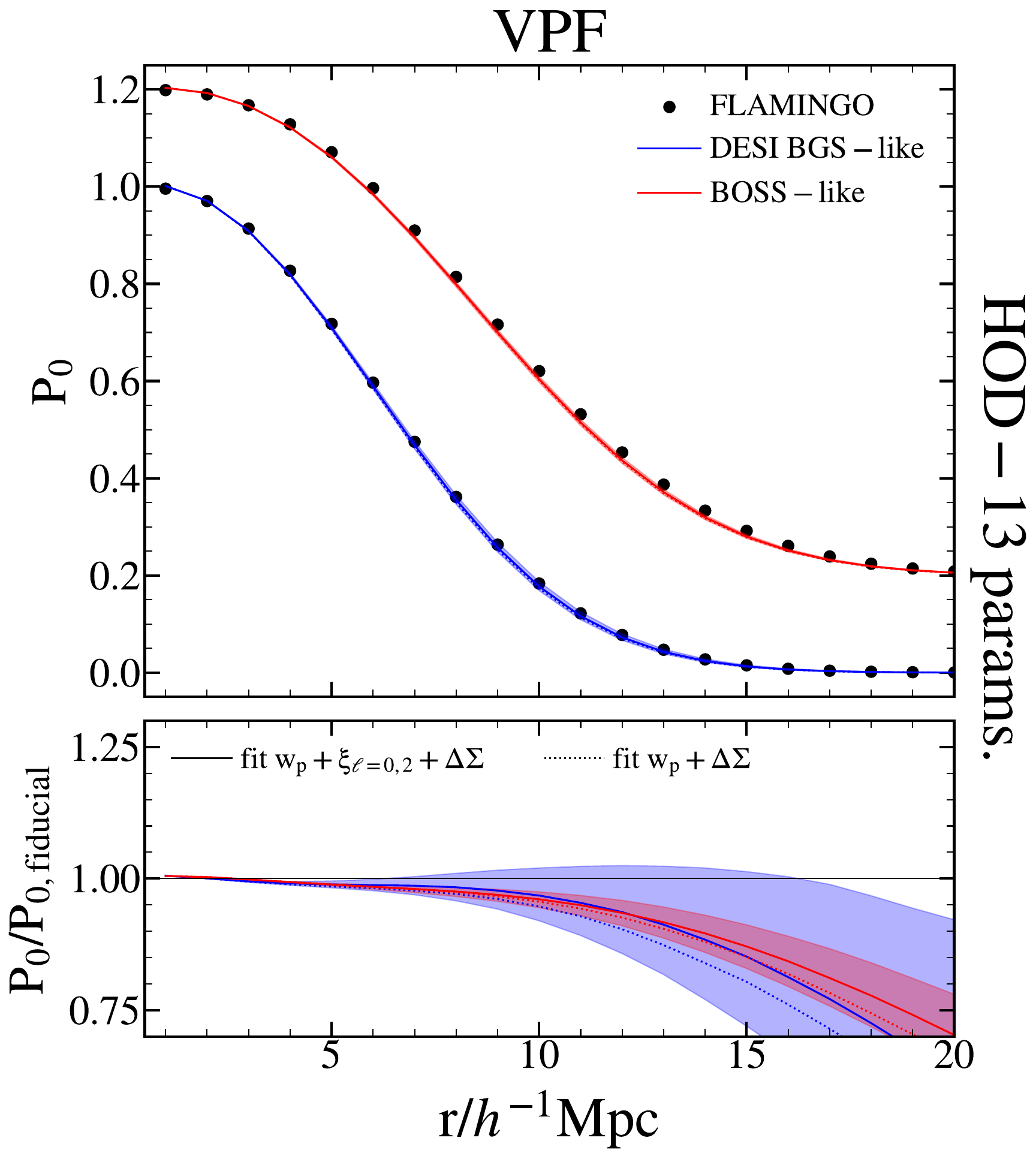}
\caption{Similar to Fig.~\ref{fig:ext1_shame}, but for the \hod~model.}  
\label{fig:ext1_hod}
\end{figure*}

\begin{figure}
\includegraphics[width=0.45\textwidth]{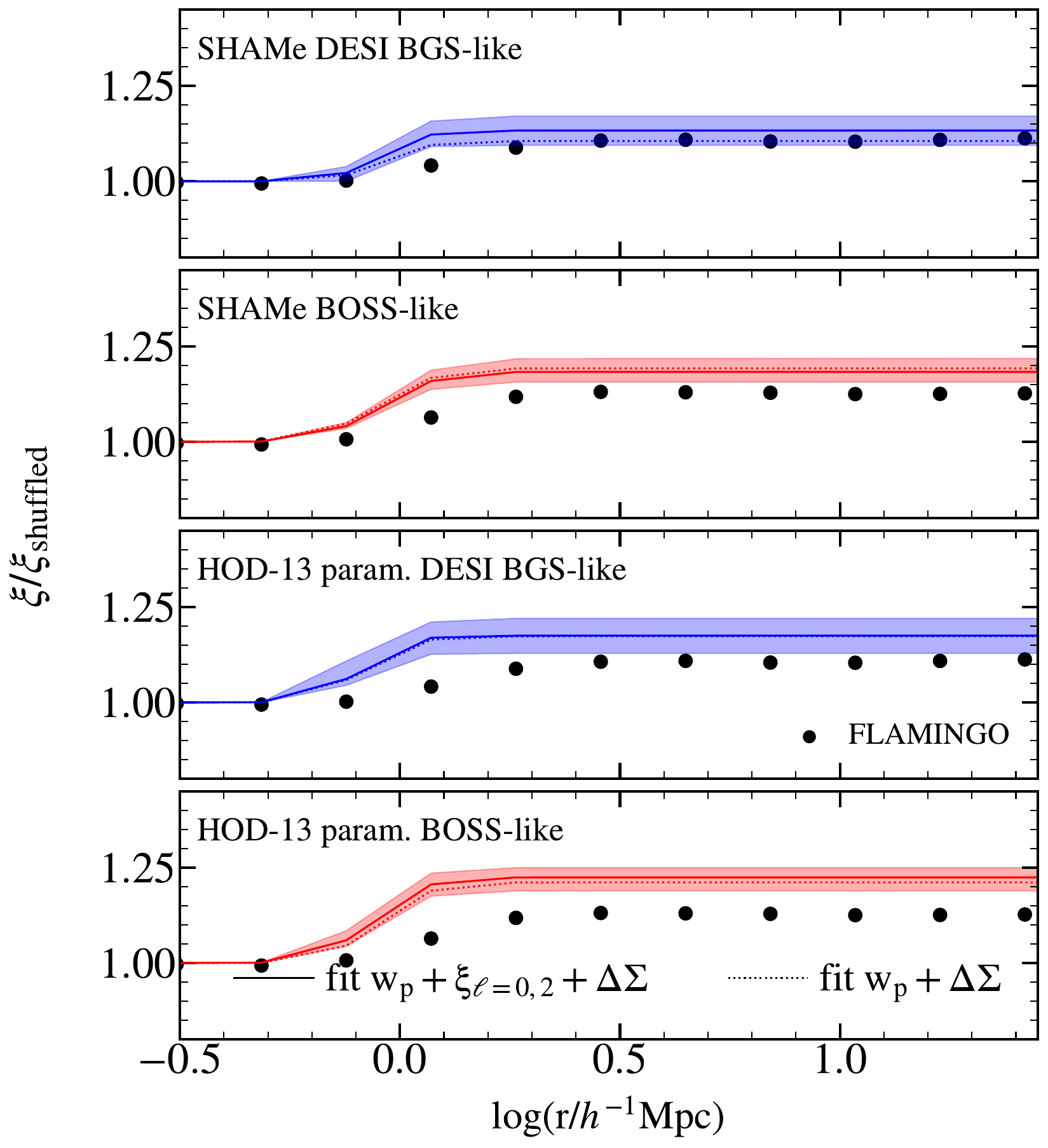}
\includegraphics[width=0.448\textwidth]{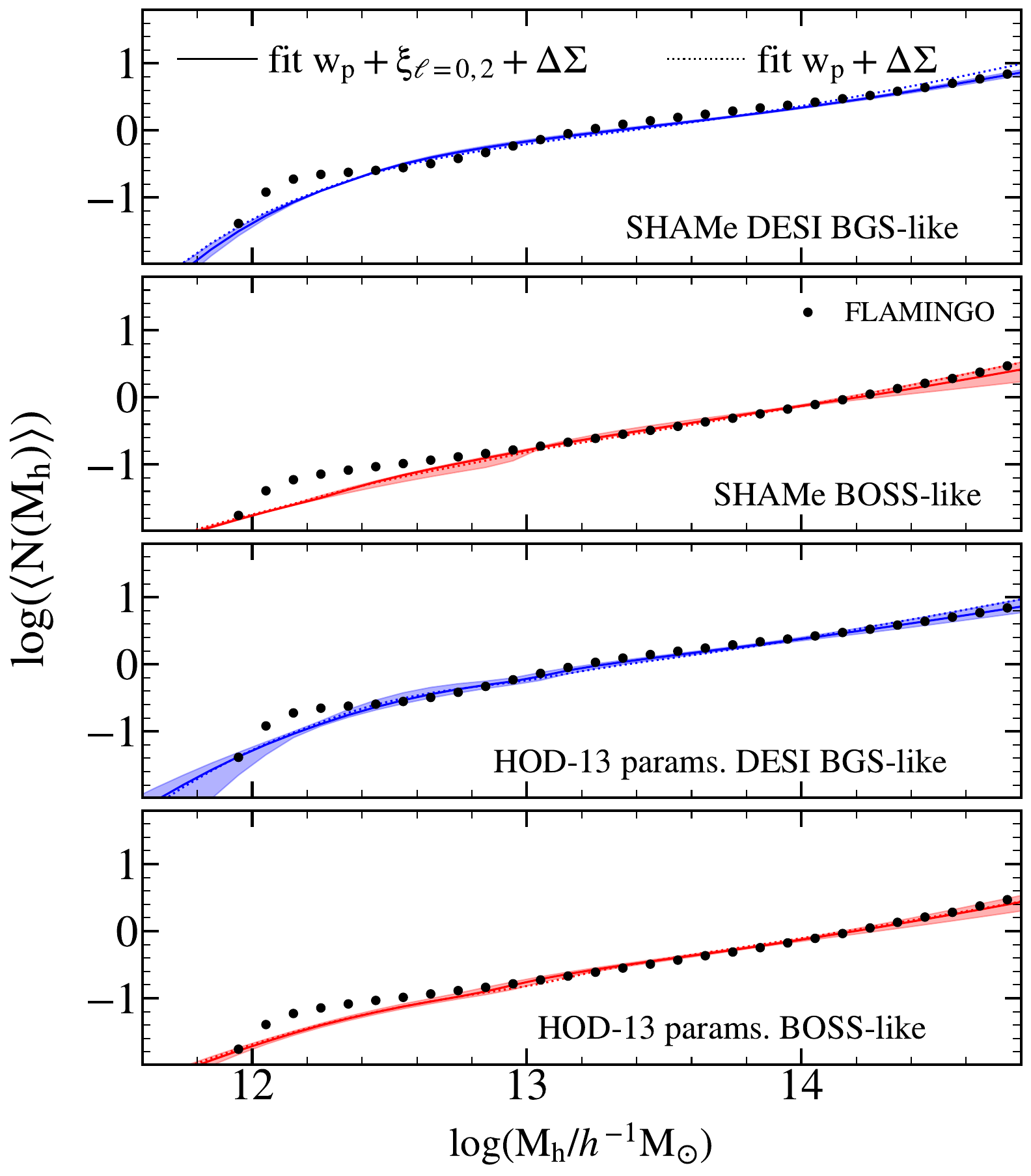}
\caption{(Top) The galaxy assembly bias of the \flamingo~simulation (dots points) and the empirical models when fitting to the galaxy clustering and galaxy-galaxy lensing predicted by \flamingo. From top to bottom, we show the \shame~predictions for the \sampleA~galaxies, the \shame~predictions for the \sampleB~galaxies, the \hod~predictions for the \sampleA~galaxies, and the \hod~predictions for the \sampleB~galaxies. The shaded region around the solid line represents the 1-$\sigma$ prediction region for the empirical models, as explained in section~\ref{sec:fit}. (Bottom) Similar to the top panel, but for the halo occupation number.}
\label{fig:gab_hod}
\end{figure}

In this section, we show the results for to the rest of the statistics. As mentioned in Sect.~\ref{sec:fit}, we predict these statistics by taking 1000 random points from the MCMC chains used to fit the galaxy clustering and galaxy-galaxy lensing and then evaluate them using \GalaxyEmu. The predictions for the kNN-CDF, counts-in-cylinder and void probability function are shown in Fig.~\ref{fig:ext1_shame} for the \shame~model and in Fig.~\ref{fig:ext1_hod} for the \hod~model. For the kNN-CDF, we only show $k=2\ \&\ 5$ and for the counts-in-cylinder only r = 1 \& 10 $\hMpc$. The lines represent the median of the 1000 random points, while the shaded region represents the 16th and 84th percentile of the distribution (only shown for the solid line, i.e., the fit for \proj,~\mono,~\quadr~\&~\lensing). These statistics, along with the galaxy clustering and galaxy-galaxy lensing measurements of the previous section, can be directly measured from observations. This means these properties provide an opportunity to improve the constraints from empirical models when used to predict cosmology or other properties from a target galaxy sample.

Overall, the predictions for both empirical models show very good agreement with the ones from \flamingo. Most of the differences in the bottom panels (the ratio between the mocks and \flamingo) occur when both curves reach small values, and the ratio diverges (notice that the \sampleB~sample is displaced along the y-axis for clarity). The ability to predict multiple statistics simultaneously demonstrates the robustness of the current generation of empirical models. These results also highlight the amount of information contained in galaxy clustering and galaxy-galaxy lensing and the advantages of having a model capable of simultaneously reproducing all these statistics. 

Contrary to the previous section, the predictions of \shame~show better agreement than the predictions of the \hod~model. In this case, we find some systematic differences between both samples (with the \sampleA~sample showing larger differences in the kNN-CDF and counts-in-cylinder, and the \sampleB~ sample for the void probability function). We also notice that the constraints for this technique are broader. We test using real mocks instead of \GalaxyEmu, finding the same discrepancies. The additional flexibility of the \hod~model may cause these differences. The larger number of free parameters of the \hod~ (2-3 times more than the \shame~model) improves the fit, as demonstrated in the previous section. Still, these fits may attempt to reproduce any minor feature of the target sample, sometimes forcing a parameter to reach unrealistic values to improve the fit.

This becomes more evident in the top panel of Fig.~\ref{fig:gab_hod}, where we show the predicted level of galaxy assembly bias of the models. For this statistic, the \hod~model does not properly constrain the level of galaxy assembly bias of \flamingo. For these cases, we believe the \hod~model requires a larger amount of assembly bias to fit the clustering of \flamingo, to compensate for any limitation of the model, which results in a larger discrepancy with the statistics we do not use to fit. The \shame~model is able to correctly constrain galaxy assembly bias within one sigma for the \sampleA~sample and two sigmas for the \sampleB~sample, consistent with previous studies \citep{C23b, C23c}. The galaxy assembly bias predictions for the \flamingo~variations exhibit a similar level of accuracy as for the largest \flamingo~simulation (see Apendix~\ref{sec:hod_gab_extra} for more details).

The last statistic we measure is the halo occupation number (bottom panel of Fig.~\ref{fig:gab_hod}). In this case, we almost perfectly reproduce the satellite occupation (massive end of the plot) but struggle to constrain the central occupation for all samples within the shaded region. It is possible that there are some discrepancies in the classification of ``centrals'' and ``satellite'' galaxies between the simulations due to the different resolutions and baryonic implementation. The different halo finder algorithms can also affect the central/satellite classification with \textsc{VELOCIraptor}, allowing satellites to be more massive than centrals on some occasions. It is also well documented that the ``sharpness'' (width) of the transition from zero to one central per halo is not strongly constrained by galaxy clustering \citep{Zehavi:2011}; nevertheless, we were surprised that this difference did not manifest itself in any of the additional clustering statistics we measure in this work. This suggests that both models still have room for improvement. The halo occupation number predictions for the \flamingo~variations exhibit a similar level of accuracy as for the largest \flamingo~simulation (see Apendix~\ref{sec:hod_gab_extra} for more details).

All in all, while there are some discrepancies between the models and \flamingo, they all performed very well. Even the poorest performance, which is arguably the galaxy assembly bias for the \hod, is still, in our opinion, an excellent prediction. Up till now, most constraints to galaxy assembly bias (most of them using HODs) have only limited themselves to predicting if the assembly bias level is consistent with zero. To our knowledge, this is the only proper validation of the \hod~predictions of assembly bias using a galaxy formation model where we can measure this signal's amplitude.

\section{Testing the performance of simplified empirical models}
\label{sec:simple}

\begin{figure*}
\includegraphics[width=1\textwidth]{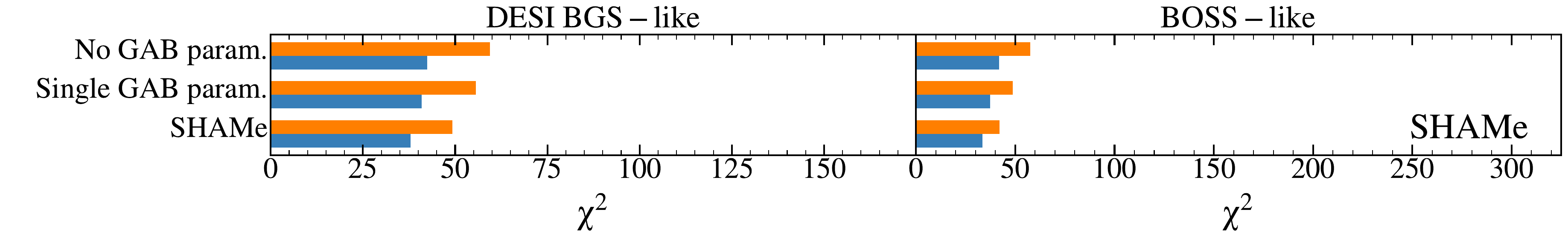}
\includegraphics[width=1\textwidth]{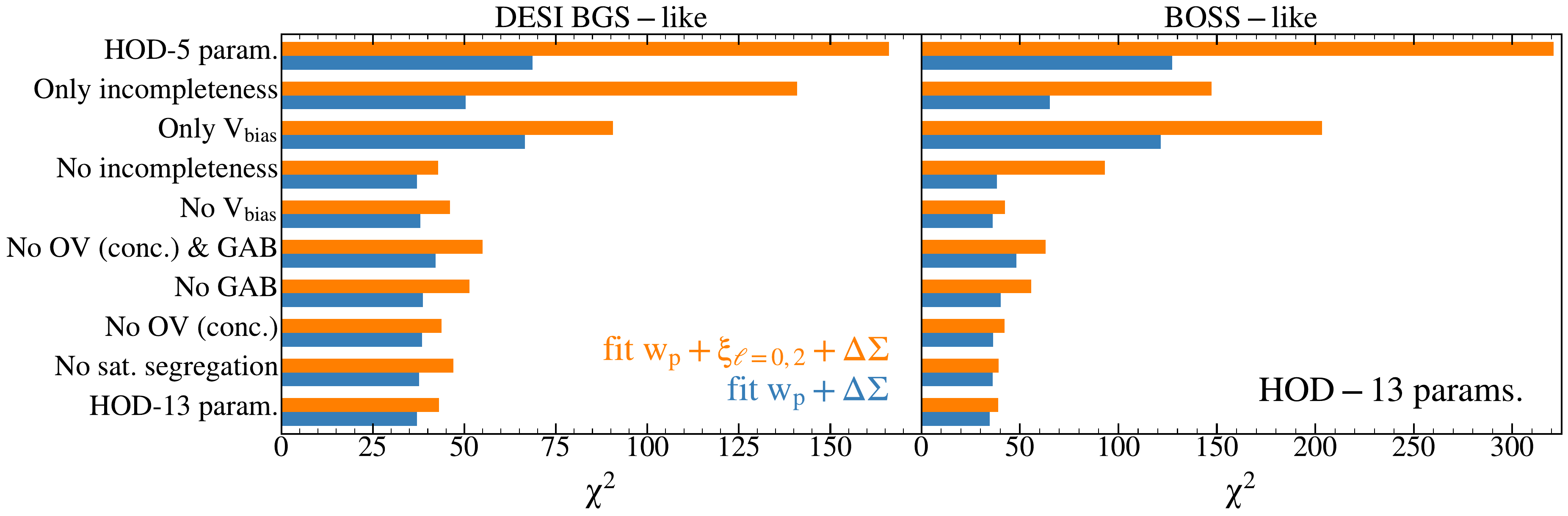}
\caption{(Top) The $\chi^2$ of the fits of the \shame~model for the \sampleA~and \sampleB~samples (left and right panels, respectively) when reproducing simultaneously: the projected correlation function and the galaxy-galaxy lensing signal (\lineB, blue bars) and the projected correlation function, the monopole and quadrupole of the correlation function and the galaxy-galaxy lensing signal (\lineC, orange bars). The bottom bar shows the $\chi^2$ for the primary model, and the following bars represent two simplifications of the models by removing either one or both of the galaxy assembly bias parameters. (Bottom) Similar to the upper panel but for the \hod~model.}  
\label{fig:chi2}
\end{figure*}

\begin{figure*}
\includegraphics[width=0.5\textwidth]{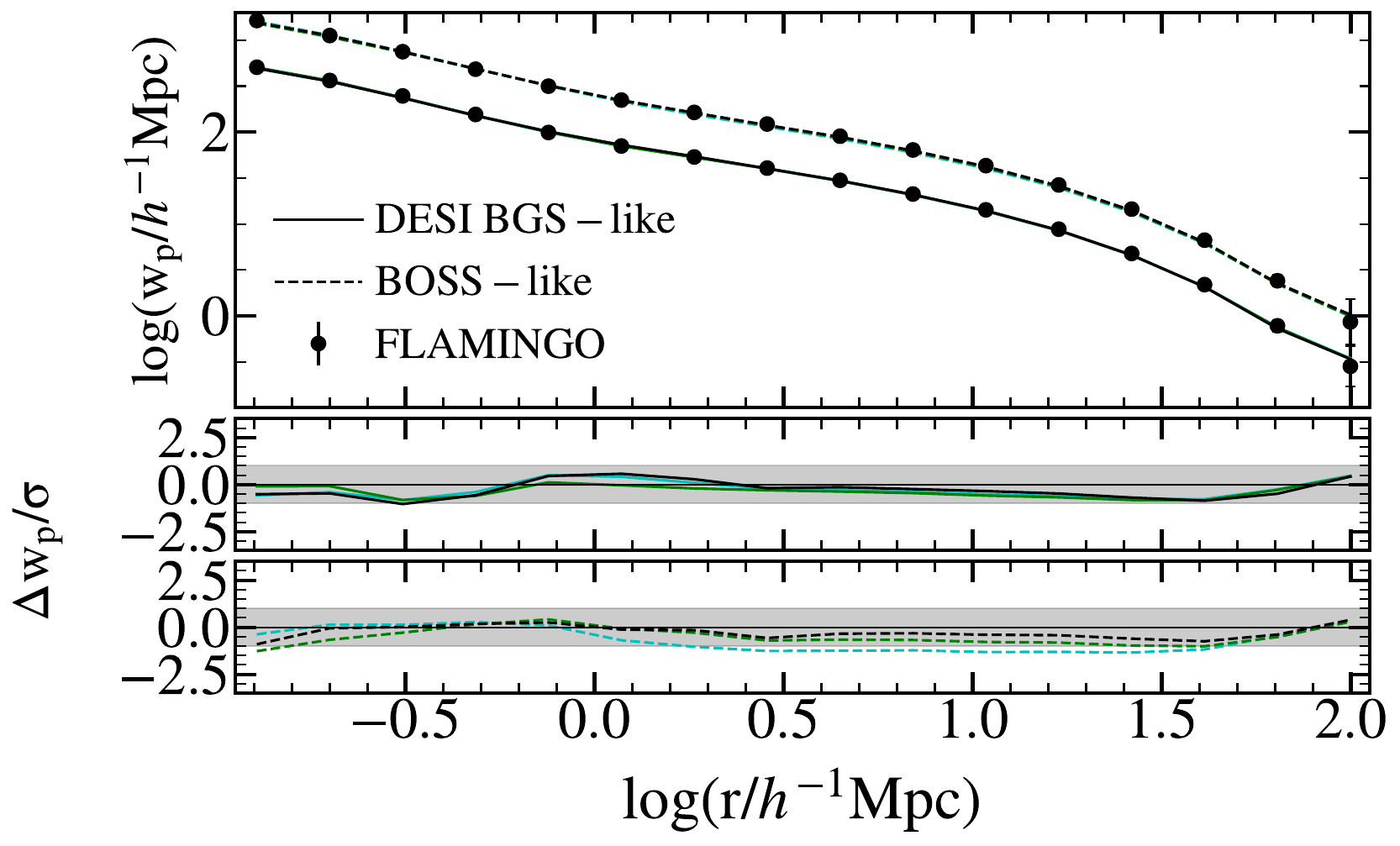}
\includegraphics[width=0.5\textwidth]{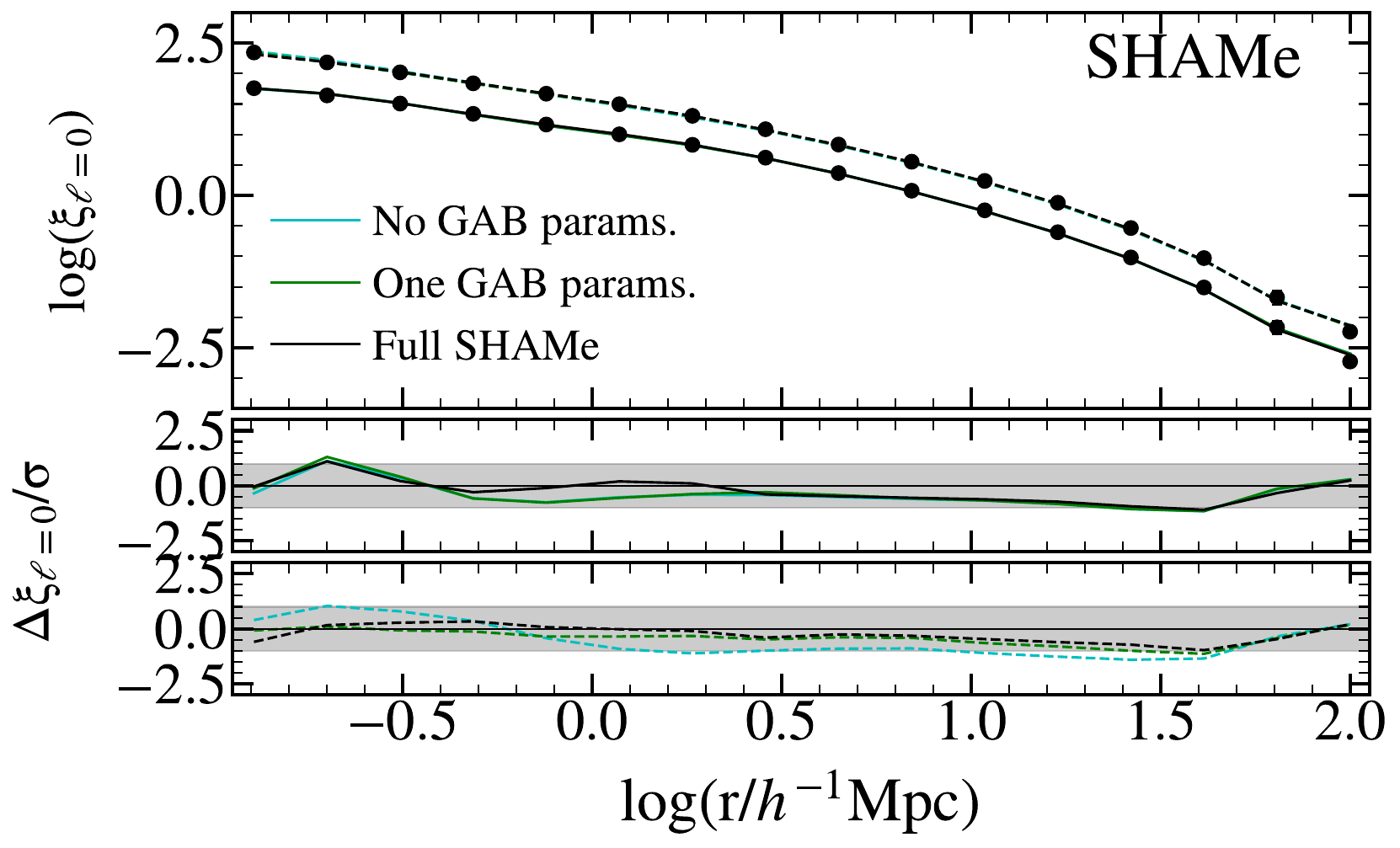}
\includegraphics[width=0.5\textwidth]{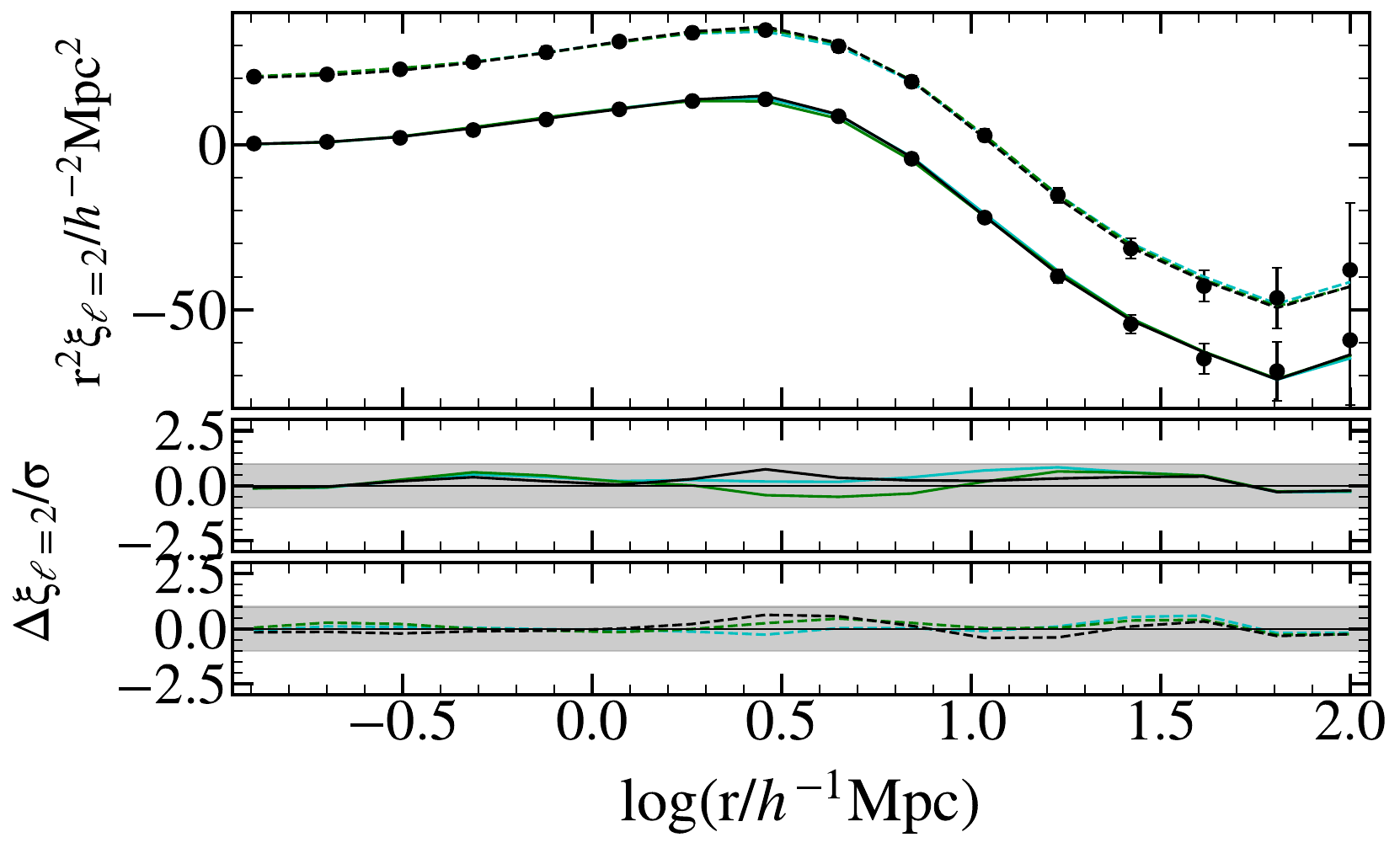}
\includegraphics[width=0.5\textwidth]{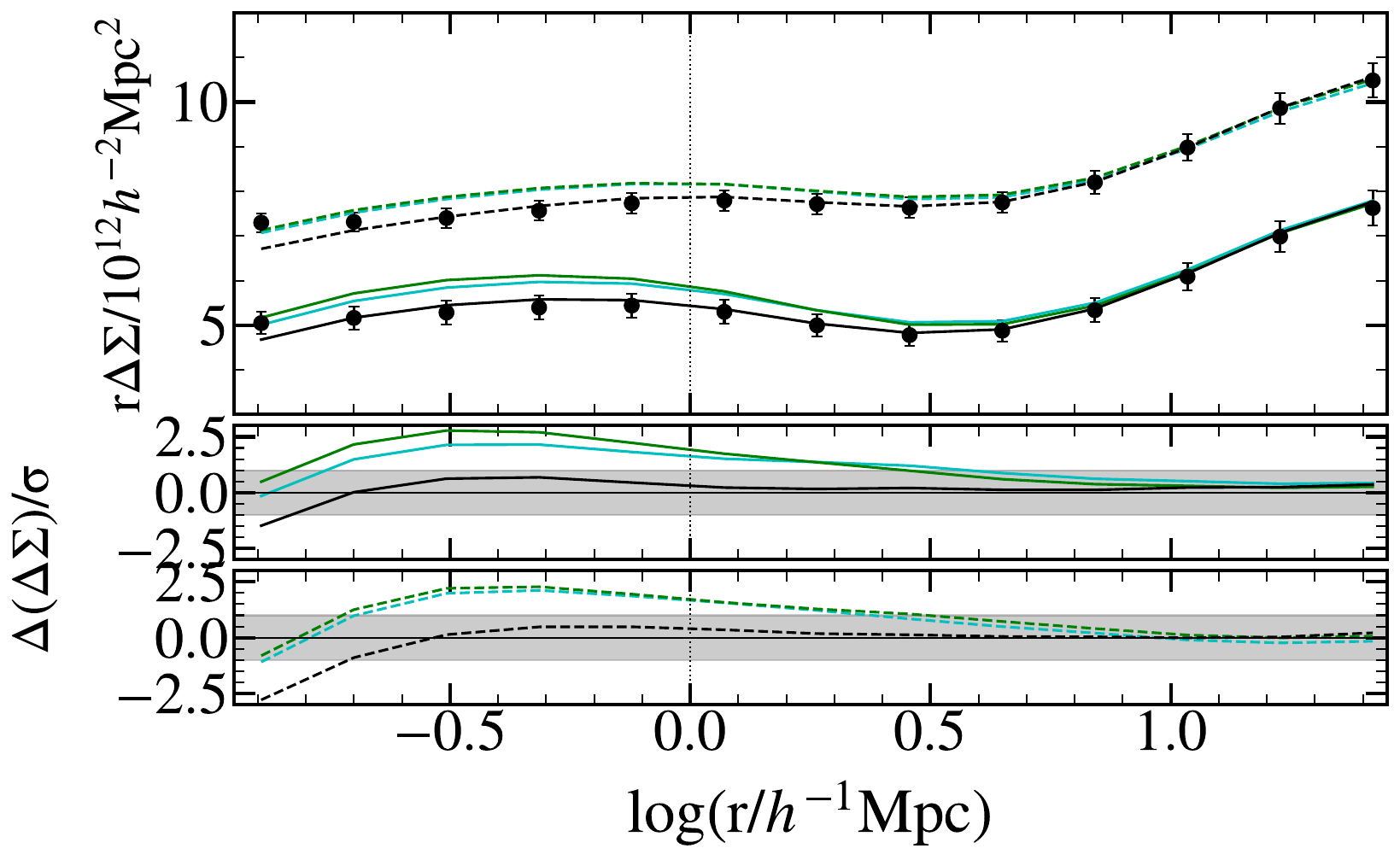}
\caption{The galaxy clustering (\proj, top left panel; \mono, top right panel; \quadr, lower left panel) and galaxy-galaxy lensing (\lensing, lower right panel) for the \flamingo~simulation (symbols), the fiducial \shame~model (black lines) and two different variations of the \shame~model (colour lines). The bottom subpanels show the difference between the empirical models and \flamingo, normalised by the error.}  
\label{fig:simple_shame}
\end{figure*}

\begin{figure*}
\includegraphics[width=0.5\textwidth]{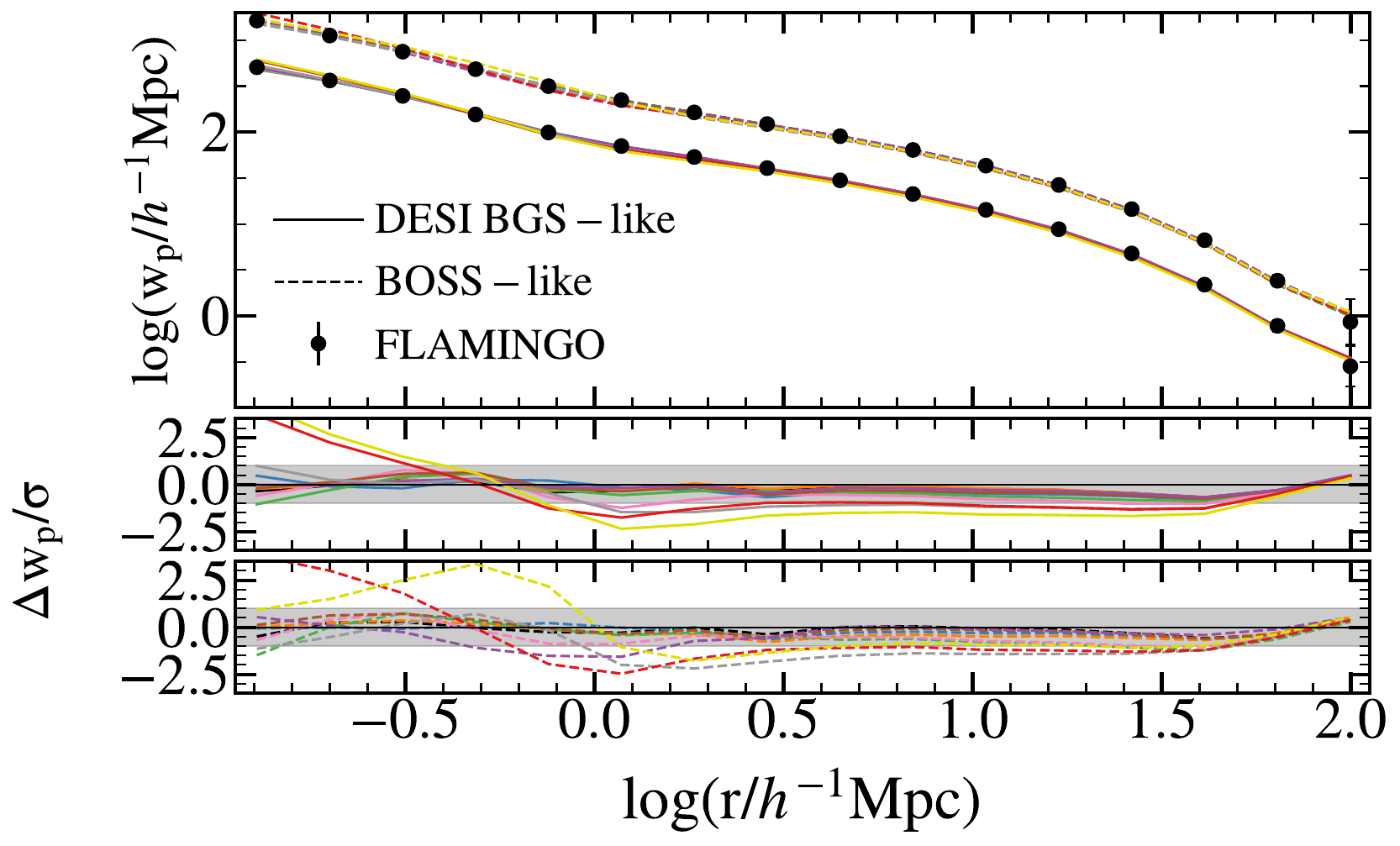}
\includegraphics[width=0.5\textwidth]{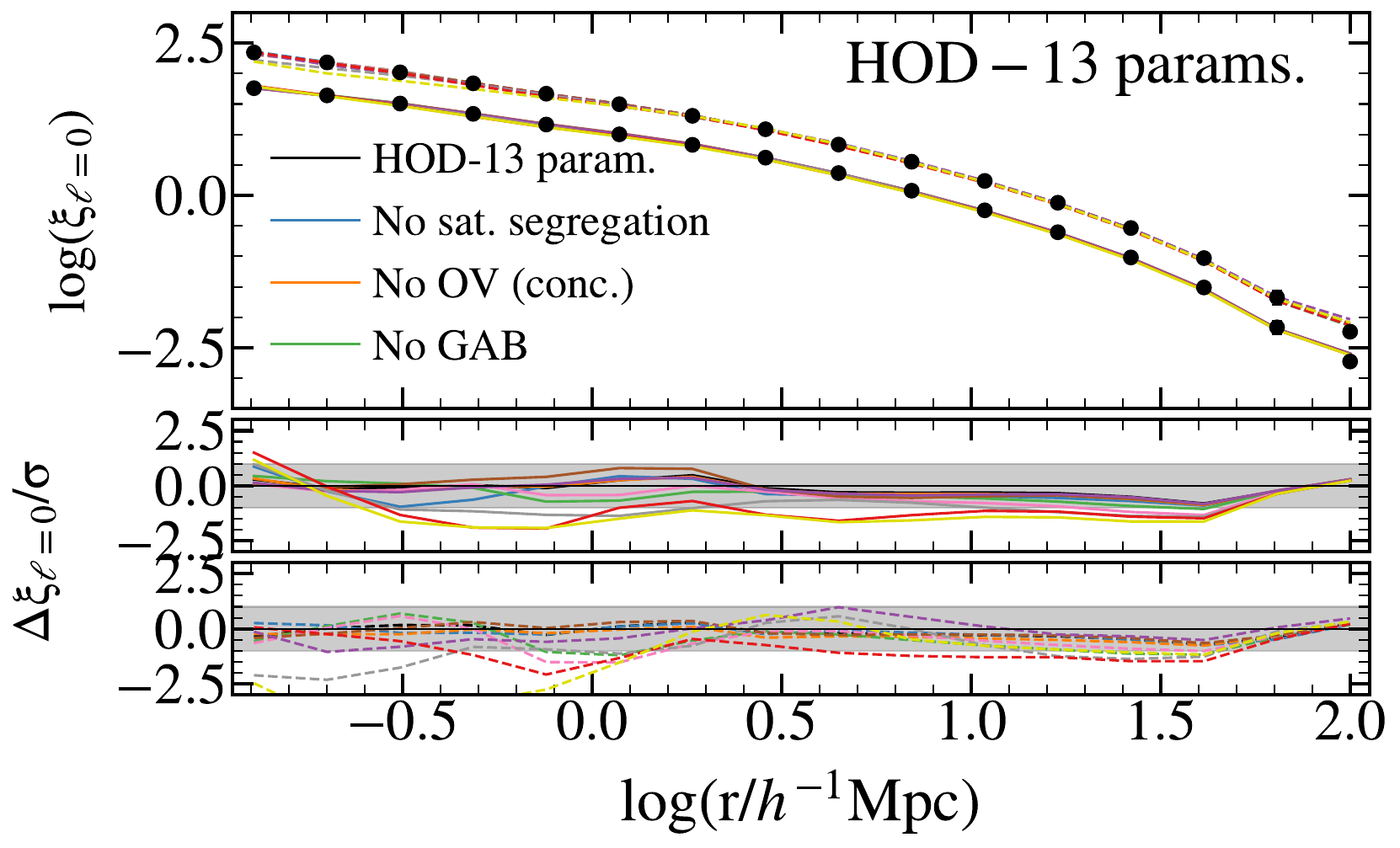}
\includegraphics[width=0.5\textwidth]{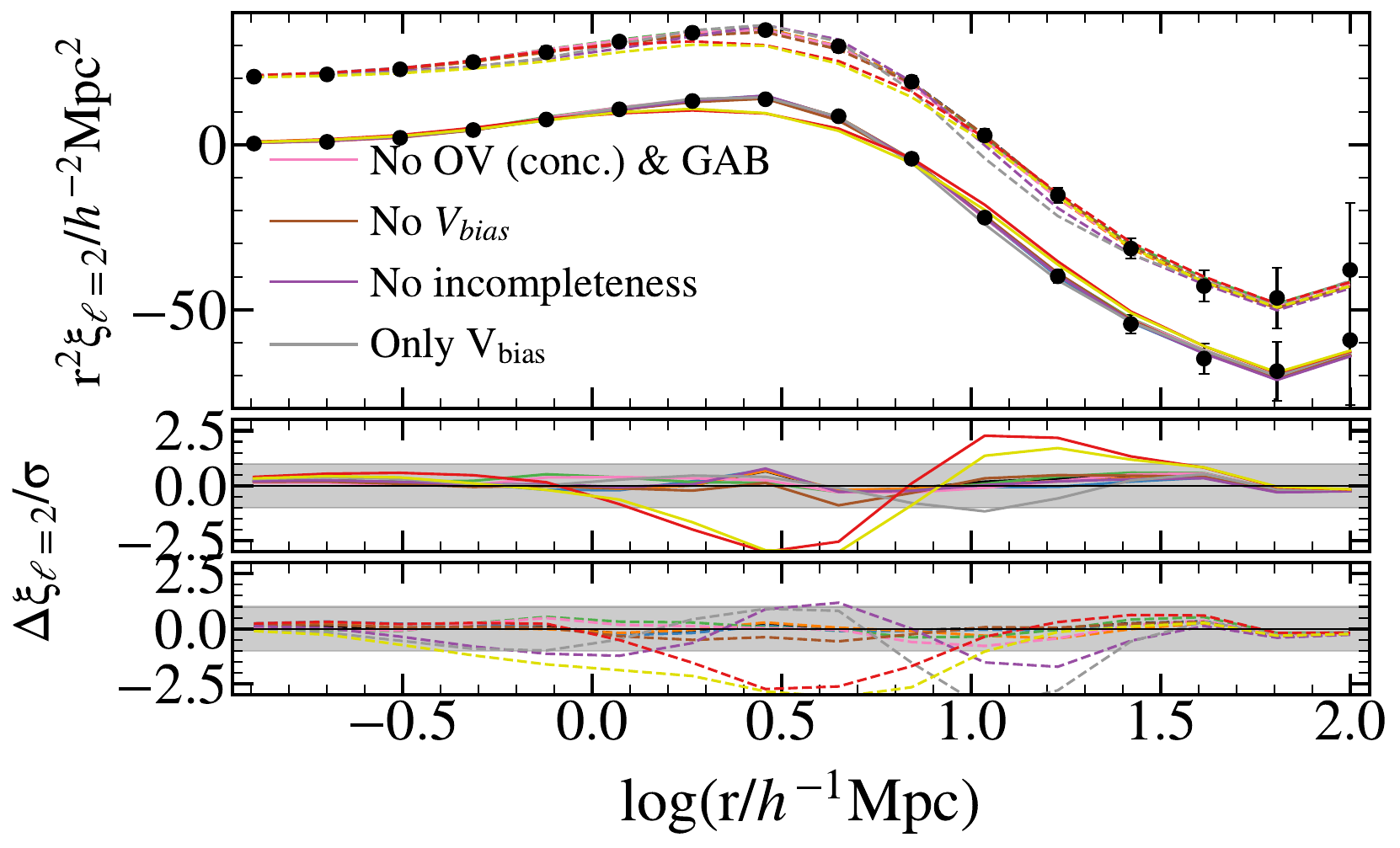}
\includegraphics[width=0.5\textwidth]{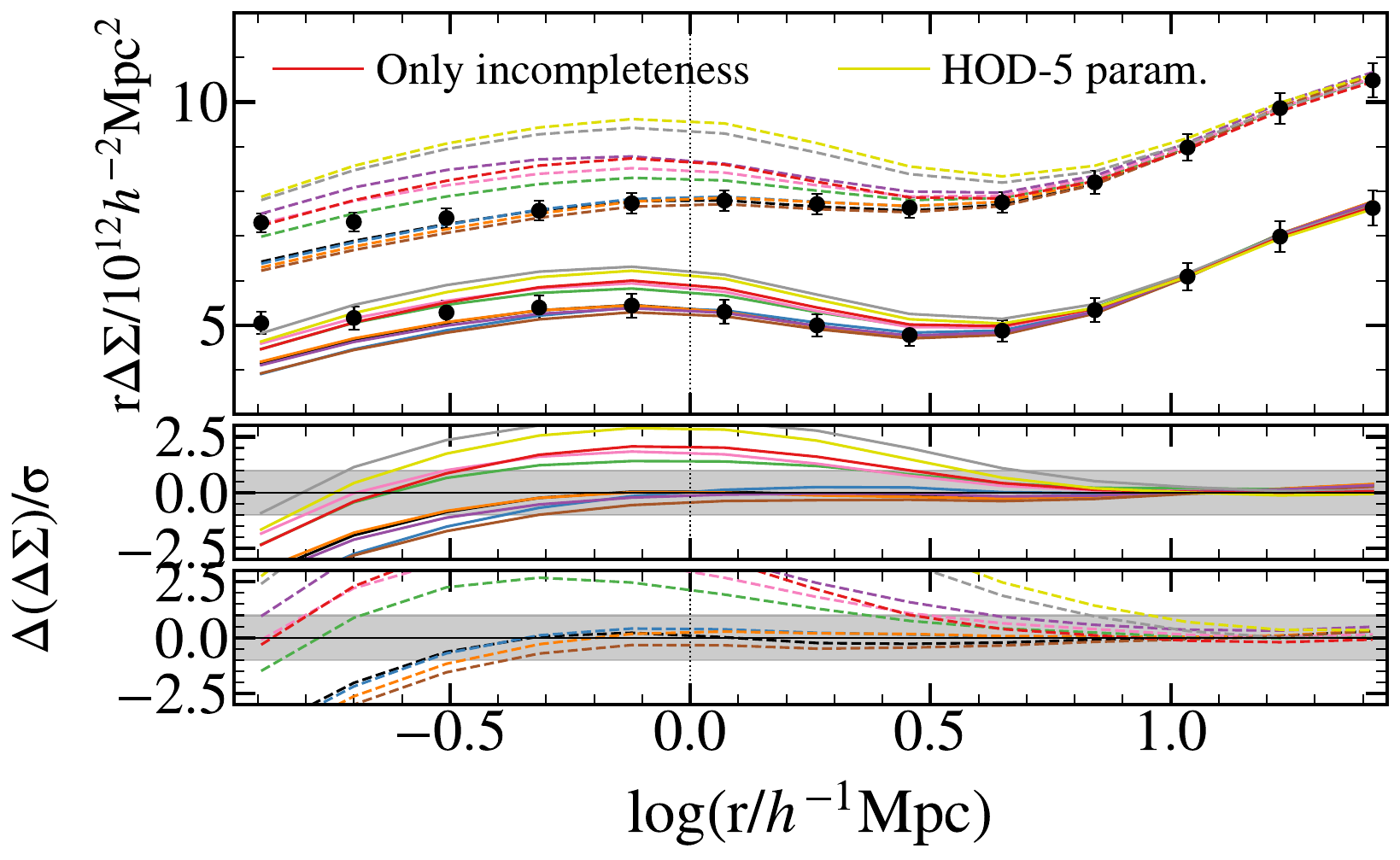}
\caption{Similar to Fig.~\ref{fig:simple_shame}, but for the \hod~model.}  
\label{fig:simple_hod}
\end{figure*}

In the previous section, we showed the performance of the \shame~and the \hod~models when reproducing galaxy clustering and galaxy-galaxy lensing, as well as their predictions for other high-order statistics. The excellent agreement we achieved for both models was partly due to the different extensions we added over the most basic version of these models, the basic SHAM and the 5-parameter HOD described in Sect.~\ref{sec:SHAMe}~\&~\ref{sec:HOD}.

We show the predictions for the \shame~simplified models in Fig.~\ref{fig:simple_shame}. We again only show the predictions when fitting all the statistics simultaneously, but show the value of the $\chi^2$ for both fittings (all the statistics and only using the projected correlation function and galaxy-galaxy lensing) in the top panel of Fig.~\ref{fig:chi2}. For this empirical model, we notice a significant improvement in performance when including the galaxy assembly bias parameters. We find that having only one assembly bias parameter is not enough to achieve a good performance.

The statistic that exhibited the most substantial enhancement upon the incorporation of the assembly bias parameter was galaxy-galaxy lensing. The lack of additional assembly bias produces an excess of signal on small scales compared to the target one, which goes in the same direction as the lensing-is-low signal. This finding aligns with the results presented in \cite{C23c}, where we show that assembly bias was key to simultaneously reproducing galaxy clustering and galaxy-galaxy lensing from the galaxies of BOSS. These findings are also consistent with those of \cite{Leauthaud:2017}, who found that the lensing-is-low problem is also present in a basic SHAM model.

For the \hod, we used nine additional models, including the full model a basic 5-parameter HOD, the basic HOD plus incompleteness (6 free parameters), the basic HOD plus velocity bias (7 free parameters), the full 13-parameters \hod~without incompleteness (12 free parameters), velocity bias (11 free parameters), galaxy assembly bias (11 free parameters), occupancy variation (11 free parameters), galaxy assembly bias and occupancy variation (9 free parameters), and satellite segregation (12 free parameters). We show the predictions of these models in Fig.~\ref{fig:simple_hod} and the $\chi^2$ in the bottom panel of Fig.~\ref{fig:chi2}.

For this empirical model, we find that all extensions are necessary, except for the occupancy variation with concentration. We see no significant improvement for any sample or combination of the fitting statistics when we include this extension. This is consistent with the results from \cite{Xu:2021b}, who show that the galaxy assembly bias of galaxy formation models is not well modelled by concentration alone, and \cite{Yuan:2022b} who show that this extension was not useful to reproduce galaxy clustering. We originally believed this extension could help to reproduce galaxy-galaxy lensing, but we find this is also not the case. Because of this, we will consider omitting this extension from the next version of \GalaxyEmu.

For the rest of the \hod~extensions, satellite segregation was the least important parameter we included, but it still provides an improvement when fitting \mono~and \quadr. Velocity bias appears to be the next least important extension. This extension is expected to be more relevant in samples with a higher number density, i.e., a larger satellite fraction. We indeed tested fitting only \proj, \mono~, and \quadr~in the \sampleA~sample, finding that velocity bias is fundamental for having a good fit (not shown here). Incompleteness is not necessary for the \sampleA~sample, which is expected since this sample is complete. Leaving out this extension for the \sampleB~sample reduces the fit quality significantly. Finally, we found that assembly bias is the most important extension of the \hod~model. We see that incompleteness and galaxy assembly bias have a significant impact when attempting to reproduce galaxy-galaxy lensing. Their absence produces an increase in signal for the galaxy-galaxy lensing at small scales, consistent with the lensing-is-low problem. We also notice that more complex galaxy samples, such as the one of BOSS, will present a more significant deviation in their galaxy-galaxy lensing measurement than the one of a complete sample, such as for our \sampleA~sample. These results are in agreement with \cite{Chaves:2023}, where we predict that these limitations of the basic HOD approach are partially responsible for the model not being able to reproduce galaxy clustering and galaxy-galaxy lensing simultaneously.

\section{Summary}
\label{sec:summary}
In this paper, we investigate the performance of two state-of-the-art galaxy population models, a 13-parameter HOD and the \shame~model, in reproducing the galaxy clustering (in redshift space), galaxy-galaxy lensing, and five other clustering statistics from the \flamingo~hydrodynamic simulation. In all cases, the agreement between \flamingo~and our mocks is excellent. Here, we summarise our most important findings.

\begin{itemize}

\item We build \GalaxyEmu, an emulator capable of reproducing the projected correlation function, the monopole, quadrupole and hexadecapole of the correlation function, the galaxy-galaxy lensing, the k-nearest neighbour cumulative distribution function, the counts-in-cylinder, the void probability function, the galaxy assembly bias and the halo occupation number. This emulator can predict these statistics for \shame~and \hod~mocks, for redshifts between 0 and 0.8 and number densities between 0.0001 $\ihMpcC$ and 0.00316 $\ihMpcC$.

\item We reproduced galaxy clustering in redshift space (\proj,~\mono, and \quadr) and galaxy-galaxy lensing (\lensing) for two different galaxy samples from the largest hydrodynamic run of the \flamingo~suite of simulations. The samples have a distinct selection function: a \sampleA~sample with a number density of 0.001 $\ihMpcC$ complete in the r-band, and the incomplete red \sampleB~sample with a number density of 0.00316 $\ihMpcC$ (Fig.~\ref{fig:smf}). The \shame~and \hod~models perform well for both target samples (figures~\ref{fig:stat_fit_shame}~\&~\ref{fig:stat_fit_hod}). The galaxy-galaxy lensing was only fit to scales above 1 $\hMpc$ to avoid scales impacted by baryonic effects.

\item We also test the performance of the two empirical models on the \flamingo~suite of simulations with different astrophysical implementations. While the clustering measurements were different between the simulations, the models were able to fit them equally well for all cases (figures~\ref{fig:MultiModel_SHAMe}~\&~\ref{fig:MultiModel_HOD}).

\item We test the ability of the empirical models to reproduce the other statistics predicted by \flamingo~when fitting only to the galaxy clustering and galaxy-galaxy lensing predictions, finding excellent agreement for most cases. In this case, the results look slightly better for the \shame~model. The differences between the models and \flamingo~are largest for the low-end part of the halo occupation number (which is usually poorly constrained by galaxy clustering) and for the galaxy assembly bias in the case of the \hod~model (figure.~\ref{fig:ext1_shame},~\ref{fig:ext1_hod}~\&~\ref{fig:gab_hod})

\item The empirical models used in this study include several extensions beyond their more conventional implementations. We explore the effects of these extensions when reproducing galaxy clustering and galaxy-galaxy lensing. We find that, for both models, incorporating galaxy assembly bias is critical for accurately reproducing galaxy clustering and galaxy-galaxy lensing. Using one galaxy assembly bias parameter instead of two yields comparable performance for the \shame~ model. For the \hod~model, we found that all extensions were useful to better reproduce galaxy clustering and galaxy-galaxy lensing, except for the occupancy variation with concentration (figures~\ref{fig:simple_shame}~\&~\ref{fig:simple_hod}).

\item An excess of the galaxy-galaxy lensing signal by empirical models on BOSS has been reported when using basic empirical models. The extensions to the empirical models are very effective at reducing the amplitude of the galaxy-galaxy lensing signal on small scales. This effect is especially noticeable in the \sampleB~sample with the \hod~model. The additional assembly bias for both empirical models and the incompleteness for the \hod~model were found to be the most effective extensions for decreasing the difference with the target lensing signal.

\end{itemize}

The empirical models we used in this work used dark matter-only simulations, which lack of the complexity of a hydrodynamic simulation as \flamingo. These mocks are fast, taking only a fraction of the time required to run a hydrodynamic simulation or a semi-analytical model, and are entirely based on our understanding of the galaxy-halo connection. The excellent performance achieved by both empirical models validates them as suitable methods for reproducing any of the statistics measured in this study. We tested these models for samples with very different and distinct selection functions, and we also tested our main results in a suite of hydrodynamic simulations with various astrophysical implementations, demonstrating that our results are robust to uncertainties in the simulation predictions.

Although the agreement between the spatial distribution of our mocks and that of \flamingo~is excellent, we observe that both models still have room for improvement. Fig.~\ref{fig:chi2} shows that the \shame~model performed less well than the \hod~model, particularly when fitting the multipoles of the correlation function. This means that better modelling of galaxies' velocity or adding a velocity bias parameter may be able to improve these results even further. On the other hand, the additional flexibility of the HOD may create models with some unphysical characteristics, such as a larger amount of assembly bias. We believe the additional assembly bias found in the \hod~mocks may have caused performance issues when reproducing the kNN-CDF, counts-in-cylinder, and the void probability function. These discrepancies can potentially be reduced by incorporating them into the fitting process, which could eventually improve the constraints on galaxy assembly bias. The predictions of these simplified models on galaxy-galaxy lensing confirm the finding of \cite{Chaves:2023} that part, if not all, of the origin of the lensing-is-low problem is caused by limitations in the galaxy modelling.

The findings in this paper provide proof of the accuracy of both empirical models. We intend to expand the reach of \GalaxyEmu~through scaled simulations \citep{Angulo:2010}, following the procedure of \cite{C23b}. In this approach, we can change not only the mock describing the galaxy-halo connection but also the cosmological parameters. This method will allow us to derive cosmological constraints (as well as assembly bias constraints) from galaxy clustering measurments. The validations provided by this work will allow us to test the additional cosmological constraints from the kNN-CDF, counts-in-cylinder, and void probability function. This work also validates the models' clustering predictions on scales spanning three orders of magnitude, which we anticipate will provide additional cosmological information. We are especially enthusiastic regarding the potential of the \shame~model. When combined with the scaling technique, its performance across all galaxy samples and statistics with only 5 free parameters could result in tighter cosmological constraints.

We want to emphasise that the conclusions of this work are only valid for the specific mock empirical models used here. We cannot guarantee that other SHAM and HOD model extensions will perform similarly well. In particular, we find that the performance of the models strongly depends on the environmental property used to mimic the assembly bias signal. We tested a more basic environmental property on both models before, finding that, while the \shame~approach produced similar results, the \hod~model was not able to simultaneously reproduce the galaxy clustering and galaxy-galaxy lensing predictions, finding a lensing-is-low like effect for the \sampleB~sample. We highly encourage the performance of basic tests on any empirical model before using them on an observational sample. This basic test includes checking its performance with a complex galaxy population model, such as a semi-analytical model or hydrodynamic simulations.

\begin{acknowledgements}
We thank Tom Abel and Eduardo Rozo for useful comments and suggestions.
SC acknowledges the support of the ``Juan de la Cierva Incorporac\'ion'' fellowship (IJC2020-045705-I).
REA \& SC acknowledge support under the grant number PID2021-128338NB-I00 from
the Spanish Ministry of Science and support from the European Research Executive Agency HORIZON-MSCA-2021-SE-01 Research
and Innovation programme under the Marie Skłodowska-Curie grant
agreement number 101086388 (LACEGAL).
REA and JCM acknowledge support from the ERC-StG number 716151 (BACCO). 
REA from the Project of Excellence Prometeo/2020/085 from the Conselleria d'Innovaci\'o, Universitats, Ci\`encia i Societat Digital de la Generalitat Valenciana and JCM from the European Union's Horizon Europe research and innovation programme (COSMO-LYA, grant agreement 101044612). IFAE is partially funded by the CERCA program of the Generalitat de Catalunya.
RK from the research programme
Athena 184.034.002 from the Dutch Research Council (NWO).
The authors also acknowledge the computer resources at MareNostrum and the technical support provided by Barcelona Supercomputing Center (RES-AECT-2019-2-0012 \& RES-AECT-2020-3-0014)
Technical and human support provided by DIPC Supercomputing Center is gratefully acknowledged.
This work used the DiRAC@Durham facility managed by the Institute for Computational Cosmology on behalf of the STFC DiRAC HPC Facility (\url{www.dirac.ac.uk}). The equipment was funded by BEIS capital funding via STFC capital grants ST/K00042X/1, ST/P002293/1, ST/R002371/1 and ST/S002502/1, Durham University and STFC operations grant ST/R000832/1. DiRAC is part of the National e-Infrastructure.
\end{acknowledgements}

\bibliographystyle{aa} 
\bibliography{aa.bib} 

\begin{appendix} 

\section{HOD extensions}
\label{sec:HOD_extra}

This section gives a detailed description of the HOD model used in this work. As mentioned in Sect.~\ref{sec:HOD}, we used the parametric form of \cite{Zheng:2005} and the HOD-extensions listed in \cite{Yuan:2022b}, plus some extra modifications we specify in the main text.
Following Zheng et al., we decompose the halo occupation into the contribution of centrals (${\rm N}_{\rm cen}$) and satellite galaxies (${\rm N}_{\rm sat}$). We defined the average halo occupation as a function of halo mass ($\rm \langle N(M_{h})\rangle$) as the sum of the average central occupation and the average satellite occupation:

\begin{equation}
 \langle {\rm N(M_{h})\rangle} ={\rm \langle N_{cen}(M_{h})\rangle} + {\rm \langle N_{sat}(M_{h})\rangle}.
\label{Eq:HOD_cen_sat}
\end{equation}

The central occupation is described as
\begin{equation}
 \langle N_{\rm cen}(M_{\rm h})\rangle = \frac{1}{2}\left[ 1 + {\rm erf} \left( \frac{\log M_{\rm h} - \log M_{\rm min}}{\sigma_{\log M}}  \right) \right],
\label{Eq:Cen_HOD}
\end{equation}
with $ {\rm erf}(x)$ is the error function,
\begin{equation}
 {\rm erf}(x) = \frac{2}{\sqrt{\pi}} \int_{0}^{x} e^{-t^2} {\rm d}t.
\end{equation}

The parameter $\Mmin$ is the halo mass where half of the haloes are occupied by a central galaxy (i.e., $\langle N_{\rm cen}(\Mmin) \rangle = 0.5$) and $\sigmaLogM$ characterize the ``sharpness'' (width) of the transition from zero to one central per halo and is linked to the scatter between the galaxy properties (e.g., luminosity, stellar mass, etc) to the host halo mass. 
The satellite occupation is described as 
\begin{equation}
 \langle N_{\rm sat}(M_{\rm h})\rangle = \left( \frac{M_{\rm h}-M_{\rm cut}}{M^*_1}\right)^\alpha,
\label{Eq:Sat_HOD}
\end{equation}
with $M_{\rm h}>\Mcut$, representing a power-law shape with a smooth cutoff at low halo masses; $\Mone - \Mcut \equiv M^{*}_1$ is the mass where there is on average one satellite galaxy per halo ($\langle N_{\rm sat}(M_1) \rangle = 1$) and $\alpha$ is the slope between the $\langle N_{\rm sat} \rangle - M_{\rm h}$ relation, and it has a value of $\approx 1$. 

By going through all the haloes of a dark matter simulation and populating these haloes with an average of $\langle N_{\rm cen}(M_{\rm h}) \rangle$ centrals and $\langle N_{\rm sat}(M_{\rm h}) \rangle$ satellites, a mock galaxy catalogue can be built in a matter of minutes, if not seconds. The probability of putting a central and satellite galaxy follows a nearest-integer distribution and a Poissonian distribution, respectively. Although some deviations from a Poissonian distribution have been detected in recent works, their impact on galaxy clustering is not expected to be significant \citep{Jimenez:2019} but could modify galaxy-galaxy lensing on sub-megaparsec scales \citep{Chaves:2023}. These scales, however, were not used during the fitting process of the galaxy-galaxy lensing in this paper. In its most basic form, central galaxies are placed in the centre of the potential of the dark matter haloes, whether the satellite galaxies are either assigned to the position of a random particle of their host halo or distributed following an NFW profile.

We now list the extensions we implement over the basic HOD model.

\subsection{Satellite segregation}
\label{sec:sat_segr}
Selection effects significantly impact the distribution of satellite galaxies within haloes. Redder galaxies tend to follow the density profile of haloes, while blue galaxies tend to accumulate in their outskirts \citep{Orsi:2018}. Other physical processes, such as baryonic effects, will also impact the satellite distribution (e.g., \citealt{vanDaalen:2014}). 

We assign the position and velocities of satellite galaxies using the dark matter particles of the halo. To select the particles for each galaxy, we follow \cite{Yuan:2022b}. In this approach, particles are first ranked by their distance to their halo centre. The ranking of each particle, $r_i$, will be 0 for the particle closest to the halo centre and $\rm N_{p}-1$ for the farthest particle (for a halo with $\rm N_{p}$ particles). The probability of a particle ``i'' hosting a galaxy is then described as:

\begin{equation}
p_{i} = \dfrac{\rm N_{\rm sat}} {\rm N_{\rm p}} \left[1+ {\rm s_{\rm segr.}}\left(1-\dfrac{2 r_i}{{\rm N_{\rm p}} - 1}\right)\right]
\label{Eq:Sat_segr}
\end{equation}
with $\rm N_{sat}$ the number of satellite galaxies of a halo and $s_{\rm segr.}$ a free parameter that can have a value within the range -1 to 1, with ``-1'' meaning that all galaxies are close to the halo centre, zero meaning the particles will be randomly chosen, and ``1'' that all galaxies will be in the outskirts of the haloes.

\subsection{Velocity bias}
\label{sec:vbias}

In addition to the positions, galaxy velocities are also not expected to be identical to the velocities of dark matter particles or the velocity of their host dark matter halo. Following \cite{Guo:2015, Guo:2015a}, we apply a correction to the velocities of centrals and satellite galaxies. For central galaxies, where the velocity is assumed to be equal to the velocity of the dark matter halo ($\vec{v_{h}}$), we add a Gaussian scatter equal to the linear velocity dispersion (i.e., $\sigma_{\rm vel}$) of the halo modulated by a free parameter, $\alpha_{c}$:
\begin{equation}
\vec{v}_{cen} = \vec{v}_{h} + \alpha_c \delta(\sigma_{\rm vel}).
\label{Eq:vbias_c}
\end{equation}
For the satellite galaxies, we modulate the relative velocity of the assigned dark matter particle to its host dark matter halo ($\vec{v_{dm.\ part.}}$) by the free parameter $\alpha_{s}$:
\begin{equation}
\vec{v}_{sat} = \alpha_s\vec{v}_{dm.\ part.}.
\label{Eq:vbias_s}
\end{equation}
While these modifications will not change the positions of galaxies and will have a low to null impact on the projected correlation function (when integrated over a long range along the line of sight) or galaxy-galaxy lensing, they will impact other statistics more susceptible to redshift space distortion effects, such as the multipoles of the correlation function.

\subsection{Galaxy assembly bias and occupancy variation}
\label{sec:HOD_GAB}

Unlike a SHAM-like model, the basic HOD does not include galaxy assembly bias or occupancy variation since the occupation number depends only on halo mass. We add additional dependences on concentration and environment following \cite{Xu:2021b}. This approach consists of having variable values of $\Mmin$ and $\Mone$ as a function of secondary halo properties. As with the SHAMe model, we use the individual bias-per-object \citep{Paranjape:2018} to characterize the large-scale environment and add an additional bias to the sample that will behave as galaxy assembly bias. We also add an additional dependence on concentration, mainly for historical reasons, since this was the most common way to add assembly bias until a few years ago \citep{Hearin:2016}, but later works have shown it does not add a significant amount. Since the dependence on concentration impacts the internal distribution of the satellite galaxies (by preferentially choosing more/less concentrated haloes) rather than assembly bias, we will refer to this effect as occupancy variation. 

The modified values of $\Mone$ and $\Mmin$, called $\Mone^{\rm mod}$ and $\Mmin^{\rm mod}$, are then computed using the ranking of the individual bias-per-object ($b^{\rm rank}_{\rm ind}$) and the halo concentration ($c^{rank}$) of haloes of similar mass (bins of 0.1 dex in halo mass). The resulting values for these parameters are:
\begin{equation}
{\rm log}(\Mmin^{\rm mod}) = {\rm log}(\Mmin) + OV_{\rm cen}\left( c^{\rm rank} - 0.5 \right) + GAB_{\rm cen}\left( b^{\rm rank}_{\rm ind} - 0.5 \right),
\label{Eq:Mmin_gab}
\end{equation}
\begin{equation}
{\rm log}(\Mone^{\rm mod}) = {\rm log}(\Mone) + OV_{\rm sat}\left( c^{\rm rank} - 0.5 \right) + GAB_{\rm sat}\left( b^{\rm rank}_{\rm ind} - 0.5 \right),
\label{Eq:Mmin_gab}
\end{equation}

with $\Acen$ and $\Asat$ the occupational variation-free parameters for the central and satellite galaxies (called $\rm A_{cent}$ and $\rm A_{sat}$ in \citealt{Yuan:2022b}) and $\Bcen$ and $\Bsat$ the occupational variation free parameters for the central and satellite galaxies (called $\rm B_{cent}$ and $\rm B_{sat}$ in \citealt{Yuan:2022b}).

\subsection{Incompleteness}
\label{sec:HOD_GAB}

The last extension we include in our HOD model is incompleteness. Due to several observational limitations, the target galaxy samples may not be complete, meaning a fraction of the galaxies expected from a given selection will not become part of the final sample. To account for this limitation, we randomly remove a fraction of all galaxies ($f_{\rm ic}$) in bins of halo mass:
\begin{equation}
 \langle N^{incompl.}_{\rm cen}(M_{\rm h})\rangle = f_{\rm ic} \langle N_{\rm cen}(M_{\rm h})\rangle,
\end{equation}
\begin{equation}
 \langle N^{incompl.}_{\rm sat}(M_{\rm h})\rangle = f_{\rm ic} \langle N_{\rm sat}(M_{\rm h})\rangle.
\end{equation}
Since we assumed that the mean number of galaxies is equal to the sum of the mean number of centrals and satellites ($\langle N_{tot} \rangle = \langle N_{cen} \rangle + \langle N_{sat} \rangle$) we need to apply this parameter on the centrals and satellites separately. This is different from other approaches, such as the one of \cite{Yuan:2022b}, that assumed ($\langle N_{tot} \rangle = \langle N_{cen} \rangle (1 + \langle N_{sat} \rangle$) and therefore only need to apply this factor to the central galaxies.

\section{Baryonic effects}
\label{sec:bar_effects}

\begin{figure}
\includegraphics[width=0.48\textwidth]{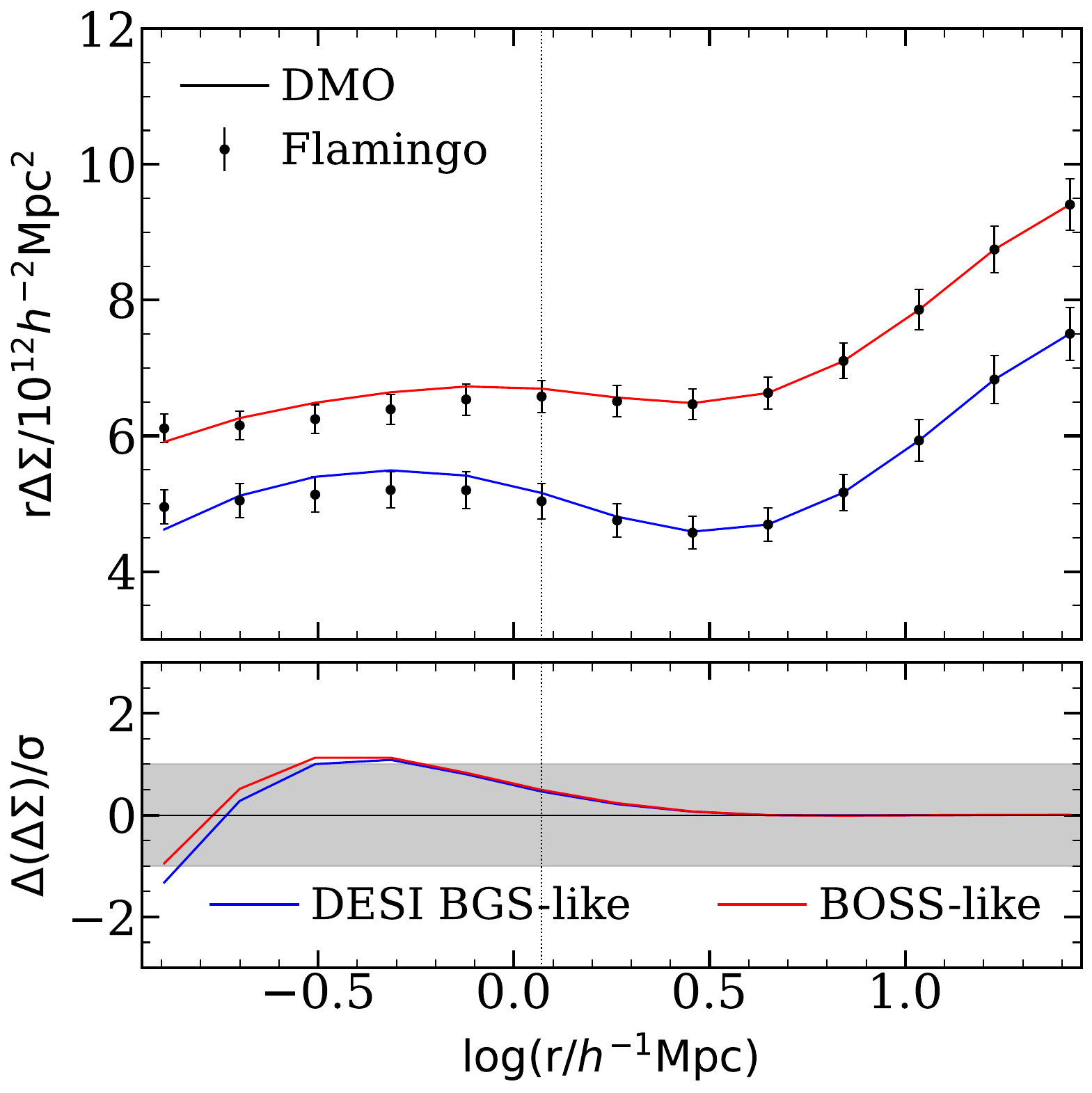}
\caption{The galaxy-galaxy lensing signal for the galaxies of \flamingo~(symbols) and for their matched subhaloes in a dark matter-only simulation (lines). The dashed vertical line indicates the lower limit where we fit the lensing signal with the empirical models. 
}  
\label{fig:ds_bar_ef}
\end{figure}

The effect of baryons on the density profile of haloes impacts galaxy-galaxy lensing, especially at smaller scales. To determine which distances are safe to use in dark matter-only simulations, we estimate the amplitude of the galaxy-galaxy lensing signal with and without baryonic effects. We estimate the lensing signal without baryonic effects by doing a cross-match of the galaxies in the two galaxy samples of \flamingo~with the subhaloes of the dark-matter-only version of this hydrodynamic simulation. This N-body simulation has the same cosmology, output redshift, and initial conditions as the original \flamingo, which allowed us to find the equivalent haloes without the impact of baryonic physics. We look at the corresponding best match bi-directionally, meaning looking at the corresponding pair from one sample to the next and the other way around. We only admit pairs that have a single match and that are consistent regardless of the direction of the matching process. We find a clean match for 95.6\% of the galaxies of the \sampleA~sample and 98.5\% for the \sampleB~sample. The galaxy-galaxy lensing for the matched samples of \flamingo~and the subhaloes of the dark matter-only simulation (DMO) are shown in Fig.~\ref{fig:ds_bar_ef}. We find that a cut of 1 $\hMpc$ ensures that for all the points the impact of the baryonic effect is less than half the size of our error bars. 

\section{Compressing the high-order statistics}
\label{sec:compr}
In this section, we describe the parametrisations done to the high-order statistics to simplify their training in our emulator described in Sect.~\ref{sec:GalaxyEmu}. We characterise most properties using a cumulative skew normal distribution:

\begin{equation}
    \rm \phi^{'}_{cum.}(x, \alpha) =  2\phi(x)\phi_{cum.}(\alpha x),
\end{equation}

\noindent where $\phi$ is a normal function, $\phi_{cum.}$ the cumulative function of a normal distribution and $\alpha$ the parameter that controls how skewed the distribution is. We also tried other more sophisticated fits, including adding kurtosis to the distribution, but we did not find any significant improvements. The way we represent each of these statistics is as follows:

\subsection*{k-Nearest Neighbour}

To describe the kNN-CDF, we used a cumulative skew normal distribution modulated by an additional free parameter:

\begin{equation}
{\rm P(N_{kNN},r)} = \delta \left[1-\phi^{'}_{\rm cum.}\left(\dfrac{{\rm r}-\beta}{\gamma},\alpha\right)\right]
\end{equation}

\noindent with $\alpha$, $\beta$, $\gamma$, \& $\delta$ free parameters of the model. To better capture the cross-correlation for different $N_{kNN}$, we train the emulators using the fitting parameters for all k's. We fit 20 parameters per galaxy sample, similar to the number of bins trained on the projected correlation function or in the multipoles of the correlation function.

\subsection*{Counts-in-cylinders}

To describe the counts-in-cylinder, we again use a cumulative skew normal distribution (without any modulation):

\begin{equation}
{\rm P}(N_{\rm CIC}) = \phi^{'}_{\rm cum.}\left(\dfrac{N_{\rm CIC}-\beta}{\gamma},\alpha\right)
\end{equation}

\noindent with $\alpha$, $\beta$, \& $\gamma$ free parameters. As with the kNN-CDF, we simultaneously trained the counts-in-cylinders for all the radii we trained (for a total of 15 free parameters per galaxy sample).

\subsection*{Void Probability Function}

To represent the void probability function, we use the differential version of the skew-normal distribution ($\phi^{'}$), forcing a value of one for a radius equal to zero. We characterise this distribution as

\begin{equation}
{\rm P_0(r)} = \dfrac{\phi^{'}\left(\dfrac{{\rm r}-\beta}{\gamma},\alpha\right)}{\phi^{'}\left(\dfrac{-\beta}{\gamma},\alpha\right)}
\end{equation}

\noindent with $\alpha$, $\beta$ \& $\gamma$ free parameters of the model.

\subsection*{Galaxy assembly bias}

The galaxy assembly bias  ($\xi/\xi_{\rm shuffled}$) is described as a smooth transition from one (since by construction, the internal distribution of galaxies is preserved during the shuffling, see Sect.~\ref{sec:gab} for more details) to a value equal to the square root of the additional bias signal. Since we mostly care about the amplitude of this ratio at large distances, we fit the galaxy assembly bias signal in a way that we can extract that value directly from the parameters. We, therefore, characterise this signal as 
\begin{equation}
{\xi(r)/\xi_{\rm shuffled}(r)} = 0.5\left(b^{1/2}_{\rm gab}-1\right)\left[{\rm erf}\left(\dfrac{r-\alpha}{\beta}\right) + 1\right]+1
\end{equation}
\noindent with $\rm erf(x)$ the error function described in Appendix~\ref{sec:HOD_extra}, $\alpha$ \& $\beta$ free parameters of the model, and $b^{1/2}_{gab}$ a free parameter equal to the square root of the additional bias of the sample.

\subsection*{Halo occupation number}

To describe the halo occupation number, we used the standard parametrization of satellite galaxies (equ.~\ref{Eq:Sat_HOD}) and an extension of equ.~\ref{Eq:Cen_HOD} for central galaxies, plus the inclusion of the incompleteness parameter $\rm f_{ic}$. We extend over the standard parametrisation of centrals since the \shame~mocks were not perfectly represented by the current parametrisation (something that has already been observed for semi-analytical models \citealt{McCullagh:2017} and hydrodynamic simulations \citealt{Chaves:2023}). We, therefore, replace the error function that describes the transition from zero to one galaxy per halo with a cumulative skew-normal distribution,
\begin{equation}
 \langle N_{\rm cen}(M_{\rm h})\rangle = {\rm f_{ic}} \phi^{'}_{\rm cum.}\left( \frac{\log M_{\rm h} - \log M_{\rm min}}{\sigma_{\log M}},  \sigma_{\log M, 2} \right)
\end{equation}
\noindent with ${\rm f_{ic}}$, $M_{\rm min}$, $\sigma_{\log M}$ and $\sigma_{\log M, 2}$ free parameters (the last one being the level of skewness). While we obtained better fits with this parametrization, we warn the reader about using it to populate mocks when constraining galaxy clustering. The constraining power on the transition from zero to one galaxy per halo is poor \citep{Zehavi:2011} since it has a really low impact on the bias of the sample, and therefore, we do not think this additional complication will benefit constraint better the occupation number from an observational sample.

\section{Additional predictions for the \flamingo~physical variations}
\label{sec:hod_gab_extra}

\begin{figure}
\includegraphics[width=0.45\textwidth]{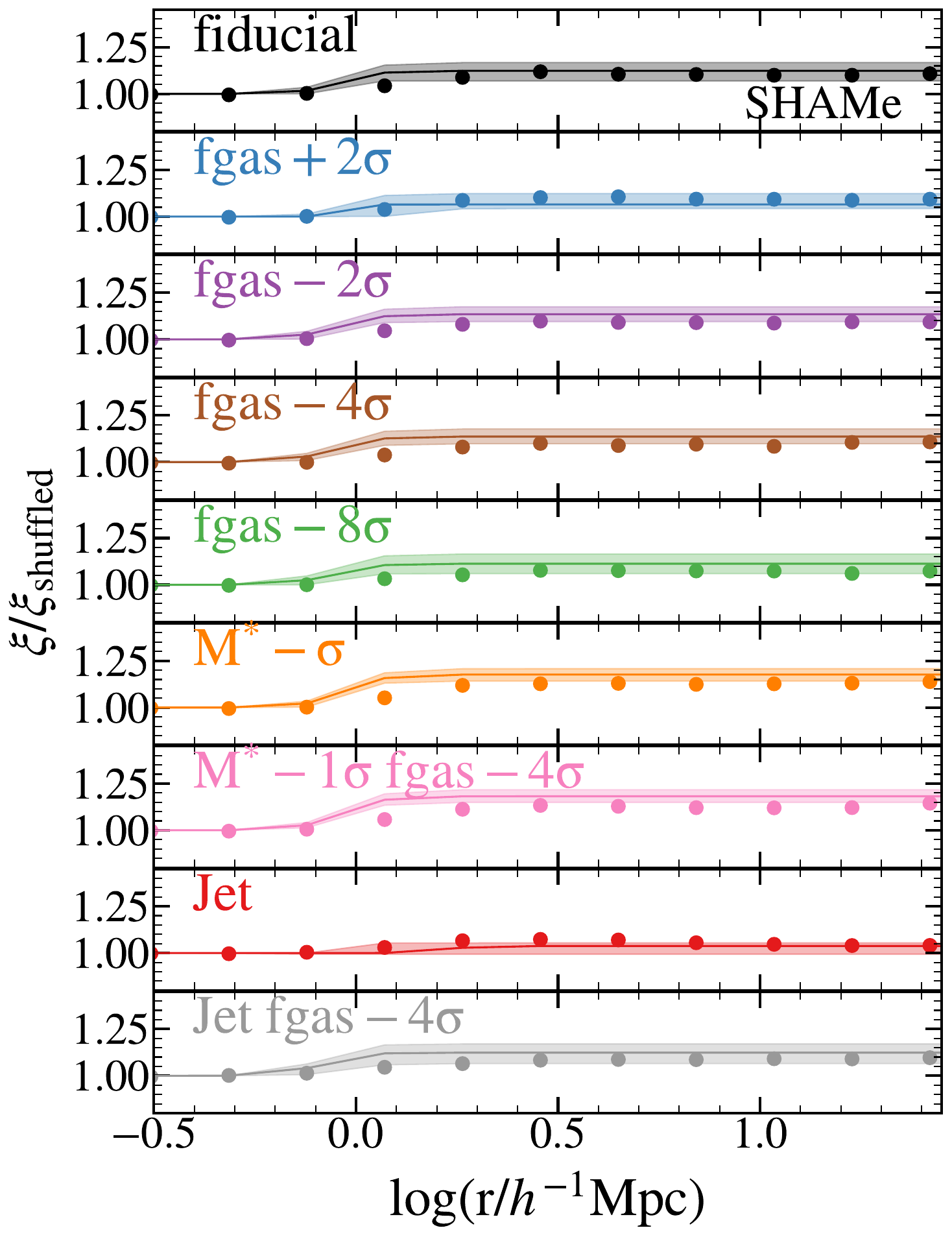}
\includegraphics[width=0.45\textwidth]{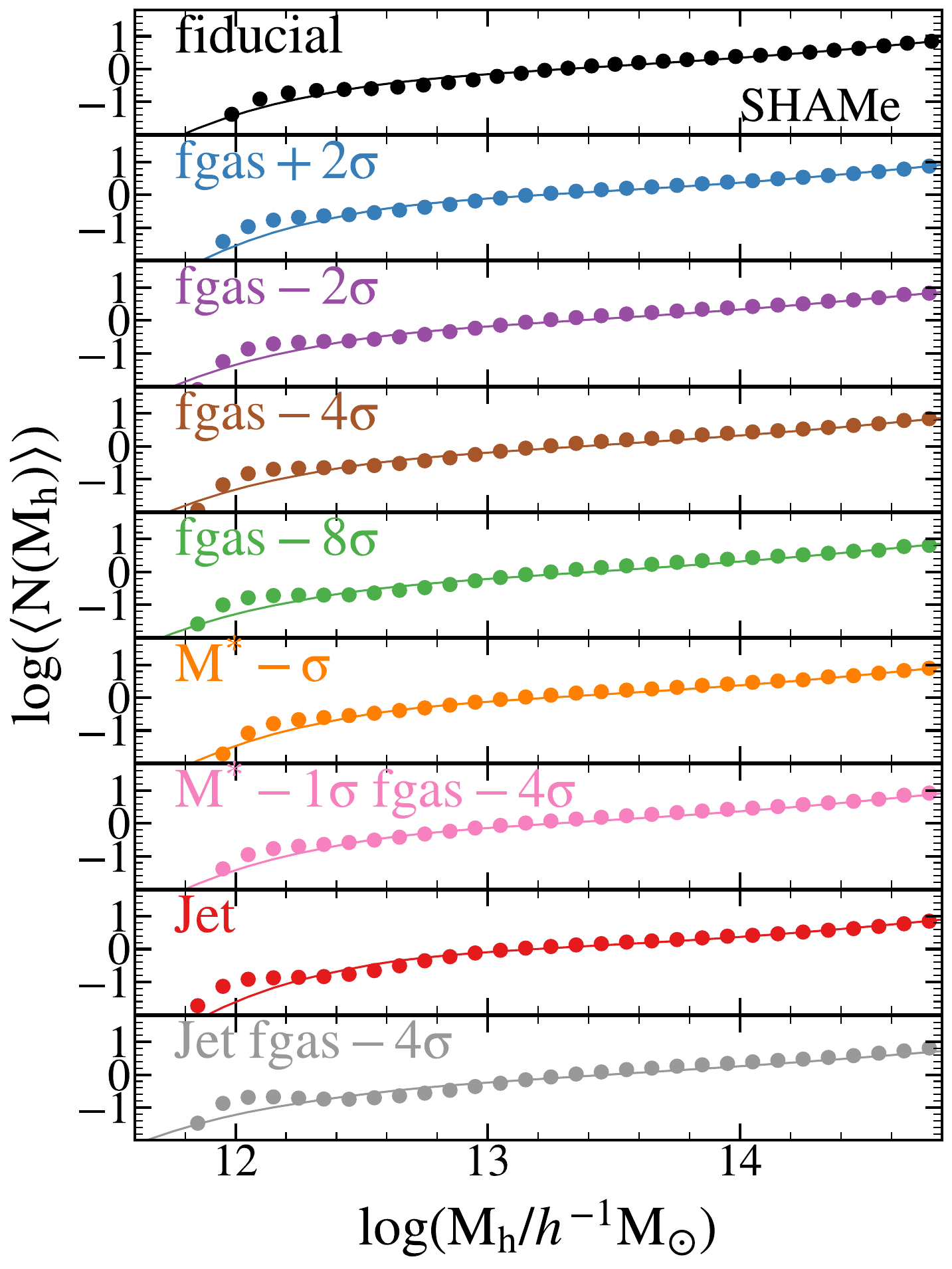}
\caption{Similar to Fig.~\ref{fig:gab_hod}, but for the \flamingo~suite of simulations with different astrophysical implementations and the \shame~model.}  
\label{fig:extra_SHAMe}
\end{figure}

\begin{figure}
\includegraphics[width=0.45\textwidth]{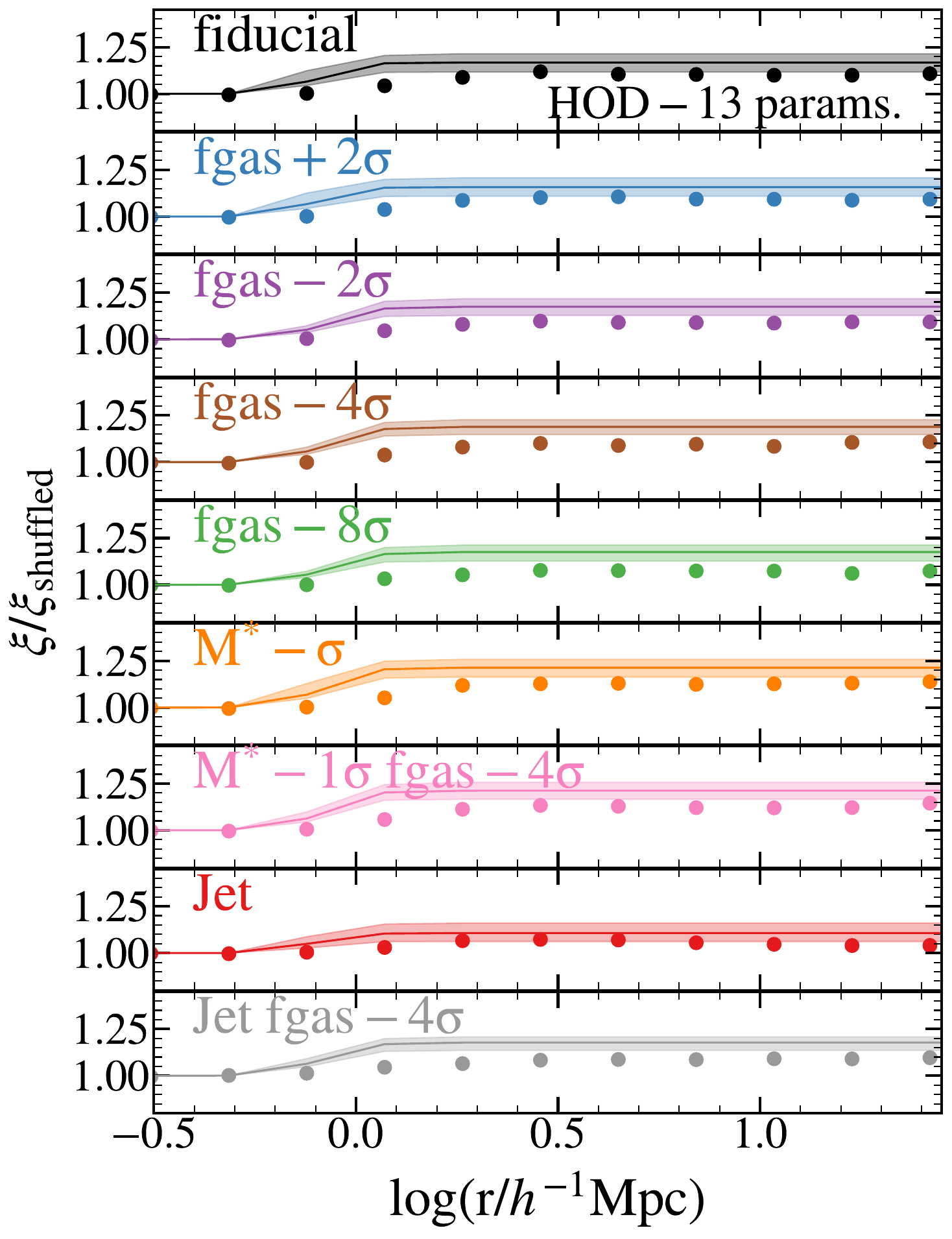}
\includegraphics[width=0.45\textwidth]{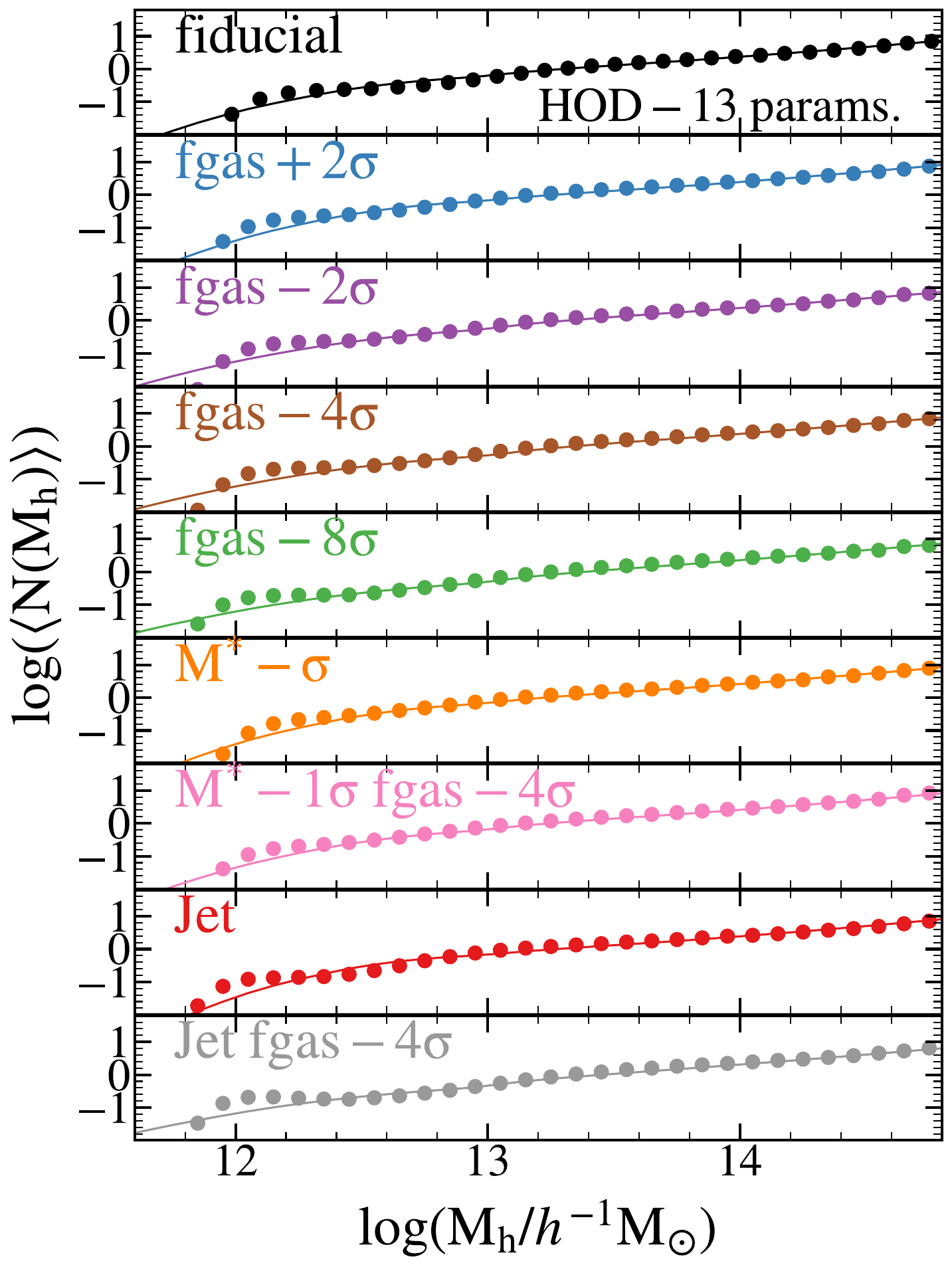}
\caption{Similar to Fig.~\ref{fig:gab_hod}, but for the \flamingo~suite of simulations with different astrophysical implementations and the \hod~model.}  
\label{fig:extra_HOD}
\end{figure}

Different from the rest of the clustering statistics we looked at in this paper, the halo occupation number and the galaxy assembly bias provide a more physical view of the empirical models. In this section, we look into the predictions for these two statistics when fitting only to the galaxy clustering (projected correlation function and multipoles of the correlation function) and galaxy-galaxy lensing for the \flamingo~physical variations. We show these predictions in Fig.~\ref{fig:extra_SHAMe} for the \shame~model and in Fig.~\ref{fig:extra_HOD} for the HOD model. In the case of the halo occupation number, both empirical models recover this statistic reasonably well, especially for the satellite occupation. These results are consistent with our constraints to the largest \flamingo~simulation. For the galaxy assembly bias, the constraints in the \shame~model are more accurate than for the HOD model, where the value is overestimated in all cases but with an overall good performance in both cases. 

For the galaxy assembly bias predictions, we notice its amplitude depends on the physical model of the simulation. This result is consistent with \cite{C21b}, where we found that this amplitude does not depend on cosmology but rather the galaxy formation physics of the model. The good performance of \shame, even when fitting models with really low galaxy assembly bias (such as the ``Jet'' physical model), suggests that it can be successfully used to constrain galaxy assembly bias from the galaxy clustering of the real Universe.

\end{appendix}

\end{document}